\pdfoutput=1
\documentclass[preprint,aps,showpacs,showkeys]{revtex4}
\usepackage{supertabular}
\usepackage{amsfonts}
\usepackage[pdftex]{graphicx}
\usepackage{amsmath}
\usepackage{amsfonts}
\usepackage{amssymb}
\newcommand\spm{\mathrel{\text{\framebox[0.9\width]{$\pm$}}}}
\newcommand\smp{\mathrel{\text{\framebox[0.9\width]{$\mp$}}}}
\newcommand\cpm{\mathrel{\text{\textcircled{\makebox{$\pm$}}}}}
\newcommand\cmp{\mathrel{\text{\textcircled{\makebox{$\mp$}}}}}
\newcommand\sminus{\boxminus}
\usepackage[pdftex]{graphicx}
\usepackage{amsmath}
\usepackage{amsxtra}
\usepackage{amstext}
\usepackage{amssymb}
\usepackage{latexsym}
\usepackage{dsfont}
\usepackage{slashed}

\newcommand{\be}{\begin{equation}}
\newcommand{\ee}{\end{equation}}
\newcommand{\bn}{\begin{eqnarray}}
\newcommand{\en}{\end{eqnarray}}

\usepackage{tikz}
\usetikzlibrary{arrows,snakes}
\begin{document}

\title{
Can spin-charge-family theory explain baryon number non conservation? 
}

\author{Norma Susana Manko\v c  
Bor\v stnik}
\affiliation{Department of Physics, FMF, University of Ljubljana,
Jadranska 19, SI-1000 Ljubljana, Slovenia}


\begin{abstract}
The {\it spin-charge-family} 
theory~\cite{NBled2013,NBled2012,JMP,pikanorma,norma92,norma93,norma94,norma95,gmdn07,gn2013,%
gn2014,gn,NscalarsweakY2014}, in which spinors,  besides the Dirac spin, also carry the second 
kind of the Clifford object - no charges - is a type of  Kaluza-Klein theories~\cite{zelenaknjiga}. 
The Dirac spinors of one Weyl representation in  $d=(13+1)$ manifest~\cite{NBled2013,pikanorma,%
JMP,gn2013,NscalarsweakY2014,HNds} in $d=(3+1)$  at low energies all the properties of quarks 
and leptons assumed by the {\it standard model}. 
The second kind of spin explains the origin of families. Spinors interact with the vielbeins and 
the two kinds of the spin connection fields, the gauge fields of the two kinds of the Clifford 
objects, which, besides the gravity and known gauge vector fields, manifest in $d=(3+1)$ also 
several scalar gauge fields. 
Scalars with the space index $s\in (7,8)$ carry the weak charge and the hyper charge ($\mp \frac{1}{2},
\pm \frac{1}{2}$, respectively), explaining the origin of the Higgs and the Yukawa couplings. 
It is demonstrated in this paper that the scalar fields with the space index $t\in (9,10,\dots,14)$
carry the triplet colour charges, causing transitions of antileptons and 
antiquarks into quarks and back, enabling the appearance and the decay of baryons. These scalar 
fields are  offering in the presence of the right handed neutrino condensate, which breaks the 
${\cal C}{\cal P}$ symmetry, 
the answer to the question about the matter-antimatter asymmetry. 
\end{abstract}

 
\maketitle

\section{Introduction } 
\label{introduction}

The {\it spin-charge-family}~\cite{NBled2013,NBled2012,JMP,pikanorma,norma92,norma93,norma94,%
norma95,gmdn07,gn2013,gn2014,gn,NscalarsweakY2014} theory is offering, as a kind of the Kaluza-Klein 
like theories, the explanation for the charges of quarks and leptons (right handed neutrinos are 
in this theory the regular members of a family) and antiquarks and antileptons~\cite{HNds,TDN2013}, 
and for the existence of the corresponding gauge vector fields. The theory explains, by using 
besides the Dirac kind of the Clifford algebra objects also the second kind of the Clifford 
algebra objects (there are only two kinds~\cite{norma92,norma93,norma94,JMP,NPLB,hn02,hn03,DKhn}, 
associated with the left and the right multiplication of any Clifford object), the origin of 
families of quarks and leptons and correspondingly the origin of the scalar gauge fields  
causing the electroweak break. These scalar fields are responsible, after gaining nonzero vacuum 
expectation values, for the masses and mixing matrices of quarks and leptons~\cite{gmdn07,gn2013,%
gn2014} and for the masses of the weak vector gauge fields. They manifest, carrying the weak 
charge and the hyper charge equal to ($\pm \frac{1}{2} $, $\mp \frac{1}{2} $, respectively)%
~\cite{NscalarsweakY2014},  as the Higgs field and the Yukawa couplings of the {\it standard model}.

The {\it spin-charge-family} theory predicts two  decoupled groups of four families~\cite{JMP,%
pikanorma,gmdn07,gn2013,gn2014}: The fourth family of the lower group is expected to be observed at the 
LHC~\cite{gn2013,gn2014}, while the lowest of the upper four families constitutes the dark 
matter~\cite{gn}. 

This theory also predicts the existence of scalar fields which carry the triplet colour
charges. All the scalars fields carry the fractional quantum numbers with respect to the scalar 
index $s\ge 5$, either the ones of the groups $SU(2)$ or the ones of the group $SU(3)$, while 
they are with respect to the  groups not connected with the space index in the adjoint 
representations. Neither these scalar fields nor the scalars causing 
the electroweak break are the supersymmetric scalar partners of the quarks and leptons, since they 
do not carry all the charges of a family member. 

These scalar fields with the triplet colour charges cause transitions of antileptons into quarks 
and antiquarks into quarks and back, offering, in the presence of the condensate of the two right 
handed neutrinos with the family quantum numbers belonging to the upper four families which breaks the 
CP symmetry, the explanation for the matter-antimatter asymmetry. This is the  topic of the 
present paper.

Let me point out that the {\it spin-charge-family} theory overlaps in many points with other 
unifying theories~\cite{GeorgiGlashow,Georgi,FM,GellMann,BEGN,Zee}, since all the unifying groups  
can be seen as the subgroups of a large enough orthogonal group, with the family  groups included. 
But there are also many differences. While the theories built on chosen groups must for their choice 
propose the Lagrange densities designed for these groups and representations (which also means that there 
must be a theory behind this effective Lagrange densities), the {\it spin-charge-family} theory 
starts with a very simple action, from where all the properties of spinors and the gauge vector and 
scalar fields follow, provided that the breaks of symmetries occur. And all the scalar and vector 
gauge fields, either directly or indirectly, manifest in the low energy regime. 

Consequently this theory differs from other unifying theories in the degrees of freedom of spinors and 
scalar and vector gauge fields which show up on different levels of the break of symmetries,
in the unification scheme, in the family degrees of freedom and correspondingly also in the 
evolution of our universe.

It will be demonstrated in this paper that one condensate of two right handed neutrinos makes all 
the scalar gauge fields and all the vector gauge fields massive on the scale of the appearance of 
the condensate, except those vector gauge fields which appear in the {\it standard model} action before
the electroweak break as massless fields. The scalar gauge fields, which cause the electroweak break
while gaining nonzero vacuum expectation values and changing their masses, then explain masses of 
quarks and leptons and of the weak bosons. 

It is an extremely encouraging fact for this theory, that one scalar condensate and nonzero vacuum 
expectation values of some scalar fields (those with the space index $s=(7,8)$ carrying the weak and 
the hyper charge equal to by the {\it standard model} required charges for the Higgs's scalar) can make a 
simple starting action in $d=(13+1)$ to 
manifest in $d=(3+1)$ in the low energy regime the observed phenomena of fermions and bosons, explaining 
the assumptions of the {\it standard model}. The theory can possibly answer also the open questions, like the 
ones of the appearance of family members, of families, of the dark matter and of the matter-antimatter 
asymmetry.

The paper leaves, however, many a question connected with the break of symmetries open. Although the 
scales of breaks of symmetries can roughly be estimated, for the trustworthy predictions  a careful 
study of the properties of fermions and bosons in the expanding universe is needed. It stays to be 
checked under which conditions in the expanding universe the starting fields (fermions with the 
two kinds of spins and the corresponding vielbeins and the two kind of the spin connection fields) 
manifest after the spontaneous breaks  of symmetries the observed phenomena. This is a 
very demanding study, a first simple step of which was done in the refs.~\cite{gn,DHN}. 
The present paper is a step towards understanding the matter-antimatter  
asymmetry within the {\it spin-charge-family} theory.  

The subsection~\ref{actionandassumptions}  presents the {\it action} and the {\it assumptions} of 
the {\it spin-charge-family} theory, with the comments added. 

In sections~(\ref{gaugefieldsQNlepbartransitions}, \ref{CPNandpropScm}, \ref{masslessandmassivebosons},
\ref{condensate}) the properties of the scalar and vector gauge fields and of the condensate are 
discussed. In appendices 
the discrete symmetries of the {\it spin-charge-family} theory and the technique used for representing 
spinors, with the one Weyl  representation of $SO(13,1)$ and the families in $SO(7,1)$ included, is 
briefly presented. 
The final discussions are presented in sect.~\ref{discussions}.

\subsection{Action of {\it spin-charge-family} theory and  assumptions}
\label{actionandassumptions}

In this subsection  all the assumptions of the {\it spin-charge-family} theory are presented and  
commented. This subsection follows to some  extend a similar subsection of the ref.~\cite{NscalarsweakY2014}.\\  
{\bf i.} $\;\;$ The space-time is $d=(13+1)$ dimensional. Spinors carry besides the internal degrees 
of freedom, determined by the Dirac $\gamma^a$'s operators, also the second kind of the Clifford algebra
operators~\cite{norma92,norma93,norma94,pikanorma}, called $\tilde{\gamma}^a$'s. \\
%
{\bf ii.} $\;\;$
In the simple action~\cite{JMP,NBled2013} fermions $\psi$ carry in $d=(13+1)$ only two kinds of 
spins, no charges,  and {\it interact correspondingly with only the two kinds of the spin connection 
gauge fields, $\omega_{ab \alpha}$ and $\tilde{\omega}_{ab \alpha}$, and the vielbeins, 
$f^{\alpha}{}_{a}$}.
\begin{eqnarray}
S            \,  &=& \int \; d^dx \; E\;{\mathcal L}_{f} +  
\nonumber\\  
               & & \int \; d^dx \; E\; (\alpha \,R + \tilde{\alpha} \, \tilde{R})\,,\nonumber\\
%
{\mathcal L}_f &=& \frac{1}{2}\, (\bar{\psi} \, \gamma^a p_{0a} \psi) + h.c., 
\nonumber\\
p_{0a }        &=& f^{\alpha}{}_a p_{0\alpha} + \frac{1}{2E}\, \{ p_{\alpha}, E f^{\alpha}{}_a\}_-, 
\nonumber\\  
   p_{0\alpha} &=&  p_{\alpha}  - 
                    \frac{1}{2}  S^{ab} \omega_{ab \alpha} - 
                    \frac{1}{2}  \tilde{S}^{ab}   \tilde{\omega}_{ab \alpha},                   
\nonumber\\ 
R              &=&  \frac{1}{2} \, \{ f^{\alpha [ a} f^{\beta b ]} \;(\omega_{a b \alpha, \beta} 
- \omega_{c a \alpha}\,\omega^{c}{}_{b \beta}) \} + h.c. \;, 
\nonumber\\
\tilde{R}      &=& \frac{1}{2}\,   f^{\alpha [ a} f^{\beta b ]} \;(\tilde{\omega}_{a b \alpha,\beta} - 
\tilde{\omega}_{c a \alpha} \tilde{\omega}^{c}{}_{b \beta}) + h.c.\;. 
\label{wholeaction}
\end{eqnarray}
Here~\footnote{$f^{\alpha}{}_{a}$ are inverted vielbeins to 
$e^{a}{}_{\alpha}$ with the properties $e^a{}_{\alpha} f^{\alpha}{\!}_b = \delta^a{\!}_b,\; 
e^a{\!}_{\alpha} f^{\beta}{\!}_a = \delta^{\beta}_{\alpha} $, $ E = \det(e^a{\!}_{\alpha}) $. 
Latin indices  
$a,b,..,m,n,..,s,t,..$ denote a tangent space (a flat index),
while Greek indices $\alpha, \beta,..,\mu, \nu,.. \sigma,\tau, ..$ denote an Einstein 
index (a curved index). Letters  from the beginning of both the alphabets
indicate a general index ($a,b,c,..$   and $\alpha, \beta, \gamma,.. $ ), 
from the middle of both the alphabets   
the observed dimensions $0,1,2,3$ ($m,n,..$ and $\mu,\nu,..$), indices from 
the bottom of the alphabets
indicate the compactified dimensions ($s,t,..$ and $\sigma,\tau,..$). 
We assume the signature $\eta^{ab} =
diag\{1,-1,-1,\cdots,-1\}$.} 
$f^{\alpha [a} f^{\beta b]}= f^{\alpha a} f^{\beta b} - f^{\alpha b} f^{\beta a}$.
$S^{ab}$ and $\tilde{S}^{ab}$ are generators~(Eqs.(\ref{sabsc}, \ref{sabtildesab})) 
of the groups $SO(13,1)$ and $\widetilde{SO}(13,1)$, 
respectively, expressible by $\gamma^a$ and $\tilde{\gamma}^a$.\\ 
{\bf iii.} $\;\;$
The manifold $M^{(13+1)}$ breaks first into $M^{(7+1)}$ times $M^{(6)}$ (which manifests as 
$SU(3)$ $\times U(1)$), affecting both internal degrees of freedom, $SO(13+1)$ and 
$\widetilde{SO}(13+1)$.  After this break there are $2^{((7+1)/2-1)}$ massless families,  all the 
rest families get heavy masses~\footnote{A toy model~\cite{DHN,DN012,HNds} 
was studied in $d=(5+1)$ with the same action as in Eq.`(\ref{wholeaction}). For a particular 
choice of vielbeins and for a class of spin connection fields the manifold $M^{5+1}$ breaks into 
$M^{(3+1)}$ times an almost $S^2$, while $2^{((3+1)/2-1)}$ families stay massless and mass protected. 
Equivalent assumption, although not yet proved that it really works, is made also in the case that  
 $M^{(13+1)}$ breaks first into $M^{(7+1)}$ $\times$ $M^{(6)}$. The study is in progress.}. \\
Both internal degrees of freedom, the ordinary $SO(13+1)$  one (where $\gamma^a$ determine 
spins and charges of spinors) and  the $\widetilde{SO}(13+1)$ (where $\tilde{\gamma}^a$ determine 
family quantum numbers),  break simultaneously with the manifolds. \\
{\bf iv.} $\;\;$ 
There are additional breaks of symmetry: The manifold $M^{(7+1)}$ breaks further into 
$M^{(3+1)} \times$  $M^{(4)}$. 
 \\
{\bf v.} $\;\;$ 
There is a scalar condensate of two right handed neutrinos with the family quantum numbers of 
the upper four families, bringing masses of the scale above the unification scale, to all the 
vector and scalar gauge fields, which interact with the condensate. \\  
{\bf vi.} $\;\;$ 
There are nonzero vacuum expectation values of the scalar fields with the scalar indices $(7,8)$, 
which cause the electroweak break and bring masses to the fermions and weak gauge bosons, 
conserving the electromagnetic and colour charge.\\

{\it Comments on the assumptions:

{\bf i.}}:  $\;\;$
There are, as already written above, two (only two) kinds of the Clifford algebra objects. The 
Dirac one~(Eq.(\ref{gammatildegamma})), $\,\gamma^a$, will be used to describe spins of spinors 
(fermions) in $d=(13+1)$, manifesting in $d=(3+1)$ the spin and all the fermion charges, 
the second one~(Eq.(\ref{gammatildegamma})), $\,\tilde{\gamma}^a$, will describe families of spinors.
The representations of $\gamma^a$'s and $\tilde{\gamma}^a$'s are orthogonal to one 
another~\footnote{One can learn in Eq.~(\ref{snmb:gammatildegamma})of appendix~(\ref{technique})
that $S^{ab}$ transforms one state of the representation into another state of the same 
representation, while $\tilde{S}^{ab}$ transforms the state into the state belonging to another
representation.}.
There are correspondingly two groups determining internal degrees of freedom of spinors  
in $d=(13+1)$: The Lorentz group $SO(13,1)$ and the group $\widetilde{SO}(13,1)$. \\
One Weyl representation of $SO(13,1)$ contains, if analysed~\cite{NBled2013,JMP,pikanorma,HNds} 
with respect to the {\it standard model} groups, all the family members, assumed by the {\it 
standard model}, with the right handed neutrinos included (the family members are presented in 
table~\ref{Table so13+1.}). It contains 
the left handed weak ($SU(2)_{I}$) charged and $SU(2)_{II}$ chargeless colour triplet quarks 
and colourless leptons (neutrinos and electrons), the right handed weak chargeless and $SU(2)_{II}$ 
charged quarks and leptons, as well as the right handed weak charged and $SU(2)_{II}$ 
chargeless colour antitriplet antiquarks and (anti)colourless antileptons, and the left handed 
weak chargeless and $SU(2)_{II}$ charged antiquarks and antileptons. The reader can easily check the 
properties of the representations  of spinors (table~\ref{Table so13+1.}), presented in the "technique"  
(appendix~\ref{technique}) way, if using Eqs.~(\ref{sabsc}, \ref{so1+3}, \ref{so42}, \ref{so64},
\ref{YQY'Q'andtilde}).\\
Each family member carries the family quantum numbers, 
originating in 
$\tilde{\gamma}^a$'s degrees of freedom. Correspondingly,  $\tilde{S}^{ab}$ change the family 
quantum numbers, leaving the family member quantum number unchanged.\\
{\bf ii.}:  $\;\;$ 
This starting action enables to represent the {\it standard model} as an effective low energy 
manifestation of the {\it spin-charge-family} theory, which explains all the {\it standard model} 
assumptions, with the families included. There are gauge vector fields, massless before the 
electroweak break: gravity, colour $SU(3)$  
octet vector gauge fields,  weak $SU(2)$ (it will be named $SU(2)_{I}$) triplet vector gauge field 
and "hyper" $U(1)$ (it will be named $U(1)_{I}$) 
singlet vector gauge fields.  All are superposition of $f^{\alpha}{}_{c}$ $\omega_{ab \alpha}$.  
There are (eight  rather than the observed three) families of quarks and leptons, massless before the 
electroweak break.  
These eight families are indeed two decoupled groups of four families, each of them in the fundamental 
representations with respect to  
$\widetilde{SU}(2)\times$ $ \widetilde{SU}(2)$ groups - the subgroups of $\widetilde{SO}(3,1)\times $  
$\widetilde{SO}(4)\in \widetilde{SO}(7,1)$.\\
The scalar gauge fields, determining the mass matrices 
of quarks and leptons, carry the scalar index $s \in (7,8)$ and correspondingly  the weak and the hyper 
charge of the scalar Higgs (sect.~\ref{CPNandpropScm}). Those among them which are superposition 
of $f^{\sigma}{}_{s}$ $\tilde{\omega}_{ab \sigma}$ carry besides the weak and the hyper charges two 
kinds of the family quantum numbers in the adjoint representations, representing two orthogonal groups: 
Each  of  the two groups contains  two triplets with respect to $\widetilde{SU}(2)_{\widetilde{SO}(3,1)}$ 
$\times$ $ \widetilde{SU}(2)_{\widetilde{SO}(4)}$) (generators of these subgroups are presented in 
Eqs.~(\ref{so1+3tilde}, \ref{so42tilde})).
The three singlet scalar fields with the space index $s=(7,8)$ and carrying the quantum numbers 
($Q,Q',Y'$) are the superposition of $f^{\sigma}{}_{s}$ $\omega_{ab\sigma}$. They again carry  
the weak and the hyper charge of the scalar Higgs. \\
One group of two triplets together with the three singlets determine, after gaining nonzero vacuum 
expectation values
at the electroweak break, the Higgs's scalar and the Yukawa couplings of the {\it standard model}.
The second group of two triplets, the three singlets and the condensate determine at the electroweak 
break  masses of the upper four families, the stable family of which is the candidate for forming the 
dark matter.\\
The starting action contains also an additional $SU(2)_{II}$ (from $SO(4)$) vector gauge field and 
the scalar fields with the 
space index $s\in (5,6)$ and  $t\in (9,10,11,12)$, as well as the auxiliary vector gauge fields 
expressible (Eqs.~(\ref{omegatildeabalpha}, \ref{omegaabalpha}) in the appendix~\ref{auxiliary})
with vielbeins. They all remain either auxiliary (if there are no spinor sources manifesting their 
quantum numbers) or become massive after the appearance of the condensate.\\ 
{\bf iii., iv.}: $\;\;$ 
The assumed break from $M^{(13+1)}$ first into $M^{(7+1)}$ times $M^{(6)}$ (manifesting the symmetry  
$SU(3)$ $\times U(1)_{II}$) explains why the left handed members of a family carry the weak charge 
while the right handed members do not: In the spinor representation of $SO(7,1)$ there are left handed 
weak charged 
quarks and leptons with the hyper charges ($\frac{1}{6}$, $-\frac{1}{2}$, respectively) and the right 
handed weak chargeless quarks  with the hyper charge either $\frac{2}{3}$ or $-\frac{1}{3}$, while the 
right handed weak chargeless leptons carry the hyper charge either equal to $0$ or to ($-1$). 
A further break from $M^{(7+1)}$ into $M^{(3+1)}$ $\times$ $M^{(4)}$ manifests  the symmetry 
$SO(3,1)$ $\times SU(2)_{I} \times SU(2)_{II}$ $\times U(1)_{II}\times $ $SU(3)$,  explaining the 
observed properties of the family members: There are the coloured quarks, left handed weak charged and 
$SU(2)_{II}$ chargeless and right handed weak chargeless and $SU(2)_{II}$ charged, and there are
colourless leptons, again left handed weak charged and $SU(2)_{II}$ chargeless and right handed weak 
chargeless and $SU(2)_{II}$ charged. Quarks carry the "spinor" charge $\frac{1}{6}$, leptons carry the 
"spinor" charge $-\frac{1}{2}$. There are the observed vector gauge fields with the corresponding 
charges in the adjoint representation and there are vector gauge fields which gain masses through 
the interaction with the condensate and are unobservable at low energies.
There are the scalar fields manifesting so far as the Higgs's scalar and Yukawa couplings,
and additional scalar fields, which through interaction with the condensate become massive.\\
Since the left handed members distinguish from the right handed partners in the weak and the hyper charges, 
the family members of all the families stay massless and mass protected up to the electroweak 
break~\footnote{As long as the 
left handed family members and their right handed partners carry different conserved charges, they can not 
behave as massive particles - they are mass protected. It is the appearance of nonzero vacuum expectation 
values  of the scalar fields, carrying the weak and the hyper charge, which cause non conservation of 
these two charges, what enables the superposition of the left and the right handed family members, 
breaking  the mass protection.}. 
Antiparticles are accessible from particles by the $\mathbb{C}_{{\cal N}}$  $\cdot {\cal P}_{{\cal N}}$, 
as explained in refs.~\cite{HNds,TDN2013} and briefly also in the appendix~\ref{discrete}. This discrete 
symmetry operator does not contain $\tilde{\gamma}^a$'s
degrees of freedom. To each family member there corresponds the antimember, with the same 
family quantum numbers.\\
{\bf v.}: $\;\;$ 
It is a condensate of two right handed neutrinos with the quantum numbers of the upper four 
families (table~\ref{Table con.}), appearing in the energy region above the unification scale 
($\ge 10^{16}$ GeV), 
which makes all the scalar gauge fields (those with the space index ($5,6,7,8$), as well as those 
with the space index $(9,\dots,14)$) and the vector gauge fields, manifesting nonzero quantum numbers 
$\tau^{4}$, $\tau^{23}$, 
$Q$ ,$Y$, $\tilde{\tau}^{4}$, $\tilde{\tau}^{23}$, $\tilde{Q}$ ,$\tilde{Y}$,$\tilde{N}^{3}_{R}$ 
(Eqs.~(\ref{so1+3}, \ref{so42}, \ref{so64}, \ref{so1+3tilde}, \ref{so42tilde}, \ref{YQY'Q'andtilde}))
massive. \\
{\bf vi.}: $\;\;$ 
At the electroweak break the scalar fields with the space index $s=(7,8)$ - twice three triplets,  
 the superposition of $\tilde{\omega}_{abs}$, Eq.~(\ref{Atildeomegas}), and the 
three singlets, the superposition of $\omega_{ts's}$, Eq.~(\ref{Aomegas}), carrying the charges 
($Q,Q',Y'$), all these scalars have the weak and the hyper charge equal to ($\mp \frac{1}{2}$, 
$\pm \frac{1}{2}$, respectively) -  get nonzero vacuum 
expectation values, changing also their own masses and breaking the weak and the hyper charge symmetry.
 These scalars determine mass matrices of twice four families, as well as the masses of the weak 
bosons.\\
The fourth family belonging to the observed three will (sooner or later) be observed at the LHC. 
Its properties are under consideration~\cite{gn2013,gn2014}, while the stable of the upper four 
families is the candidate for the dark matter constituents. 
\\

The above assumptions enable that the starting action (Eq.~(\ref{wholeaction})) 
manifests  effectively in $d=(3+1)$ in the low energy regime fermion and boson fields 
as assumed by the {\it standard model}. The starting action offers the explanation also for 
the dark matter content and for the matter-antimatter asymmetry in the universe.

To see that the action~(Eq.(\ref{wholeaction})) really manifests  in $d=(3+1)$ by the {\it standard 
model} required degrees of freedom of fermions and bosons~\cite{JMP,NBled2013,pikanorma,norma92,%
norma93,norma94,norma95,NBled2012,gmdn07,gn2013,gn,NscalarsweakY2014}, 
let us formally rewrite the Lagrange density for a Weyl spinor of~(Eq.(\ref{wholeaction})), 
which includes also families, as follows  
\begin{eqnarray}
{\mathcal L}_f &=&  \bar{\psi}\gamma^{m} (p_{m}- \sum_{Ai}\; g^{A}\tau^{Ai} A^{Ai}_{m}) \psi 
+ \nonumber\\
               & &  \{ \sum_{s=7,8}\;  \bar{\psi} \gamma^{s} p_{0s} \; \psi \} 
+ \nonumber\\ 
& & \{ \sum_{t=5,6,9,\dots, 14}\;  \bar{\psi} \gamma^{t} p_{0t} \; \psi \}\,,
\nonumber\\
p_{0s} &=&  p_{s}  - \frac{1}{2}  S^{s' s"} \omega_{s' s" s} - 
                    \frac{1}{2}  \tilde{S}^{ab}   \tilde{\omega}_{ab s}\,,\nonumber\\
p_{0t} &=&  p_{t}  - \frac{1}{2}  S^{t' t"} \omega_{t' t" t} 
                   - \frac{1}{2}  \tilde{S}^{ab}   \tilde{\omega}_{ab t}\,,                    
\label{faction}
\end{eqnarray}
where $ m \in (0,1,2,3)$, $s \in {7,8},\, (s',s") \in (5,6,7,8)$, $(a,b)$ (appearing in $\tilde{S}^{ab}$) 
run within $\in (0,1,2,3)$ and $\in (5,6,7,8)$, $t \in (5,6,9,\dots,13,14)$, $(t',t") \in (5,6,7,8)$ and 
$\in (9,10,\dots,14)$.  
$\psi$ represents all family members of all the families.
The generators of the charge groups $\tau^{Ai} $ (expressed in Eqs.~(\ref{tau}), (\ref{so42}), 
(\ref{so64}) in terms of $S^{ab}$) fulfil the commutation relations 
%
\begin{eqnarray}
\tau^{Ai} = \sum_{a,b} \;c^{Ai}{ }_{ab} \; S^{ab}\,,
\nonumber\\ 
\{\tau^{Ai}, \tau^{Bj}\}_- = i \delta^{AB} f^{Aijk} \tau^{Ak}\,.
\label{tau}
\end{eqnarray}
%
The spin generators are  defined in Eq.~(\ref{so1+3}).  These group generators determine all the 
internal degrees of freedom of one family members as seen from the 
point of view of $d=(3+1)$: The spin is determined by the group $SO(3,1)$, the colour charge is determined 
by the group $SU(3)$ (originating in $SO(6)$) and with the generators $\vec{\tau}^{3}$, the "spinor charge" 
is determined by $U(1)_{II}$ (originating in $SO(6)$) and with the generator $\tau^{4}$, the weak charge 
is determined by the group $SU(2)_{I}$ (originating in $SO(4)$) and with the generators $\vec{\tau}^{1}$, 
and the second $SU(2)_{II}$ charge ($SU(2)_{II}$ originating in $SO(4)$) has the generators 
$\vec{\tau}^{2}$. The group $SU(2)_{II}$ breaks~\cite{JMP} in the presence of the condensate into 
$ U(1)_{I}$. The generators $\tau^{23}$ define together with $\tau^{4}$  the hyper charge $Y $ 
$(= \tau^{23}+\tau^{4}$). 

The condensate of two right handed neutrinos with the family quantum numbers of the upper four families 
bring masses (of the unifying scale $\ge 10^{16}$ GeV) to all of the scalar gauge fields and those 
vector gauge fields which are not observed at currently measurable energies, since the observed vectors do not 
couple to the condensate. 

The scalar fields  with the scalar index $s=(7,8)$ bring masses, when gaining nonzero vacuum expectation 
values at the electroweak phase transition, to twice four families and to the weak 
bosons. 
I shall comment all these fields in what follows.

The first line of Eq.~(\ref{faction}) describes~\cite{NBled2013,JMP} before the electroweak break 
the dynamics of eight families of massless fermions in interaction with the massless colour 
$\vec{A}^{3}_{m}$, weak $\vec{A}^{1}_{m}$ 
and hyper $A^{Y}_{m}$ ($=\sin \vartheta_{2}\,A^{23}_{m} +\cos \vartheta_{2}\,A^{4}_{m}$) gauge 
fields, all are the superposition of $\omega_{abm}$~\footnote{These superposition can easily be found
by using Eqs.~(\ref{so64}, \ref{so42}). They are explicitly written in the ref.~\cite{JMP}. The 
interaction with the condensate makes the fields $A^{Y'}_{m}$, Eq.~(\ref{YQY'Q'andtilde}), $A^{21}_{m}$ 
and $A^{22}_{m}$ very  massive (at the scale of the condensate).}.

The second line of the same equation (Eq.~(\ref{faction})) determines the mass term, which 
after the electroweak break brings masses to all the family members of the eight families and 
to the weak bosons. 
The scalar fields responsible - after gaining nonzero vacuum expectation values - for masses of 
the family members and of the weak bosons are namely included in the second line of Eq.~(\ref{faction}) 
as ($-\frac{1}{2} S^{s's"} \omega_{s's" s}- $ $ \frac{1}{2}  \tilde{S}^{\tilde{a}\tilde{b}} $ 
$\tilde{\omega}_{\tilde{a}\tilde{b} s}$, $\,s\in(7,8)\,$, $(s',s") \in(5,6,7,8)$, 
$(\tilde{a},\tilde{b}) \in(\tilde{0},\tilde{1}, \dots,\tilde{8})$~\footnote{To point out that $S^{ab}$
and $\tilde{S}^{ab}$ belong to two different kinds of the Clifford algebra objects are   
the indices $(a,b)$ in $\tilde{S}^{ab}$ in this paragraph written as $(\tilde{a},\tilde{b})$. 
Normally only $(a,b)$ will be used for $S^{ab}$ and $\tilde{S}^{ab}$.}). 

The properties of these scalar fields  with the scalar index $s=(7,8)$ are discussed in 
sect.~(\ref{CPNandpropScm}), where the 
proof is presented that they all carry the weak charge and the hyper charge as the 
{\it standard model} Higgs's scalar, while they are  either triplets with respect to the family quantum 
numbers or singlets with respect to the charges $Q,Q'$ and $Y'$. While the two triplets 
($\vec{\tilde{A}}^{1}_{s}$, $\vec{\tilde{A}}^{\tilde{N}_{L}}_{s}$) interact with the lower four 
families, ($\vec{\tilde{A}}^{2}_{s}$, $\vec{\tilde{A}}^{\tilde{N}_{R}}_{s}$) interact with the upper 
four families. These twice two triplets are superposition of $\frac{1}{2}\,\tilde{S}^{\tilde{a}\tilde{b}}$ 
$\tilde{\omega}_{\tilde{a}\tilde{b} s}$, $\,s=\in(7,8)$, Eq.~(\ref{Atildeomegas}).
The three singlets ($A^{Q}_{s}$, $A^{Q'}_{s}$ and $A^{Y'}_{m}$) are superposition of $\omega_{s's"s}$,
Eq.(\ref{Aomegas}). They interact with the family members of all the families, "seeing"  charges of the family 
members.

The third line of Eq.~(\ref{faction}) represents fermions in interaction with all the rest scalar fields.
Scalar fields become massive after interacting with the condensate. Those, which do not gain nonzero vacuum 
expectation values, keep the heavy  masses of the order of the scale of the condensate up to low energies. 

The massive scalars with the space index $t\in(5,6)$ transform (table~\ref{Table so13+1.}) $u_{R}$-quarks 
into $d_{L}$-quarks and 
$\nu_{R}$-leptons into $e_{L}$-leptons and back, as well as $\bar{u}_{R}$-antiquarks into $\bar{d}_{L}$-%
antiquarks and back and $\bar{\nu}_{R}$-antileptons into $\bar{e}_{L}$-antileptons and back.

Those scalar fields with the space index $t = (9,10,\cdots,14)$ transform antileptons into quarks and 
antiquarks into quarks and back. They are offering, in the presence of the scalar condensate breaking 
the ${\cal C} {\cal P}$ symmetry, the explanation for the observed matter-antimatter asymmetry, as we 
shall show in sect.~\ref{gaugefieldsQNlepbartransitions}. 

Let us write down the third line of Eq.~(\ref{faction}), the part of the fermion action  which in the 
presence of the condensate offers the explanation for the observed matter-antimatter asymmetry.
\begin{eqnarray}
{\mathcal L}_{f'}&=&  \psi^{\dagger} \,\gamma^0 \,\gamma^{t}\,\{ \sum_{t=(9,10,\dots 14)}\,  
 \Large{[}p_{t} - (\,\frac{1}{2}\, S^{s' s"} \,\omega_{s' s" t}+ \,
 \frac{1}{2} \, S^{t' t''} \,\omega_{t' t" t}  \nonumber\\
 &+&\frac{1}{2} \,\tilde{S}^{ab}\, \tilde{\omega}_{ab t}\,)  
\Large{]} \} \, \; \psi\,,
\label{factionMaM0}
\end{eqnarray}
where  
$(s',s") \in (5,6,7,8)$, $(t, t',t") \in (9,10,\dots,14)$ and $(a,b) \in (0,1,2,3)$ and 
$ \in(5,6,7,8)$, in agreement with the assumed breaks in sect.~\ref{introduction}.
 Again the operators $\tilde{S}^{ab}$ determine family quantum numbers  and $S^{ab}$ determine 
 family members  quantum numbers.  
Correspondingly the superposition of the scalar fields $ \tilde{\omega}_{abt}$ and the  
superposition of the scalar fields $ \omega_{abt}$ carry the quantum numbers determined by 
either the superposition of $\tilde{S}^{ab}$ or by the superposition of $S^{ab}$ in the adjoint 
representations, while they all carry  the 
colour charge, determined by the space index $t \in (9,10,\dots,14)$, in the triplet 
representation of the $SU(3)$ group, as we shall see. Similarly the scalars 
with the space index $s \in (7,8)$ carry the weak and the hyper charge in the doublets 
representations~\footnote{Although  there are three  scalar fields, each with the colour charge of 
one of the triplet representation and there are also two scalar fields, each with the 
weak charge of one of the doublet representation,  neither a triplet nor a doublet 
has all the other quantum numbers equal
to any of the observed spinors to be a candidate for its supersymmetric partner.}.

The condensate  of two right handed neutrinos with the family quantum numbers of the upper 
four families carries (table~\ref{Table con.}) $\tau^{4}=1$, $\tau^{23}=-1$, $\tau^{13}=0$, $Y=0$, 
$Q=0$, and the family quantum numbers of the upper four families and gives masses to scalar and 
vector gauge fields with the nonzero corresponding quantum numbers. The only vector gauge fields 
which stay massless up to the electroweak break are the hyper charge field ($A^{Y}_m$), the 
weak charge field ($\vec{A}^{1}_m$) and the colour charge field ($\vec{A}^{3}_m$) (besides the gravity).

\subsubsection{{\it Standard model} subgroups of  $SO(13+1)$ and $\widetilde{SO}(13+1)$
groups  and corresponding gauge fields}
\label{generatorsandfields}

This section follows  the refs.~\cite{JMP,NscalarsweakY2014}.
To calculate quantum numbers of one Weyl representation presented in table~\ref{Table so13+1.} 
in terms of the generators of the {\it standard model} charge groups $\tau^{Ai}$ ($ = \sum_{a,b} $ 
$c^{Ai}{ }_{ab} \; S^{ab}$) one must look for the coefficients $c^{Ai}{ }_{ab} $ (Eq.~(\ref{tau})). 
Similarly also the spin and the family degrees of freedom have to be expressed as superposition of 
$S^{ab}$, or $\tilde{S}^{ab}$. 

The same coefficients $c^{Ai}{ }_{ab}$ determine operators which apply either on spinors or on vectors.  
The difference among the three kinds of operators - vector and two kinds of spinor  operators - lies in the 
difference among $S^{ab}$, $\tilde{S}^{ab}$  and ${\cal S}^{ab}$. 

While $S^{ab}$ for  spins of spinors is equal to
\begin{eqnarray}
\label{sabsc}
S^{ab} &=& \,\frac{i}{4} (\gamma^a\, \gamma^b - \gamma^b\, \gamma^a)\,,\quad 
\{\gamma^a\,, \gamma^b\}_{+} = 2 \eta^{ab}\,,
\end{eqnarray}
and $\tilde{S}^{ab}$ for families of spinors is equal to 
\begin{eqnarray}
\label{tildesabsc}
\tilde{S}^{ab} &=& \,\frac{i}{4} (\tilde{\gamma}^a\, \tilde{\gamma}^b - 
\tilde{\gamma}^b\, \tilde{\gamma}^a)\,,
\quad \{\tilde{\gamma}^a\,, \tilde{\gamma}^b\}_{+} = 2 \eta^{ab}\,,\nonumber\\
& & \{\gamma^a\,, \tilde{\gamma}^b\}_{+} = 0\,,
\end{eqnarray}
one must take, when ${\cal S}^{ab}$ apply on the spin connections $\omega_{bd e}$ ($= f^{\alpha}{}_{e}\, 
$ $\omega_{bd \alpha}$) and $\tilde{\omega}_{\tilde{b} \tilde{d} e}$ ($= f^{\alpha}{}_{e}\, 
$ $\tilde{\omega}_{\tilde{b} \tilde{d} \alpha}$), on either the space index $e$ or the indices 
$(b,d,\tilde{b},\tilde{d})$, the operator
\begin{eqnarray}
\label{bosonspin}
({\cal S}^{ab})^{c}{}_{e}A^{d\dots e \dots g} &=& i(\eta^{ac}\delta^{b}_{e}- \eta^{bc}\delta^{a}_{e})\, 
A^{d\dots e \dots g}\,.
\end{eqnarray}
This means that the space index ($e$) of $\omega_{bd e}$ transforms according to the requirement 
of Eq.~(\ref{bosonspin}), and so do   ($b,d$) and ($\tilde{b},\tilde{d}$). Here I used again the notation 
($\tilde{b},\tilde{d}$) to point out that $S^{ab}$ and $\tilde{S}^{ab}$ 
($ = \tilde{S}^{\tilde{a} \tilde{b}}$) are the generators of two independent groups\cite{NscalarsweakY2014}.

One finds~\cite{NBled2013,JMP,pikanorma,norma92,norma93,norma94,norma95,NBled2012} for the 
generators of the spin and the charge groups, which are the subgroups of $SO(13,1)$, the expressions: 
\begin{eqnarray}
\label{so1+3}
\vec{N}_{\pm}(= \vec{N}_{(L,R)}): &=& \,\frac{1}{2} (S^{23}\pm i S^{01},S^{31}\pm i S^{02}, 
S^{12}\pm i S^{03} )\,,
\end{eqnarray}
where the generators $\vec{N}_{\pm}$ determine representations of the two $SU(2)$ invariant 
subgroups of $SO(3,1)$, the generators $\vec{\tau}^{1}$  and $\vec{\tau}^{2}$, 
 \begin{eqnarray}
 \label{so42}
 \vec{\tau}^{1}:&=&\frac{1}{2} (S^{58}-  S^{67}, \,S^{57} + S^{68}, \,S^{56}-  S^{78} )\,,\;\;
 \vec{\tau}^{2}:= \frac{1}{2} (S^{58}+  S^{67}, \,S^{57} - S^{68}, \,S^{56}+  S^{78} )\,,\,\;\nonumber\\
 \end{eqnarray}
 determine representations of the $SU(2)_{I}\times$ $SU(2)_{II}$ invariant subgroups of the group 
 $SO(4)$, which is further the  subgroup of $SO(7,1)$ ($SO(4), SO(3,1)$ are subgroups of $SO(7,1)$),
 and the generators $\vec{\tau}^{3}$,   $\tau^{4}$ and $\tilde{\tau}^{4}$
  \begin{eqnarray}
 \label{so64}
 \vec{\tau}^{3}: = &&\frac{1}{2} \,\{  S^{9\;12} - S^{10\;11} \,,
  S^{9\;11} + S^{10\;12} ,\, S^{9\;10} - S^{11\;12} ,\nonumber\\
 && S^{9\;14} -  S^{10\;13} ,\,  S^{9\;13} + S^{10\;14} \,,
  S^{11\;14} -  S^{12\;13}\,,\nonumber\\
 && S^{11\;13} +  S^{12\;14} ,\, 
 \frac{1}{\sqrt{3}} ( S^{9\;10} + S^{11\;12} - 
 2 S^{13\;14})\}\,,\nonumber\\
 \tau^{4}: = &&-\frac{1}{3}(S^{9\;10} + S^{11\;12} + S^{13\;14})\,,\;\;\nonumber\\
 \tilde{\tau}^{4}: = &&-\frac{1}{3}(\tilde{S}^{9\;10} + \tilde{S}^{11\;12} + \tilde{S}^{13\;14})\,,
 \end{eqnarray}
determine representations of $SU(3) \times U(1)$,  originating in $SO(6)$, and of $\tilde{\tau}^{4}$
originating in $\widetilde{SO}(6)$. 
 
One correspondingly finds the generators of the subgroups of $\widetilde{SO}(7,1)$,
\begin{eqnarray}
\label{so1+3tilde}
\vec{\tilde{N}}_{L,R}:&=& \,\frac{1}{2} (\tilde{S}^{23}\pm i \tilde{S}^{01},
\tilde{S}^{31}\pm i \tilde{S}^{02}, \tilde{S}^{12}\pm i \tilde{S}^{03} )\,,
\end{eqnarray}
which determine representations of the two $\widetilde{SU}(2)$ invariant subgroups of $\widetilde{SO}(3,1)$, 
while 
 \begin{eqnarray}
 \label{so42tilde}
 \vec{\tilde{\tau}}^{1}:&=&\frac{1}{2} (\tilde{S}^{58}-  \tilde{S}^{67}, \,\tilde{S}^{57} + 
 \tilde{S}^{68}, \,\tilde{S}^{56}-  \tilde{S}^{78} )\,,\;\;\nonumber\\
 \vec{\tilde{\tau}}^{2}:&=&\frac{1}{2} (\tilde{S}^{58}+  \tilde{S}^{67}, \,\tilde{S}^{57} - 
 \tilde{S}^{68}, \,\tilde{S}^{56}+  \tilde{S}^{78} )\,, 
 \end{eqnarray}
 determine representations of $\widetilde{SU}(2)_{I}\times $ $\widetilde{SU}(2)_{II}$ of $\widetilde{SO}(4)$. 
 Both, $\widetilde{SO}(3,1)$ and $\widetilde{SO}(4)$, 
 are the subgroups of $\widetilde{SO}(7,1)$.

 One further finds~\cite{JMP}
 \begin{eqnarray}
  \label{YQY'Q'andtilde}
 & & Y = \tau^{4} + \tau^{23}\,,\quad Y'= -\tau^{4}\tan^2\vartheta_2 + \tau^{23}\,,\quad
  Q =  \tau^{13} + Y\,,\quad Q'= -Y \tan^2\vartheta_1 + \tau^{13}\,,\nonumber\\
 & & \tilde{Y}= \tilde{\tau}^{4} + \tilde{\tau}^{23}\,,\quad \tilde{Y'}= -\tilde{\tau}^{4} 
  \tan^2 \tilde{\vartheta}_2 + \tilde{\tau}^{23}\,,\quad
  \tilde{Q}= \tilde{Y} + \tilde{\tau}^{13}\,,\quad \tilde{Q'}= -\tilde{Y} \tan^2\tilde{\vartheta}_1 
  + \tilde{\tau}^{13}\,.
  \end{eqnarray}

The scalar fields from the second line of Eq.~(\ref{faction}) (let me remind you that they are 
responsible~\cite{NBled2013,NBled2012,JMP} after gaining in the electroweak break nonzero vacuum 
expectation values for the masses of the family members and of the weak bosons) are expressible 
in terms of $\omega_{abc}$ fields and $\tilde{\omega}_{abc}$ fields as presented in Eqs.~(\ref{Aomegas}, 
\ref{Atildeomegas}).
One can find the below expressions by taking into account Eqs.~(\ref{so42}, \ref{so64}, 
\ref{so1+3tilde}, \ref{so42tilde}) and Eq.~(\ref{YQY'Q'andtilde}).
\begin{eqnarray}
\label{Aomegas}
-\frac{1}{2} S^{s's"}\, \omega_{s's"s}&=& - (g^{23}\, \tau^{23}\, A^{23}_{s} + 
                                             g^{13}\, \tau^{13}\, A^{13}_{s} + 
                                             g^{4 }\, \tau^{4} \, A^{4}_{s})\,,
\nonumber\\
g^{13}\,\tau^{13}\, A^{13}_{s} + g^{23}\,\tau^{23}\, A^{23}_{s} + g^{4}\,\tau^{4}\, A^{4}_{s} &=& 
g^{Q}\, Q A^{Q}_{s} + g^{Q'}\, Q' A^{Q'}_{s} + g^{Y'}\,Y'\, A^{Y'}_{s}\,, \nonumber\\
A^{4}_{s} &=& -(\omega_{9\,10\,s} + \omega_{11\,12\,s} + \omega_{13\,14\,s})\,,\nonumber\\
A^{13}_{s}&=&(\omega_{56 s}- \omega_{78 s})\,, \quad A^{23}_{s}=(\omega_{56 s}+
\omega_{78 s})\,,\nonumber\\
A^{Q}_{s} &=& \sin \vartheta_{1} \,A^{13}_{s} + \cos \vartheta_{1} \,A^{Y}_{s}\,,\quad
A^{Q'}_{s}  = \cos \vartheta_{1} \,A^{13}_{s} - \sin \vartheta_{1} \,A^{Y}_{s}\,,\nonumber\\
A^{Y}_{s} & =& \sin \vartheta_{2} \,A^{23}_{s} + \cos \vartheta_{2} \,A^{4}_{s}\,,\nonumber\\
A^{Y'}_{s}&=&\cos \vartheta_{2} \,A^{23}_{s} - \sin \vartheta_{2} \,A^{4}_{s}\,,\nonumber\\
          & &(s\in (7,8))\,.
\end{eqnarray}
In Eq.~(\ref{Aomegas}) the coupling 
constants are explicitly written to see the analogy with the gauge fields in the {\it standard 
model}.
 \begin{eqnarray}
 \label{Atildeomegas}
 -\frac{1}{2}  \tilde{S}^{\tilde{a} \tilde{b}}\,\tilde{\omega}_{\tilde{a} \tilde{b} s} &=&
 - (\vec{\tilde{\tau}}^{\tilde{1}}\, \vec{\tilde{A}}^{\tilde{1}}_{s} + 
  \vec{\tilde{N}}_{\tilde{L}}\, \vec{\tilde{A}}^{\tilde{N}_{\tilde{L}}}_{s} 
   + \vec{\tilde{\tau}}^{\tilde{2}}\, \vec{\tilde{A}}^{\tilde{2}}_{s} + 
  \vec{\tilde{N}}_{\tilde{R}}\, \vec{\tilde{A}}^{\tilde{N}_{\tilde{R}}}_{s})\,, \nonumber\\
  \vec{\tilde{A}}^{\tilde{1}}_{s} &=& (\tilde{\omega}_{\tilde{5} \tilde{8}s}-
  \tilde{\omega}_{\tilde{6} \tilde{7}s},\, \tilde{\omega}_{\tilde{5} \tilde{7}s}+
  \tilde{\omega}_{\tilde{6} \tilde{8}s}, \,\tilde{\omega}_{\tilde{5} \tilde{6}s}-
  \tilde{\omega}_{\tilde{7} \tilde{8}s})\,, \nonumber\\
\vec{\tilde{A}}^{\tilde{N}_{\tilde{L}}}_{s} &=& (\tilde{\omega}_{\tilde{2} \tilde{3}s}+i\,
  \tilde{\omega}_{\tilde{0} \tilde{1}s}, \,    \tilde{\omega}_{\tilde{3} \tilde{1}s}+i\,
  \tilde{\omega}_{\tilde{0} \tilde{2}s},  \,   \tilde{\omega}_{\tilde{1} \tilde{2}s}+i\,
  \tilde{\omega}_{\tilde{0} \tilde{3}s})\,, \nonumber\\
 \vec{\tilde{A}}^{\tilde{2}}_{s} &=& (\tilde{\omega}_{\tilde{5} \tilde{8}s}+
  \tilde{\omega}_{\tilde{6} \tilde{7}s}, \,\tilde{\omega}_{\tilde{5} \tilde{7}s}-
  \tilde{\omega}_{\tilde{6} \tilde{8}s},\, \tilde{\omega}_{\tilde{5} \tilde{6}s}+
  \tilde{\omega}_{\tilde{7} \tilde{8}s})\,, \nonumber\\
\vec{\tilde{A}}^{\tilde{N}_{\tilde{R}}}_{s} &=& (\tilde{\omega}_{\tilde{2} \tilde{3}s}-i\,
  \tilde{\omega}_{\tilde{0} \tilde{1}s}, \,    \tilde{\omega}_{\tilde{3} \tilde{1}s}-i\,
  \tilde{\omega}_{\tilde{0} \tilde{2}s}, \,    \tilde{\omega}_{\tilde{1} \tilde{2}s}-i\,
  \tilde{\omega}_{\tilde{0} \tilde{3}s})\,,\nonumber\\
                                             & &(s\in (7,8))\,. 
\end{eqnarray}
Scalar fields from Eq.~(\ref{Atildeomegas}) couple to fermions due to the family quantum numbers, 
while those from Eq.~(\ref{Aomegas})  distinguish among family members. 

The vector gauge fields $\vec{A}^{3}_{m}$ (the colour octet), $\vec{A}^{1}_{m}$ (the weak $SU(2)_{I}$ 
triplet), $\vec{A}^{2}_{m}$ (the $SU(2)_{II}$ triplet) and $\vec{A}^{4}_{m}$ (the $U(1)_{II}$ singlet
originating in $SO(6)$) can be expressed in terms of the spin connection fields and the vielbeins by 
taking  into account Eqs.~(\ref{so42}, \ref{so64}). Equivalently one finds the vector gauge fields 
in the "tilde" sector.

\section{Properties of scalar and vector gauge fields, contributing to transitions of 
antileptons into quarks}
\label{gaugefieldsQNlepbartransitions}

In this - the main  part of the present paper -  we study the properties, quantum numbers, and  discrete 
symmetries of those scalar and vector gauge fields appearing in the action~(Eqs.(\ref{wholeaction}, 
\ref{faction}), \ref{factionMaM0}) of the {\it spin-charge-family} 
theory~\cite{NBled2013,NBled2012,JMP,pikanorma,norma92,norma93,norma94,norma95,gmdn07,gn} which cause 
transitions of antileptons into quarks and back, and antiquarks into quarks and back.

These scalar gauge fields  carry the triplet or antitriplet colour 
charge (see table~\ref{Table bosons.}) and the fractional hyper and electromagnetic charge.

The Lagrange densities from Eqs.~(\ref{wholeaction}, \ref{faction}, \ref{factionMaM0}) manifest 
$\mathbb{C}_{{ \cal N}} \cdot {\cal P}_{{\cal N}}$ invariance (appendix~\ref{discrete}).  
All the vector and the spinor gauge fields are before the appearance of the condensate massless 
and  reactions creating particles from antiparticles and back go in both directions equivalently. 
Correspondingly there is no matter-antimatter 
asymmetry. 

The {\it spin-charge-family} theory breaks the matter-antimatter symmetry by the appearance 
of the condensate~(sect.~\ref{condensate}) and also by nonzero vacuum expectation values of  
the scalar fields causing the electroweak phase transition~(sect.~\ref{CPNandpropScm}).   
I shall show that there is the condensate of two right handed neutrinos which breaks this symmetry,
giving masses to all the scalar gauge fields and to all those vector gauge fields which would be 
in contradiction with the observations.

Let us start by analysing the Lagrange density presented in Eq.~(\ref{factionMaM0}) before the 
appearance of the condensate. 
The term $\gamma^{t}\, \frac{1}{2}\, S^{s' s"} \,\omega_{s' s" t}$ in Eq.~(\ref{factionMaM0}) can 
be rewritten, if taking into account Eq.~(\ref{signaturegamma}), as follows
\begin{eqnarray}
\gamma^{t}\,\frac{1}{2}\, S^{s' s"} \,\omega_{s' s" t} &=&
\sum_{+,-}\,\sum_{(t\,t')} \,\,\stackrel{t t'}{(\cpm)}\,\frac{1}{2}\, S^{s' s"} \, 
\omega_{\scriptscriptstyle{s" s" \stackrel{t t'}{(\cpm)}}}\,,\nonumber\\
\omega_{\scriptscriptstyle{s" s" \stackrel{t t'}{(\cpm)}}}: &=& 
\omega_{\scriptscriptstyle{s" s" \stackrel{t t'}{(\pm)}}} = 
(\omega_{s' s" t}\,\mp \,i \,\omega_{s' s" t'})\,, \nonumber\\ 
\stackrel{t t'}{(\cpm)}: &=& \stackrel{t t'}{(\pm)} = \frac{1}{2}\, 
(\gamma^{t} \pm \gamma^{t'})\,, \nonumber\\
(t\,t') &\in& ((9\,10), (11\,12),(13\,14))\,. 
\label{factionMaMpart10}
\end{eqnarray}
I introduced the notations $\stackrel{t t'}{(\cpm)}$ and  
$\omega_{\scriptscriptstyle{s" s" \stackrel{t t'}{(\cpm)}}}$ to distinguish among different superposition 
of states in equations below.

Using Eqs.~(\ref{so42}, \ref{so64}, \ref{Aomegas}) the expression  $\stackrel{t t'}{(\cpm)}\, \frac{1}{2}\, S^{s' s"} \,$  
$\omega_{\scriptscriptstyle{s" s" \,\stackrel{t t'}{(\cpm)}}}$ can be further rewritten 
as follows
\begin{eqnarray}
&&\stackrel{t t'}{(\cpm)}\,\frac{1}{2}\, S^{s' s"} \,  
\omega_{\scriptscriptstyle{s" s" \,\stackrel{t t'}{(\cpm)}}}=\nonumber\\
&& \stackrel{t t'}{(\cpm)}\,\{\, \tau^{2+}\,A^{2+}_{\scriptscriptstyle{\stackrel{t t'}{(\cpm)}}} 
+ \tau^{2-}\,A^{2-}_{\scriptscriptstyle{\stackrel{t t'}{(\cpm)}}} + \tau^{23}\,
A^{23}_{\scriptscriptstyle{\stackrel{t t'}{(\cpm)}}}
+ \tau^{1+}\,A^{1+}_{\scriptscriptstyle{\stackrel{t t'}{(\cpm)}}} + 
\tau^{1-}\,A^{1-}_{\scriptscriptstyle{\stackrel{t t'}{(\cpm)}}} + \tau^{13}\,
A^{13}_{\scriptscriptstyle{\stackrel{t t'}{(\cpm)}}}\, \} \,,\nonumber\\
A^{2\spm}_{\scriptscriptstyle{\stackrel{t t'}{(\cpm)}}} &=& 
(\omega_{\scriptscriptstyle{58 \stackrel{t t'}{(\cpm)}}}+ 
\omega_{\scriptscriptstyle{67 \stackrel{t t'}{(\cpm)}}})\,\smp \, i 
(\omega_{\scriptscriptstyle{57 \stackrel{t t'}{(\cpm)}}}- 
\omega_{\scriptscriptstyle{68 \stackrel{t t'}{(\cpm)}}})\,,
\quad A^{23}_{\scriptscriptstyle{\stackrel{t t'}{(\cpm)}}}= 
(\omega_{\scriptscriptstyle{56 \stackrel{t t'}{(\cpm)}}}+ 
\omega_{\scriptscriptstyle{78 \stackrel{t t'}{(\cpm)}}})\,, 
\nonumber\\
A^{1\spm}_{\scriptscriptstyle{\stackrel{t t'}{(\cpm)}}} &=& 
(\omega_{\scriptscriptstyle{58 \stackrel{t t'}{(\cpm)}}}- 
\omega_{\scriptscriptstyle{67 \stackrel{t t'}{(\cpm)}}})\smp \, i 
(\omega_{\scriptscriptstyle{57 \stackrel{t t'}{(\cpm)}}}+ 
\omega_{\scriptscriptstyle{68 \stackrel{t t'}{(\cpm)}}})\,,\quad 
A^{13}_{\scriptscriptstyle{\stackrel{t t'}{(\cpm)}}}= 
(\omega_{\scriptscriptstyle{56 \stackrel{t t'}{(\cpm)}}}- 
\omega_{\scriptscriptstyle{78 \stackrel{t t'}{(\cpm)}}})\,.
\label{factionMaMpart11}
\end{eqnarray}
Equivalently one  expresses the term $\gamma^{t}$ 
$\frac{1}{2}\, \tilde{S}^{ab} \,\tilde{\omega}_{ab t}$   in Eq.~(\ref{factionMaM0}), 
by using Eqs.~(\ref{so1+3tilde}, \ref{so42tilde}, \ref{Atildeomegas}), as
\begin{eqnarray}
&&\gamma^{t} \frac{1}{2}\, \tilde{S}^{ab} \,\tilde{\omega}_{ab t} = 
\stackrel{t t'}{(\cpm)} \, \frac{1}{2}\, \tilde{S}^{a b} \, 
\tilde{\omega}_{\scriptscriptstyle{ab \,\stackrel{t t'}{(\cpm)}}}= \nonumber\\
&&\stackrel{t t'}{(\cpm)} \,
\{\,\tilde{\tau}^{2+}\,\tilde{A}^{2+}_{\scriptscriptstyle{\stackrel{t t'}{(\cpm)}}} + 
\tilde{\tau}^{2-}\,\tilde{A}^{2-}_{\scriptscriptstyle{\stackrel{t t'}{(\cpm)}}} + 
\tilde{\tau}^{23}\,\tilde{A}^{23}_{\scriptscriptstyle{\stackrel{t t'}{(\cpm)}}} + \nonumber\\
& & \tilde{\tau}^{1+}\,\tilde{A}^{1+}_{\scriptscriptstyle{\stackrel{t t'}{(\cpm)}}} + 
\tilde{\tau}^{1-}\,\tilde{A}^{1-}_{\scriptscriptstyle{\stackrel{t t'}{(\cpm)}}} + 
\tilde{\tau}^{13}\,\tilde{A}^{13}_{\scriptscriptstyle{\stackrel{t t'}{(\cpm)}}} + \nonumber\\
& & \tilde{N}^{+}_{R}\,\tilde{A}^{N_{R}+}_{\scriptscriptstyle{\stackrel{t t'}{(\cpm)}}} + 
\tilde{N}^{-}_{R}\,\tilde{A}^{N_{R}-}_{\scriptscriptstyle{\stackrel{t t'}{(\cpm)}}} + 
\tilde{N}^{3}_{R}\,\tilde{A}^{N_{R}3}_{\scriptscriptstyle{\stackrel{t t'}{(\cpm)}}} + \nonumber\\
& &
\tilde{N}^{+}_{L}\,\tilde{A}^{N_{L}+}_{\scriptscriptstyle{\stackrel{t t'}{(\cpm)}}} +
\tilde{N}^{-}_{L}\,\tilde{A}^{N_{L}-}_{\scriptscriptstyle{\stackrel{t t'}{(\cpm)}}} + 
\tilde{N}^{3}_{L}\,\tilde{A}^{N_{L}3}_{\scriptscriptstyle{\stackrel{t t'}{(\cpm)}}}\,\}
\,,\nonumber\\ 
\tilde{A}^{N_{R}\spm}_{\scriptscriptstyle{\stackrel{t t'}{(\cpm)}}} &=& 
(\tilde{\omega}_{\scriptscriptstyle{23 \stackrel{t t'}{(\cpm)}}}- i\,
 \tilde{\omega}_{\scriptscriptstyle{01 \stackrel{t t'}{(\cpm)}}})\smp \, i 
(\tilde{\omega}_{\scriptscriptstyle{31 \stackrel{t t'}{(\cpm)}}}- i\,
 \tilde{\omega}_{\scriptscriptstyle{02 \stackrel{t t'}{(\cpm)}}})\,,\quad 
\tilde{A}^{N_{R}3}_{\scriptscriptstyle{\stackrel{t t'}{(\cpm)}}}= 
(\tilde{\omega}_{\scriptscriptstyle{12 \,\stackrel{t t'}{(\cpm)}}}- i\,
\tilde{\omega}_{\scriptscriptstyle{03 \stackrel{t t'}{(\cpm)}}})\,, \nonumber\\
\tilde{A}^{N_{L}\spm}_{\scriptscriptstyle{\stackrel{t t'}{(\cpm)}}} &=& 
(\tilde{\omega}_{\scriptscriptstyle{23 \stackrel{t t'}{(\cpm)}}}+ i\,
\tilde{\omega}_{\scriptscriptstyle{01 \stackrel{t t'}{(\cpm)}}})\smp \, i 
(\tilde{\omega}_{\scriptscriptstyle{31 \stackrel{t t'}{(\cpm)}}}+ i\,
 \tilde{\omega}_{\scriptscriptstyle{02 \stackrel{t t'}{(\cpm)}}})\,,\quad 
\tilde{A}^{N_{R}3}_{\scriptscriptstyle{\stackrel{t t'}{(\cpm)}}}= 
(\tilde{\omega}_{\scriptscriptstyle{12 \stackrel{t t'}{(\cpm)}}}+ i\,
 \tilde{\omega}_{\scriptscriptstyle{03 \stackrel{t t'}{(\cpm)}}})\,. 
\label{factionMaMpart20}
\end{eqnarray}
The expressions for $\tilde{A}^{2\spm}_{\scriptscriptstyle{\stackrel{t t'}{(\cpm)}}}$, 
$\tilde{A}^{23}_{\scriptscriptstyle{\stackrel{t t'}{(\cpm)}}}$, 
$\tilde{A}^{1\spm}_{\scriptscriptstyle{\stackrel{t t'}{(\cpm)}}}$ and 
$\tilde{A}^{1 3}_{\scriptscriptstyle{\stackrel{t t'}{(\cpm)}}}$ can easily be obtained 
from Eq.(\ref{factionMaMpart11}) by replacing in expressions  for 
$A^{2\spm}_{\scriptscriptstyle{\stackrel{t t'}{(\cpm)}}}$, 
$A^{23}_{\scriptscriptstyle{\stackrel{t t'}{(\cpm)}}}$, 
$A^{1\spm}_{\scriptscriptstyle{\stackrel{t t'}{(\cpm)}}}$ 
and $A^{1 3}_{\scriptscriptstyle{\stackrel{t t'}{(\cpm)}}}$, respectively, 
$\omega_{\scriptscriptstyle{s' s" \,\stackrel{t t'}{(\cpm)}}}$ by 
$\tilde{\omega}_{\scriptscriptstyle{s' s" \,\stackrel{t t'}{(\cpm)}}}$. 

There is the additional term in Eq.~(\ref{factionMaM0}): $\gamma^t$ $\frac{1}{2}\,
S^{t' t"}\,\omega_{t' t" t}$. 
This term 
can be rewritten  with respect to the generators $S^{t' t"}$
as one colour octet scalar field and one $U(1)_{II}$ singlet scalar field (Eq.~\ref{so64})
\begin{eqnarray}
\gamma^{t}\,\frac{1}{2}\, S^{t" t'"} \,\omega_{t" t'" t} &=&
\sum_{+,-}\,\sum_{(t\,t')}  \, \stackrel{t t'}{(\cpm)}\, 
\{\,\vec{\tau}^{3}\cdot \vec{A}^{3}_{\scriptscriptstyle{\stackrel{t t'}{(\cpm)}}}\, + \tau^{4}\cdot
A^{4}_{\scriptscriptstyle{\stackrel{t t'}{(\cpm)}}}\,\}\,, \nonumber\\
(t\,t') &\in& ((9\,10), 11\,12),13\,14))\,. 
\label{factionMaMpart21}
\end{eqnarray}

 Taking all above equations~(\ref{factionMaMpart10}, \ref{factionMaMpart11}, \ref{factionMaMpart20},
 \ref{factionMaMpart21})  into account Eq.~(\ref{factionMaM0}) can be rewritten, 
if we leave out $p_{\scriptscriptstyle{\stackrel{t t'}{(\cpm)}}}$ 
since in the low energy limit the momentum does not play any role, as follows
\begin{eqnarray}
{\mathcal L}_{f" }&=&  \psi^{\dagger} \,\gamma^0 (-)\,\{  
\sum_{+,-}\,\sum_{(t\,t')} \,\stackrel{t t'}{(\cpm)}\,{\bf \cdot}\nonumber\\
&&[\,
\tau^{2+}\,A^{2+}_{\scriptscriptstyle{\stackrel{t t'}{(\cpm)}}} 
+ \tau^{2-}\,A^{2-}_{\scriptscriptstyle{\stackrel{t t'}{(\cpm)}}} + \tau^{23}\,
A^{23}_{\scriptscriptstyle{\stackrel{t t'}{(\cpm)}}}  \nonumber\\
&+&
\tau^{1+}\,A^{1+}_{\scriptscriptstyle{\stackrel{t t'}{(\cpm)}}} + 
\tau^{1-}\,A^{1-}_{\scriptscriptstyle{\stackrel{t t'}{(\cpm)}}} + \tau^{13}\,
A^{13}_{\scriptscriptstyle{\stackrel{t t'}{(\cpm)}}} \nonumber\\
&+& 
\tilde{\tau}^{2+}\,\tilde{A}^{2+}_{\scriptscriptstyle{\stackrel{t t'}{(\cpm)}}} + 
\tilde{\tau}^{2-}\,\tilde{A}^{2-}_{\scriptscriptstyle{\stackrel{t t'}{(\cpm)}}} + 
\tilde{\tau}^{23}\,\tilde{A}^{23}_{\scriptscriptstyle{\stackrel{t t'}{(\cpm)}}}\nonumber\\
&+& 
\tilde{\tau}^{1+}\,\tilde{A}^{1+}_{\scriptscriptstyle{\stackrel{t t'}{(\cpm)}}} + 
\tilde{\tau}^{1-}\,\tilde{A}^{1-}_{\scriptscriptstyle{\stackrel{t t'}{(\cpm)}}} + 
\tilde{\tau}^{13}\,\tilde{A}^{13}_{\scriptscriptstyle{\stackrel{t t'}{(\cpm)}}} \nonumber\\
&+&
\tilde{N}^{+}_{R}\,\tilde{A}^{N_{R}+}_{\scriptscriptstyle{\stackrel{t t'}{(\cpm)}}} + 
\tilde{N}^{-}_{R}\,\tilde{A}^{N_{R}-}_{\scriptscriptstyle{\stackrel{t t'}{(\cpm)}}} + 
\tilde{N}^{3}_{R}\,\tilde{A}^{N_{R}3}_{\scriptscriptstyle{\stackrel{t t'}{(\cpm)}}} \nonumber\\
&+&
\tilde{N}^{+}_{L}\,\tilde{A}^{N_{L}+}_{\scriptscriptstyle{\stackrel{t t'}{(\cpm)}}} +
\tilde{N}^{-}_{L}\,\tilde{A}^{N_{L}-}_{\scriptscriptstyle{\stackrel{t t'}{(\cpm)}}} + 
\tilde{N}^{3}_{L}\,\tilde{A}^{N_{L}3}_{\scriptscriptstyle{\stackrel{t t'}{(\cpm)}}} \nonumber\\
&+& 
\tau^{3i}\,A^{3i}_{\scriptscriptstyle{\stackrel{t t'}{(\cpm)}}}\, + 
\tau^{4}\, A^{4}_{\scriptscriptstyle{\stackrel{t t'}{(\cpm)}}}\,] 
%
%
\,\}\, \psi\,,
%
%
\label{factionMaM1}
\end{eqnarray}
where $(t,t')$ run in pairs over $[(9,10),\dots(13,14)]$ and the summation must go over $+$ and $-$ 
of ${}_{\scriptscriptstyle{\stackrel{t t'}{(\cpm)}}}$.

Let us calculate now quantum numbers of the scalar and vector gauge fields appearing in 
Eq.~(\ref{factionMaM1}) by taking into account that the spin of gauge fields is determined 
according to Eq.~(\ref{bosonspin})
%
($(S^{ab})^{c}{}_{e}A^{d\dots e \dots g} = i(\eta^{ac}\delta^{b}_{d}- \eta^{bc}\delta^{a}_{d})\, 
A^{d\dots e \dots g}$,
%
for each index ($ \in (d \dots g)$) of a bosonic field $A^{d\dots g}$ separately).  
We must take into account also the relation among $S^{ab}$ and the charges (the relations are, of course, 
the same for bosons and fermions) 
(Eqs.~(\ref{so1+3}, \ref{so42}, \ref{so64}, \ref{so1+3tilde}, \ref{so42tilde}, \ref{YQY'Q'andtilde})).

On table~\ref{Table bosons.} properties of the  scalar gauge fields appearing in 
Eq.~(\ref{factionMaM1}) are presented. 
 \begin{table}
 \begin{tiny}
 \begin{center}
 \begin{tabular}{|c|c|c|c|c|c|c|c|c|c|c|c|c|c|}
 \hline
 ${\rm field}$&prop. & $\tau^4$&$\tau^{13}$&$\tau^{23}$&($\tau^{33},\tau^{38}$)&$Y$&$Q$&$\tilde{\tau}^4$ 
 &$\tilde{\tau}^{13}$&$\tilde{\tau}^{23}$&$\tilde{N}_{L}^{3}$ &$\tilde{N}_{R}^{3}$  \\
 \hline
 $A^{1\spm}_{\scriptscriptstyle{\stackrel{9\,10}{(\cpm)}}}$& scalar&  $\cmp \frac{1}{3}$&$\spm 1$&$0$ 
 & ($\cpm\frac{1}{2},$ $\cpm \frac{1}{2\sqrt{3}}$)& $\cmp \frac{1}{3}$&$\cmp \frac{1}{3}+ \smp 1$&$0$
 &$0$&$0$&$0$&$0$\\ 
 $A^{13}_{\scriptscriptstyle{\stackrel{9\,10}{(\cpm)}}}$   & scalar&  $\cmp \frac{1}{3}$&$0$&$0$ 
 & ($\cpm\frac{1}{2},$ $\cpm \frac{1}{2\sqrt{3}}$)& $\cmp \frac{1}{3}$&$\cmp \frac{1}{3}$&$0$&$0$&$0$&$0$&$0$\\ 
 $A^{1\spm}_{\scriptscriptstyle{\stackrel{11\,12}{(\cpm)}}}$& scalar&  $\cmp \frac{1}{3}$&$\smp 1$&$0$ 
  & ($\cmp\frac{1}{2},$ $\cpm \frac{1}{2\sqrt{3}}$)& $\cmp \frac{1}{3}$&$\cmp \frac{1}{3}+ \smp 1$&$0$
  &$0$&$0$&$0$&$0$\\ 
  $A^{13}_{\scriptscriptstyle{\stackrel{11\,12}{(\cpm)}}}$   & scalar&  $\cmp \frac{1}{3}$&$0$&$0$ 
 & ($\cmp\frac{1}{2},$ $\cpm \frac{1}{2\sqrt{3}}$)& $\cmp \frac{1}{3}$&$\cmp \frac{1}{3}$&$0$&$0$&$0$&$0$&$0$\\ 
 $A^{1\spm}_{\scriptscriptstyle{\stackrel{13\,14}{(\cpm)}}}$& scalar&  $\cmp \frac{1}{3}$&$\smp 1$&$0$ 
   & ($0,$ $\cmp \frac{1}{\sqrt{3}}$)& $\cmp \frac{1}{3}$&$\cmp \frac{1}{3}+ \smp 1$&$0$
   &$0$&$0$&$0$&$0$\\ 
   $A^{13}_{\scriptscriptstyle{\stackrel{13\,14}{(\cpm)}}}$   & scalar&  $\cmp \frac{1}{3}$&$0$&$0$ 
 & ($0,$ $\cmp \frac{1}{\sqrt{3}}$)& $\cmp \frac{1}{3}$&$\cmp \frac{1}{3}$&$0$&$0$&$0$&$0$&$0$\\ 
 \hline
 $A^{2\spm}_{\scriptscriptstyle{\stackrel{9\,10}{(\cpm)}}}$& scalar&  $\cmp \frac{1}{3}$&$0$&$\spm 1$ 
  & ($\cpm\frac{1}{2},$ $\cpm \frac{1}{2\sqrt{3}}$)& $\cmp \frac{1}{3}+ \smp 1$&$\cmp \frac{1}{3}+ \smp 1$&$0$
  &$0$&$0$&$0$&$0$\\  
  $A^{23}_{\scriptscriptstyle{\stackrel{9\,10}{(\cpm)}}}$   & scalar&  $\cmp \frac{1}{3}$&$0$&$0$ 
 & ($\cpm\frac{1}{2},$ $\cpm \frac{1}{2\sqrt{3}}$)& $\cmp \frac{1}{3}$&$\cmp \frac{1}{3}$&$0$&$0$&$0$&$0$&$0$\\ 
 $\cdots$&&&&&&&&&&&&\\
 \hline
 $\tilde{A}^{1 \spm}_{\scriptscriptstyle{\stackrel{9 10}{(\cpm)}}}$& scalar& $\cmp \frac{1}{3}$&$0$&$0$ 
 & ($\cpm\frac{1}{2},$ $\cpm \frac{1}{2\sqrt{3}}$)& $\cmp \frac{1}{3}$&$\cmp \frac{1}{3}$&$0$
 &$\spm 1$&$0$&$0$&$0$\\ 
 $\tilde{A}^{13}_{\scriptscriptstyle{\stackrel{9 10}{(\cpm)}}}$& scalar& $\cmp \frac{1}{3}$&$0$&$0$ 
  & ($\cpm\frac{1}{2},$ $\cpm \frac{1}{2\sqrt{3}}$)& $\cmp \frac{1}{3}$&$\cmp \frac{1}{3}$&$0$
 &$0$&$0$&$0$&$0$\\ 
  $\cdots$&&&&&&&&&&&&\\
 \hline
 $\tilde{A}^{2 \spm}_{\scriptscriptstyle{\stackrel{9 10}{(\cpm)}}}$& scalar& $\cmp \frac{1}{3}$&$0$&$0$ 
  & ($\cpm\frac{1}{2},$ $\cpm \frac{1}{2\sqrt{3}}$)& $\cmp \frac{1}{3}$&$\cmp \frac{1}{3}$&$0$
  &$0$&$\spm 1$&$0$&$0$\\ 
  $\tilde{A}^{23}_{\scriptscriptstyle{\stackrel{9 10}{(\cpm)}}}$& scalar& $\cmp \frac{1}{3}$&$0$&$0$ 
   & ($\cpm\frac{1}{2},$ $\cpm \frac{1}{2\sqrt{3}}$)& $\cmp \frac{1}{3}$&$\cmp \frac{1}{3}$&$0$
  &$0$&$0$&$0$&$0$\\ 
  $\cdots$&&&&&&&&&&&&\\
 \hline
 $\tilde{A}^{N_{L} \spm}_{\scriptscriptstyle{\stackrel{9 10}{(\cpm)}}}$& scalar& $\cmp \frac{1}{3}$&$0$&$0$ 
   & ($\cpm\frac{1}{2},$ $\cpm \frac{1}{2\sqrt{3}}$)& $\cmp \frac{1}{3}$&$\cmp \frac{1}{3}$&$0$
   &$0$&$0$&$\spm 1$&$0$\\ 
   $\tilde{A}^{N_{L} 3}_{\scriptscriptstyle{\stackrel{9 10}{(\cpm)}}}$& scalar& $\cmp \frac{1}{3}$&$0$&$0$ 
    & ($\cpm\frac{1}{2},$ $\cpm \frac{1}{2\sqrt{3}}$)& $\cmp \frac{1}{3}$&$\cmp \frac{1}{3}$&$0$
   &$0$&$0$&$0$&$0$\\ 
   $\cdots$&&&&&&&&&&&&\\
 \hline
 $\tilde{A}^{N_{R} \spm}_{\scriptscriptstyle{\stackrel{9 10}{(\cpm)}}}$& scalar& $\cmp \frac{1}{3}$&$0$&$0$ 
    & ($\cpm\frac{1}{2},$ $\cpm \frac{1}{2\sqrt{3}}$)& $\cmp \frac{1}{3}$&$\cmp \frac{1}{3}$&$0$
    &$0$&$0$&$0$&$\spm 1$\\ 
    $\tilde{A}^{N_{R} 3}_{\scriptscriptstyle{\stackrel{9 10}{(\cpm)}}}$& scalar& $\cmp \frac{1}{3}$&$0$&$0$ 
     & ($\cpm\frac{1}{2},$ $\cpm \frac{1}{2\sqrt{3}}$)& $\cmp \frac{1}{3}$&$\cmp \frac{1}{3}$&$0$
    &$0$&$0$&$0$&$0$\\ 
    $\cdots$&&&&&&&&&&&&\\
 \hline
 $A^{3 i}_{\scriptscriptstyle{\stackrel{9\,10}{(\cpm)}}}$& scalar&  $\cmp \frac{1}{3}$&$0$&$0$ 
  & ($\spm 1 + \cpm\frac{1}{2},$ $\cpm \frac{1}{2\sqrt{3}}$)& $\cmp \frac{1}{3}$&$\cmp \frac{1}{3}$&$0$
 &$0$&$0$&$0$&$0$\\ 
    $\cdots$&&&&&&&&&&&&\\
  \hline
 $A^{4}_{\scriptscriptstyle{\stackrel{9\,10}{(\cpm)}}}$& scalar&  $\cmp \frac{1}{3}$&$0$&$0$ 
  & ($ \cpm\frac{1}{2},$ $\cpm \frac{1}{2\sqrt{3}}$)& $\cmp \frac{1}{3}$&$\cmp \frac{1}{3}$&$0$
 &$0$&$0$&$0$&$0$\\ 
    $\cdots$&&&&&&&&&&&&\\   
\hline 
$\vec{A}^{3}_{m}$& vector&  $0$&$0$&$0$ 
  & octet& $0$&$0$&$0$
 &$0$&$0$&$0$&$0$\\ 
\hline
 $A^{4}_{m}$& vector&  $0$&$0$&$0$ 
  & $0$& $0$&$0$&$0$
 &$0$&$0$&$0$&$0$\\ 
\hline   
 \end{tabular}
 \end{center}
 \end{tiny}
 \caption{\label{Table bosons.}%
 Quantum numbers of the scalar gauge fields carrying the space index $t =(9,10,\cdots,14)$, 
 appearing in  Eq.~(\ref{factionMaM1}), are presented. To the colour charge of all these scalar 
 fields the space degrees  of freedom contribute one of the triplets values. These scalars are 
 with respect to the two $SU(2)$ charges, ($\tau^{13}$  and $\vec{\tau}^2$), and the two  
 $\widetilde{SU}(2)$  charges, ($\vec{\tilde{\tau}}^1$  and $\vec{\tilde{\tau}}^2$), triplets 
 (that is in the adjoint representations of the  corresponding groups), and they all carry twice the 
 "spinor" number ($\tau^{4}$) of the quarks.  The quantum numbers of the two vector gauge fields, 
 the colour and the  $U(1)_{II}$ ones, are added. 
 }
  \end{table}

The scalar fields  with the scalar index $s=(9,10,\cdots,14)$, presented in table~\ref{Table bosons.}, 
carry one of the triplet colour charges and the "spinor" charge equal to twice the quark "spinor" charge,
or the antitriplet colour charges and the anti "spinor" charge. 
They carry in addition the quantum numbers of the adjoint representations originating in $S^{ab}$ or in
$\tilde{S}^{ab}$. Although carrying the colour charge in  one of the triplet or antitriplet states, these  
fields can not be interpreted as superpartners of the quarks 
as required by, let say, the $N=1$ supersymmetry.
The hyper charges  and the electromagnetic charges are namely not those required by the supersymmetric 
partners to the family members.

Let us have a look what do the scalar fields, appearing in Eq.~(\ref{factionMaM1}) and in 
table~\ref{Table bosons.}, do when applying on the left handed members of the Weyl 
representation presented on table~\ref{Table so13+1.}, 
containing quarks and leptons and antiquarks and antileptons~\cite{pikanorma,Portoroz03,HNds}.
Let us choose the $57^{th}$ line of table~\ref{Table so13+1.}, which represents in the 
spinor technique the left handed positron,~$\bar{e}^{+}_{L}$. If we make, let say, the choice of the 
term ($\gamma^0 \stackrel{9 10}{(+)}\, \tau^{2 \sminus}\,$)  
 $A^{2\sminus}_{\scriptscriptstyle{\stackrel{9\,10}{(\oplus)}}} $ (the scalar field 
$ A^{2\sminus}_{\scriptscriptstyle{\stackrel{9\,10}{(\oplus)}}} $ is presented  in the $7^{th}$ 
line in table~\ref{Table bosons.} and in the second line of Eq.~(\ref{factionMaM1})), 
the family quantum numbers will not be affected and 
they can be any. The state carries the "spinor" (lepton) 
number  $\tau^{4}=\frac{1}{2}$, the weak charge $\tau^{13} =0$, the second $SU(2)_{II}$ 
charge $\tau^{23} =\frac{1}{2}$ and the colour charge $(\tau^{33},\tau^{38})=(0,0)$. Correspondingly, 
its hyper charge ($Y(=\tau^{4}+\tau^{23})$) is $1$ and the electromagnetic charge 
$Q(=Y + \tau^{13})$  is $1$. 

So, what does the term $\gamma^0 \stackrel{9 10}{(+)}\, \tau^{2 \sminus}\,$  
 $A^{2\sminus}_{\scriptscriptstyle{\stackrel{9\,10}{(\oplus)}}} $ 
make on this spinor $\bar{e}^{+}_{L}$? 
Making use of Eqs.~(\ref{snmb:gammatildegamma}, \ref{graphbinoms}, \ref{plusminus})  of 
appendix~\ref{technique} one easily finds that operator $\gamma^0 \stackrel{9 10}{(+)}\,$ 
$ \tau^{2 -}\,$ transforms the left handed  positron  into $\stackrel{03}{(+i)}\,\stackrel{12}{(+)}|
\stackrel{56}{[-]}\,\stackrel{78}{[-]}||\stackrel{9 \;10}{(+)}\;\;
\stackrel{11\;12}{(-)}\;\;\stackrel{13\;14}{(-)} $, which is $d_{R}^{c1} $, presented on line $3$
of  table~\ref{Table so13+1.}. Namely, $\gamma^0$ transforms $\stackrel{03}{[-i]}$ into 
$\stackrel{03}{(+i)}$, $\stackrel{9 10}{(+)}$ transforms $\stackrel{9 \;10}{[-]}$ into 
$\stackrel{9 \;10}{(+)}$, while $\tau^{2 -}$ ($= -\stackrel{56}{(-)}$ $\stackrel{78}{(-)}$)
transforms $\stackrel{56}{(+)}$ $\stackrel{78}{(+)}$ into $\stackrel{56}{[-]}$ $\stackrel{78}{[-]}$. 
The state $d_{R}^{c1} $ carries the "spinor" (quark) number  $\tau^{4}=\frac{1}{6}$, the weak charge 
$\tau^{13} =0$, the second $SU(2)_{II}$ charge $\tau^{23} =-\frac{1}{2}$ and the colour charge 
$(\tau^{33},\tau^{38})=(\frac{1}{2},\frac{1}{2\sqrt{3}})$. Correspondingly its hyper charge  is
($Y=\tau^{4}+\tau^{23}=$) $-\frac{1}{3}$ and the electromagnetic charge ($Q=Y + \tau^{13}=$) 
$ -\frac{1}{3}$. The scalar field $A^{2\sminus}_{\scriptscriptstyle{\stackrel{9\,10}{(\oplus)}}} $
carries just  the needed quantum numbers as we can see in the $7^{th}$ line of table~\ref{Table bosons.}.

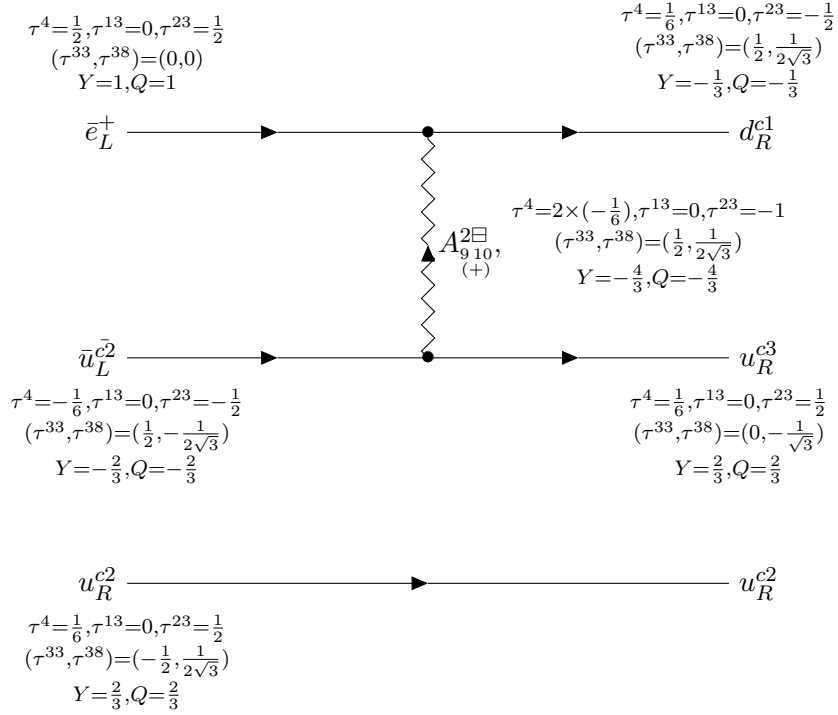
\begin{figure}
\begin{tikzpicture}[>=triangle 45]
\draw [->] (0,-3) node [anchor=east] {$u_R^{c2}$} 
node [anchor=north] {$\begin{smallmatrix} 
                                     { }\\ {}\\
                                     \tau^4=\frac{1}{6},
                                     \tau^{13}=0,
                                     \tau^{23}=\frac{1}{2}\\
                                     (\tau^{33},\tau^{38})=(-\frac{1}{2},\frac{1}{2\sqrt{3}})\\
                                      Y=\frac{2}{3}, Q=\frac{2}{3}
                                      \end{smallmatrix}$}  -- (4,-3);
\draw (4,-3) -- (8,-3) node [anchor=west] {$u_R^{c2}$};
\draw [->] (0,0) node [anchor=east] {$\bar{u}_L^{\bar{c2}}$}
node [anchor=north] {$\begin{smallmatrix} 
                                     {}\\ {}\\
                                     \tau^4=-\frac{1}{6},
                                     \tau^{13}=0,
                                     \tau^{23}=-\frac{1}{2}\\
                                     (\tau^{33},\tau^{38})=(\frac{1}{2},-\frac{1}{2\sqrt{3}})\\
                                      Y=-\frac{2}{3}, Q=-\frac{2}{3}
                                      \end{smallmatrix}$}  -- (2,0);
\draw (2,0) -- (4,0);
\draw [->] (4,0) -- (6,0);
\draw (6,0) -- (8,0) node [anchor=west] {$u_R^{c3}$}
node [anchor=north] {$\begin{smallmatrix} 
                                     {}\\ {}\\
                                     \tau^4=\frac{1}{6},
                                     \tau^{13}=0,
                                     \tau^{23}= \frac{1}{2}\\
                                     (\tau^{33},\tau^{38})=(0,-\frac{1}{\sqrt{3}})\\
                                      Y=\frac{2}{3}, Q=\frac{2}{3}
                                      \end{smallmatrix}$} ;
\draw [->] (0,3) node [anchor=east] {$\bar{e}_L^+$} 
node [anchor=south] {$\begin{smallmatrix} 
                                     \tau^4=\frac{1}{2},
                                     \tau^{13}=0,
                                     \tau^{23}=\frac{1}{2}\\
                                     (\tau^{33},\tau^{38})=(0,0)\\
                                      Y=1, Q=1\\
                                      { \ } \\ { \ }
                                      \end{smallmatrix}$} -- (2,3);
\draw (2,3) -- (4,3);
\draw [->] (4,3) -- (6,3);
\draw (6,3) -- (8,3) node [anchor=west] {$d_R^{c1}$} 
node [anchor=south] {$\begin{smallmatrix} 
                                     \tau^4=\frac{1}{6},
                                     \tau^{13}=0,
                                     \tau^{23}=-\frac{1}{2}\\
                                     (\tau^{33},\tau^{38})=(\frac{1}{2},\frac{1}{2\sqrt{3}})\\
                                      Y=-\frac{1}{3}, Q=-\frac{1}{3}\\
                                     { \ } \\ {\ }
                                      \end{smallmatrix}$};
\draw [->,snake]
(4,0) node {$\bullet$}-- (4,1.5) 
node [anchor=west]
{$A^{2\sminus}_{\scriptscriptstyle{\stackrel{9\,10}{(+)}}}, 
\begin{smallmatrix} 
                                     \tau^4=2\times(-\frac{1}{6}),
                                     \tau^{13}=0,
                                     \tau^{23}=-1\\
                                     (\tau^{33},\tau^{38})=(\frac{1}{2},\frac{1}{2\sqrt{3}})\\
                                      Y=-\frac{4}{3}, Q=-\frac{4}{3}
                                      \end{smallmatrix} $};
\draw [snake]
(4,1.5) -- (4,3) node {$\bullet$};
\end{tikzpicture}
\caption{\label{proton is born1.} The birth of a "right handed proton" out of an positron  $\bar{e}^{+}_{L}$, 
antiquark $\bar{u}_L^{\bar{c2}}$ and quark (spectator) $u_{R}^{c2}$.  
The family quantum number can be any.} 
\end{figure}

If the antiquark $ \bar{u}_{L}^{\bar{c2}}$, from the line $43$ (it is not presented, but one can 
very easily construct it) in table~\ref{Table so13+1.},  with
the "spinor" charge $\tau^{4}=-\frac{1}{6}$, the weak charge $\tau^{13} =0$, 
the second $SU(2)_{II}$ charge $\tau^{23} =-\frac{1}{2}$, the colour charge 
$(\tau^{33},\tau^{38})=(\frac{1}{2},-\frac{1}{2\sqrt{3}})$, the hyper charge 
$Y(=\tau^{4}+\tau^{23}=$) $-\frac{2}{3}$ and the electromagnetic charge $Q (\,=Y + \tau^{13}=$) 
$ -\frac{2}{3}$ submits the $A^{2 \sminus}_{\scriptscriptstyle{\stackrel{9\,10}{(\oplus)}}} $ scalar 
field, it transforms into $u_{R}^{c3}$ from the line $17$ of table~\ref{Table so13+1.}, carrying the 
quantum numbers $\tau^{4}=\frac{1}{6}$, $\tau^{13} =0$, $\tau^{23} =\frac{1}{2}$, 
$(\tau^{33},\tau^{38})=(0,-\frac{1}{\sqrt{3}})$, $Y=\frac{2}{3}$ and $Q=\frac{2}{3} $ . 
These two quarks, $d_{R}^{c1} $ and $u_{R}^{c3}$ can bind  together 
with $u_{R}^{c2}$ from the $9^{th}$ line of the same table (at low enough energy, after the electroweak 
transition, and if they belong in a superposition with the left handed partners to the first family)
into the colour chargeless baryon - a proton.
This transition is presented in figure~\ref{proton is born1.}.

The opposite transition at low energies would make the proton decay.

Let us look at one more example. The $63^{th}$ line of table~\ref{Table so13+1.}  
represents in the 
spinor technique the right handed positron, $\bar{e}^{+}_{R}$. Since we shall again not have a look 
on a transition, in which scalar fields with the nonzero family quantum numbers are involved, 
the family quantum number of this positron is not important.  The state carries the "spinor" (lepton) 
number  $\tau^{4}=\frac{1}{2}$, the weak charge $\tau^{13} =\frac{1}{2}$, the second $SU(2)_{II}$ 
charge $\tau^{23} =0$ and the colour charge $(\tau^{33},\tau^{38})=(0,0)$. Correspondingly, 
its hyper charge ($Y=\tau^{4}+\tau^{23}$) is $\frac{1}{2}$ and the electromagnetic charge $Q=Y + \tau^{13}$  
is $1$.

What does, let say, the term $\gamma^0 \stackrel{9 10}{(+)}\, \tau^{1 \sminus}\,$  
$A^{1\sminus}_{\scriptscriptstyle{\stackrel{9\,10}{(\oplus)}}} $ (the scalar field
$A^{1\sminus}_{\scriptscriptstyle{\stackrel{9\,10}{(\oplus)}}} $ is 
presented in the first line of table~\ref{Table bosons.}) make on $\bar{e}^{+}_{R}$? Making use of  
Eqs.~(\ref{snmb:gammatildegamma}, \ref{graphbinoms}, \ref{plusminus})  of appendix~\ref{technique} 
one easily finds that the right handed  positron transforms under the application of $\gamma^0 $ 
$\tau^{1-}$ $\stackrel{9 10}{(+)}$ into $\stackrel{03}{[-i]}\,\stackrel{12}{(+)}|
\stackrel{56}{[-]}\,\stackrel{78}{(+)}||\stackrel{9 \;10}{(+)}\;\;
\stackrel{11\;12}{(-)}\;\;\stackrel{13\;14}{(-)} $, which is $d_{L}^{c1} $ presented on line $5$
of  table~\ref{Table so13+1.}. Namely, $\gamma^0$ transforms $\stackrel{03}{(+i)}$ into 
$\stackrel{03}{[-i]}$, $\stackrel{9 10}{(+)}$ transforms $\stackrel{9 \;10}{[-]}$ into 
$\stackrel{9 \;10}{(+)}$, while $\tau^{1 \sminus}$ ($= \stackrel{56}{(-)}$ $\stackrel{78}{(+)}$)
transforms $\stackrel{56}{(+)}$ $\stackrel{56}{[-]}$ into $\stackrel{56}{[-]}$ $\stackrel{56}{(+)}$. 
The state $d_{L}^{c1} $  carries the "spinor" (quark) number  $\tau^{4}=\frac{1}{6}$, the weak charge 
$\tau^{13} = - \frac{1}{2}$, the second $SU(2)_{II}$ charge $\tau^{23} =0$ and the colour charge 
$(\tau^{33},\tau^{38})=(\frac{1}{2},\frac{1}{2\sqrt{3}})$. Correspondingly its hyper charge  is
($Y=\tau^{4}+\tau^{23}=$) $\frac{1}{6}$ and the electromagnetic charge ($Q=Y + \tau^{13}=$) 
$ -\frac{1}{3}$.  The scalar field $A^{1\sminus}_{\scriptscriptstyle{\stackrel{9\,10}{(\oplus)}}} $ 
carries all the needed quantum numbers, as one can see in figure~\ref{proton is born1.}.

\begin{figure}
\begin{tikzpicture}[>=triangle 45]
\draw [->] (0,-3) node [anchor=east] {$u_L^{c2}$} 
node [anchor=north] {$\begin{smallmatrix} 
                                     { }\\ {}\\
                                     \tau^4=\frac{1}{6},
                                     \tau^{13}=\frac{1}{2},
                                     \tau^{23}=0\\
                                     (\tau^{33},\tau^{38})=(-\frac{1}{2},\frac{1}{2\sqrt{3}})\\
                                      Y=\frac{1}{6}, Q=\frac{2}{3}
                                      \end{smallmatrix}$}  -- (4,-3);
\draw (4,-3) -- (8,-3) node [anchor=west] {$u_L^{c2}$};
\draw [->] (0,0) node [anchor=east] {$\bar{u}_R^{\bar{c2}}$}
node [anchor=north] {$\begin{smallmatrix} 
                                     {}\\ {}\\
                                     \tau^4=-\frac{1}{6},
                                     \tau^{13}=-\frac{1}{2},
                                     \tau^{23}=0\\
                                     (\tau^{33},\tau^{38})=(\frac{1}{2},-\frac{1}{2\sqrt{3}})\\
                                      Y=-\frac{1}{6}, Q=-\frac{2}{3}
                                      \end{smallmatrix}$}  -- (2,0);
\draw (2,0) -- (4,0);
\draw [->] (4,0) -- (6,0);
\draw (6,0) -- (8,0) node [anchor=west] {$u_L^{c3}$}
node [anchor=north] {$\begin{smallmatrix} 
                                     {}\\ {}\\
                                     \tau^4=\frac{1}{6},
                                     \tau^{13}=\frac{1}{2},
                                     \tau^{23}= 0\\
                                     (\tau^{33},\tau^{38})=(0,-\frac{1}{\sqrt{3}})\\
                                      Y=\frac{1}{6}, Q=\frac{2}{3}
                                      \end{smallmatrix}$} ;
\draw [->] (0,3) node [anchor=east] {$\bar{e}_R^+$} 
node [anchor=south] {$\begin{smallmatrix} 
                                     \tau^4=\frac{1}{2},
                                     \tau^{13}=\frac{1}{2},
                                     \tau^{23}=0\\
                                     (\tau^{33},\tau^{38})=(0,0)\\
                                      Y=\frac{1}{2}, Q=1\\
                                      { \ } \\ { \ }
                                      \end{smallmatrix}$} -- (2,3);
\draw (2,3) -- (4,3);
\draw [->] (4,3) -- (6,3);
\draw (6,3) -- (8,3) node [anchor=west] {$d_L^{c1}$} 
node [anchor=south] {$\begin{smallmatrix} 
                                     \tau^4=\frac{1}{6},
                                     \tau^{13}=-\frac{1}{2},
                                     \tau^{23}=0\\
                                     (\tau^{33},\tau^{38})=(\frac{1}{2},\frac{1}{2\sqrt{3}})\\
                                      Y=\frac{1}{6}, Q=-\frac{1}{3}\\
                                     { \ } \\ {\ }
                                      \end{smallmatrix}$};
\draw [->,snake]
(4,0) node {$\bullet$}-- (4,1.5) 
node [anchor=west]
{$A^{1 \sminus}_{\scriptscriptstyle{\stackrel{9\,10}{(+)}}}\,\,, 
\begin{smallmatrix} 
                                     \tau^4=2\times(-\frac{1}{6}),
                                     \tau^{13}=-1,
                                     \tau^{23}=0\\
                                     (\tau^{33},\tau^{38})=(\frac{1}{2},\frac{1}{2\sqrt{3}})\\
                                      Y=-\frac{1}{3}, Q=-\frac{4}{3}
                                      \end{smallmatrix} $};
\draw [snake]
(4,1.5) -- (4,3) node {$\bullet$};
\end{tikzpicture}
\caption{\label{proton is born2.} The birth of a " left handed proton" out of an positron  $\bar{e}^{+}_{R}$, 
antiquark $\bar{u}_R^{\bar{c2}}$ and quark (spectator) $u_{L}^{c2}$.  
The family quantum number can be any.} 
\end{figure}
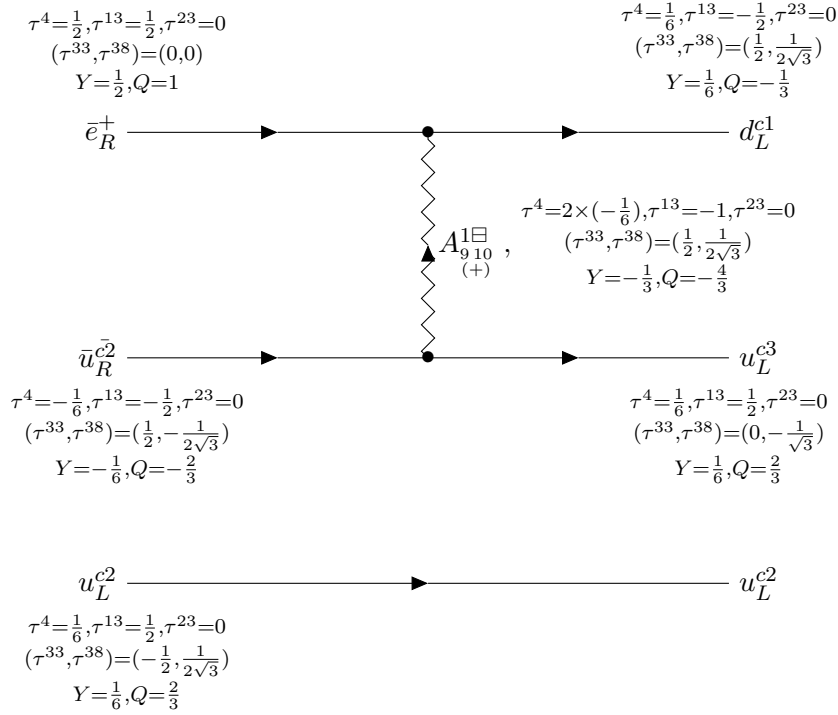

If the antiquark $ \bar{u}_{R}^{\bar{c2}}$, from the line $47$ in table~\ref{Table so13+1.} (the 
reader can easily find the expression  $\stackrel{03}{(+i)}\,\stackrel{12}{(+)}|
\stackrel{56}{[-]}\,\stackrel{78}{(+)}||\stackrel{9 \;10}{(+)}\;\;
\stackrel{11\;12}{(-)}\;\;\stackrel{13\;14}{[+]} $),  with
the "spinor" charge $\tau^{4}=-\frac{1}{6}$, the weak charge $\tau^{13} =-\frac{1}{2}$, 
the second $SU(2)_{II}$ charge $\tau^{23} =0$, the colour charge 
$(\tau^{33},\tau^{38})=(\frac{1}{2},-\frac{1}{2\sqrt{3}})$, the hypercharge 
($Y=\tau^{4}+\tau^{23}=$) $-\frac{1}{6}$ and the electromagnetic charge ($Q=Y + \tau^{13}=$) 
$ -\frac{2}{3}$, submits the $A^{1 \sminus}_{\scriptscriptstyle{\stackrel{9\,10}{(\oplus)}}} $ scalar 
field, it transforms into $u_{L}^{c3}$ from the line $23$ of table~\ref{Table so13+1.}  
($\stackrel{03}{[-i]}\,\stackrel{12}{(+)}|\stackrel{56}{(+)}\,\stackrel{78}{[-]}||\stackrel{9 \;10}{[-]}\;\;
\stackrel{11\;12}{(-)}\;\;\stackrel{13\;14}{[+]} $), carrying the 
quantum numbers $\tau^{4}=\frac{1}{6}$, $\tau^{13} =\frac{1}{2}$, $\tau^{23} =0$, 
$(\tau^{33},\tau^{38})=(0,-\frac{1}{\sqrt{3}})$, $Y=\frac{1}{6}$ and $Q=\frac{2}{3}$. 
These two quarks, $d_{L}^{c1} $ and $u_{L}^{c3}$, can bind (at low enough energy, when making after the 
electroweak transition the superposition with the right handed partners) together 
with $u_{L}^{c2}$ from the $15^{th}$ line of the same table, into the colour chargeless baryon - a proton.
This transition is presented in  figure~\ref{proton is born2.}. 

The opposite transition would make the proton decay.

Similar transitions go also with other scalars from  Eq.~(\ref{factionMaM1}) and 
table~\ref{Table bosons.}. 
The $\vec{\tilde{A}}^{1}_{\stackrel{t'\,t"}{(+)}}$, $\vec{\tilde{A}}^{2}_{\stackrel{t'\,t"}{(+)}}$,
$\vec{\tilde{A}}^{N_{L}}_{\stackrel{t'\,t"}{(+)}}$ and $\vec{\tilde{A}}^{N_{L}}_{\stackrel{t'\,t"}{(+)}}$
fields cause transitions among the family members, changing a particular member into the antimember 
of another colour and of another family. The term $\gamma^0 \stackrel{9 10}{(+)}\, \tilde{N}^{-}_{R}\,$  
$A^{\tilde{N}_{R}-}_{\scriptscriptstyle{\stackrel{9\,10}{(\oplus)}}} $  transforms $\bar{e}^{+}_{R}$ 
into $u^{c1}_{L}$, changing the family quantum numbers.

The action from Eqs.~(\ref{wholeaction}, \ref{faction}, \ref{factionMaM0}) manifests 
$\mathbb{C}_{{ \cal N}} \cdot {\cal P}_{{\cal N}}$ invariance.  All the vector and 
the spinor gauge fields are massless. 

Since none of the scalar fields from table~\ref{Table bosons.} have been observed and also no  
vector gauge fields like $\vec{A}^{2}_{m}$, $A^{4}_{m}$ and other scalar and vector fields, we 
shall discuss this topic in sect.~\ref{masslessandmassivebosons}, it must exist a mechanism, which 
makes the non observed scalar and vector gauge fields massive enough.

Scalar fields from table~\ref{Table bosons.} carry the colour and the electromagnetic charge. 
Therefore their nonzero vacuum expectation values would not be in agreement with the observed 
phenomena. 
One, however, notices that all the scalar gauge fields from table~\ref{Table bosons.} and several 
other scalar and vector gauge fields (see sect.~\ref{masslessandmassivebosons}) couple to the 
condensate with the nonzero quantum number $\tau^{4}$ and $\tau^{23}$ and nonzero family quantum 
numbers. 

It is not difficult to recognize that the desired condensate must have spin zero, $Y=\tau^4 + $ 
$\tau^{23}=0$, $Q=Y + \tau^{13}=0$ and $\vec{\tau}^{1}=0$ in order that in the low energy 
limit the {\it spin-charge-family} theory would manifest effectively as the {\it standard model}.

I make a choice of the two right handed neutrinos of the $VIII^{th}$ family 
coupled into a scalar, with $\tau^4=-1$, $\tau^{23}=1$, correspondingly $Y=0$, $Q=0$ and 
$\vec{\tau}^{1}=0$, and with family quantum numbers~(Eqs. (\ref{so42tilde}, \ref{so1+3tilde})) 
$\tilde{\tau}^4=-1 $, $\tilde{\tau}^{23}=1 $, $\tilde{N}_{R}^{3}=1$, and correspondingly with 
$\tilde{Y}=$ $\tilde{\tau}^4 + \tilde{\tau}^{23}=0$, $\tilde{Q}=\tilde{Y} + \tilde{\tau}^{13}=0$, 
and $\vec{\tilde{\tau}}^{1}=0$. 
The condensate carries the family quantum numbers of the upper four families.

The condensate made out of spinors couples to spinors differently than to antispinors -  
"anticondensate" would namely carry $\tau^4=1$, and $\tau^{23}=-1$ - breaking 
correspondingly the $\mathbb{C}_{{ \cal N}} \cdot $ ${\cal P}_{{\cal N}}$ symmetry:  
The reactions creating particles from antiparticles  are not any longer symmetric to those 
creating antiparticle from particles.
 
Such  a condensate leaves the hyper field $A^{Y}_{m}$ ($= \sin \vartheta_{2}\,A^{23}_{m} +
\cos \vartheta_{2}\,A^{4}_{m}$) (for the choice  that $\sin \vartheta_{2}= \cos \vartheta_{2}$ 
and $g^{4}= g^{2}$, there is no justification for such a choice, $A^{Y}_{m} = \frac{1}{\sqrt{2}}\,
(A^{23}_{m} + A^{4}_{m})$ follows ) massless, while it gives masses to $A^{2\pm}_m$  and $A^{Y'}_m$ 
($=\frac{1}{\sqrt{2}}\,(A^{4}_{m} - A^{23}_{m})$ for $\sin \vartheta_{2}= \cos \vartheta_{2}$) 
and it gives masses also to all the scalar gauge fields from table~\ref{Table bosons.}, since they all couple 
to the condensate through $\tau^{4}$. 

The weak vector gauge fields, $\vec{A}^{1}_m$, the hyper charge vector gauge fields, 
$A^{Y}_{m}$, and the colour vector gauge fields, $\vec{A}^{1}_m$, stay massless. 

The scalar fields with the scalar space index $s=(7,8)$ (there are three singlets which couple to all eight 
families,  two triplets which couple only to the upper four families and another two triplets  which
couple only to the lower four families) -  carrying the weak and the hyper charges of the 
Higgs's scalar - wait for gaining 
nonzero vacuum  expectation values  to change their masses while causing the electroweak break. 

The condensate does  what is needed so that in the low energy regime
the {\it spin-charge-family} manifests as an effective theory which agrees with the 
{\it standard model} to such an extent that it is in agreement with the observed phenomena, 
explaining the {\it standard model} assumptions and predicting new fermion and boson 
fields. 

It also may hopefully explain  the observed matter-antimatter asymmetry if the 
conditions in the expanding universe would be appropriate~(\ref{Sakharovconditions}). The work needed 
to check these conditions in the expanding universe within the {\it spin-charge-family} theory 
is very demanding. Although we do have some experience with following the history of the 
expanding universe~\cite{gn}, this study needs much more efforts, not only in  calculations, 
but also in understanding the mechanism of the condensate appearance, relations among the velocity 
of the expansion, the temperature and the dimension of space-time in the period of the appearance 
of the condensate.
This study has not yet been really started.

\section{Properties of the condensate}
\label{condensate}

In table~\ref{Table con.} the properties of the condensate of the two right handed neutrinos 
($|\nu_{R}^{VIII}>_{1}|\nu_{R}^{VIII}>_{2}\,$), 
one with spin up and another with spin down (table~\ref{Table so13+1.}, line 25 and 26), carrying 
the family quantum numbers of the $VIII^{th}$ 
family (table~\ref{Table III.} ), are presented. 
The  condensate carries the quantum numbers of $SU(2)_{II}$, $\tau^{23}=1$ (Eq.~(\ref{so42})), 
of $U(1)_{II}$ originating in $SO(6)$, $\tau^{4}=-1$ (Eq.\ref{so64}), correspondingly $Y=0$, $Q=0$, 
and the family quantum numbers (table~\ref{Table III.}) 
$\tilde{\tau}^4=-1$ (Eq.~(\ref{so64})), $\tilde{\tau}^{23}=1$ (Eq.~(\ref{so42tilde})), and 
$\tilde{N}_{R}^{3}=1$ (Eq.~(\ref{so1+3tilde})). 
Each of the two neutrinos could belong to a different family of the  upper four families. In this 
case the family quantum numbers of the condensate change.

The condensate is presented in the first line of table~\ref{Table con.}  as a member of a triplet 
of the group $SU(2)_{II}$ with the generators $\tau^{2i}$. Correspondingly the condensate 
couples to all the vector gauge fields which 
carry nonzero $\tau^{2i}$, $\tau^{4}$, $\tilde{\tau}^{2i}$, $\tilde{N}_{R}^{i}$
and $\tilde{\tau}^{4}$. The fields $A^{Y}_{m}, \vec{A}^{3}_{m}$ and $\vec{A}^{1}_{m}$ stay massless. 
The condensate 
couples also to all the scalar 
gauge fields with the scalar indices $s \in (5,6,7,8,9,\dots,14)$, since they all carry nonzero 
either $\tau^{4}$ or $\tau^{23}$.

%
 \begin{table}
 \begin{center}
 \begin{tabular}{|c|c c c c c c c |c c c c c c c|}
 \hline
 state & $S^{03}$& $ S^{12}$ & $\tau^{13}$& $\tau^{23}$ &$\tau^{4}$& $Y$&$Q$&$\tilde{\tau}^{13}$&
 $\tilde{\tau}^{23}$&$\tilde{\tau}^4$&$\tilde{Y} $& $\tilde{Q}$&$\tilde{N}_{L}^{3}$& $\tilde{N}_{R}^{3}$
 \\
 \hline
 ${\bf (|\nu_{1 R}^{VIII}>_{1}\,|\nu_{2 R}^{VIII}>_{2})}$
 & $0$& $0$& $0$& $1$& $-1$ & $0$& $0$& $0$ &$1$& $-1$& $0$& $0$& $0$& $1$\\ 
 \hline
 $ (|\nu_{1 R}^{VIII}>_{1}|e_{2 R}^{VIII}>_{2})$
 & $0$& $0$& $0$& $0$& $-1$ & $-1$& $-1$ & $0$ &$1$& $-1$& $0$& $0$& $0$& $1$\\ 
 $ (|e_{1 R}^{VIII}>_{1}|e_{2 R}^{VIII}>_{2})$
 & $0$& $0$& $0$& $-1$& $-1$ & $-2$& $-2$ & $0$ &$1$& $-1$& $0$& $0$& $0$& $1$\\ 
 \hline 
 \end{tabular}
 \end{center}
\caption{\label{Table con.} 
The condensate of the two right handed neutrinos $\nu_{R}$,  with the $VIII^{th}$ 
family quantum number, coupled to spin zero and belonging to a triplet with 
respect to the generators $\tau^{2i}$, is presented, together with its two partners. 
The condensate carries $\vec{\tau}^{1}=0$, $\tau^{23}=1$, 
$\tau^{4}=-1$ and $Q=0=Y$. The triplet carries $\tilde{\tau}^4=-1$, $\tilde{\tau}^{23}=1$ and 
$\tilde{N}_{R}^3 = 1$, $\tilde{N}_{L}^3 = 0$,  
$\tilde{Y}=0 $, $\tilde{Q}=0$. The family quantum numbers are presented in table~\ref{Table III.}. 
}
 \end{table}

Coupling of the scalar gauge fields to the condensate is proportional to
\begin{eqnarray}
\label{scalcouplecondensate}
&&(<\nu_{1 R}^{VIII}|_{1}\,<\nu_{2 R}^{VIII}|) \,(\gamma^0\,\stackrel{t t'}{(\cpm)}\,\tau^{Ai}\,
A^{Ai}_{\scriptscriptstyle{\stackrel{t t'}{(\cpm)}}})^{\dagger} 
(\gamma^0\,\stackrel{t t'}{(\cpm)}\,\tau^{Ai}\,
A^{Ai}_{\scriptscriptstyle{\stackrel{t t'}{(\cpm)}}})(|\nu_{1 R}^{VIII}>_{1}\,|\nu_{2 R}^{VIII}>)
\,\nonumber\\
&&\propto \,(A^{Ai}_{\scriptscriptstyle{\stackrel{t t'}{(\cpm)}}})^{\dagger}\,
(A^{Ai}_{\scriptscriptstyle{\stackrel{t t'}{(\cpm)}}})\, ,\nonumber\\
&& (t t') \in [(56),(78),(9\,10),\dots,(13\,14)]\,.
\end{eqnarray}

The condensate does break the $\mathbb{C}_{{ \cal N}} \cdot {\cal P}_{{\cal N}}$ symmetry. (The 
"anticondensate" would namely carry $\tau^{23}=-1$ and $\tau^{4}=1$). 

The condensate gives masses to all the scalars from table~\ref{Table bosons.}, either because they 
couple to the condensate due to $\tau^{4}$ or due to $\tau^{4}$ and $\tau^{23}$ quantum numbers. 
It gives masses also to all the scalar fields with $s \in (5,6,7,8)$, since they couple to the condensate 
due to the nonzero $\tau^{23}$. The scalar fields with the quantum numbers of the upper four families 
couple in addition through their family quantum numbers. 

The condensate couples also to all the vector gauge fields except  to the gauge colour octet field 
$\vec{A}^{3}_{m}$, 
the hyper charge vector fields $A^{Y}_{m}$ 
and the weak charge vector triplet fields $\vec{A}^{1}_{m}$,  
since they  carry zero $\tau^{23}$, $\tau^{4}$ and $Y$ quantum numbers. 

The spin connection fields, of either "tilde" ($\tilde{S}^{ab}$) or "nontilde" ($S^{ab}$) origin, 
which do not couple to the 
spinor condensate, are auxiliary fields, expressible with vielbeins fields (appendix~\ref{auxiliary}).

Below the scalar and vector gauge fields are presented, which get masses through the interaction 
with the condensate.
\begin{eqnarray}
&& A^{2\spm}_{\scriptscriptstyle{\stackrel{t t'}{(\cpm)}}}\,,\, \;
   A^{23}_{\scriptscriptstyle{\stackrel{t t'}{(\cpm)}}}\, ,\,\quad
A^{1\spm}_{\scriptscriptstyle{\stackrel{t t'}{(\cpm)}}}\,, \,\;
   A^{13}_{\scriptscriptstyle{\stackrel{t t'}{(\cpm)}}}\, ,\quad 
    \vec{A}^{3}_{\scriptscriptstyle{\stackrel{t t'}{(\cpm)}}}\,,\nonumber\\
&&\tilde{A}^{2 \spm}_{\scriptscriptstyle{\stackrel{t t'}{(\cpm)}}}\,,\,\;
   \tilde{A}^{2 3}_{\scriptscriptstyle{\stackrel{t t'}{(\cpm)}}}\,, \quad 
 \tilde{A}^{1 \spm}_{\scriptscriptstyle{\stackrel{t t'}{(\cpm)}}}\,,\;
   \tilde{A}^{1 3}_{\scriptscriptstyle{\stackrel{t t'}{(\cpm)}}}\,, \quad %
   \nonumber\\  
&&\tilde{A}^{N_{L} \spm}_{\scriptscriptstyle{\stackrel{t t'}{(\cpm)}}}\,,\;
   \tilde{A}^{N_{L} 3}_{\scriptscriptstyle{\stackrel{t t'}{(\cpm)}}}\,, \quad 
 \tilde{A}^{N_{R} \spm}_{\scriptscriptstyle{\stackrel{t t'}{(\cpm)}}}\,,\;
   \tilde{A}^{N_{R} 3}_{\scriptscriptstyle{\stackrel{t t'}{(\cpm)}}}\,, \quad 
\;\; \nonumber\\
&& (t t')\in[(9\,10), (11\,12),(13\,14)]\,, \nonumber\\
&& A^{2\spm}_{\scriptscriptstyle{\stackrel{s s'}{(\cpm)}}}\,, \quad 
A^{Y'}_{\scriptscriptstyle{\stackrel{s s'}{(\cpm)}}}= 
\frac{1}{\sqrt{2}}\,(A^{23}_{\scriptscriptstyle{\stackrel{s s'}{(\cpm)}}}-
A^{4}_{\scriptscriptstyle{\stackrel{s s'}{(\cpm)}}})\,,\nonumber\\
&& (s s')\in[(5 6), (7 8)]\,, \nonumber\\ 
&& A^{2\spm}_{m}\,, \,\; A^{Y'}_{m}= \frac{1}{\sqrt{2}}\, (A^{23}_{m}-A^{4}_{m})\,,\;\; 
    \nonumber\\
&& \vec{\tilde{A}}^{2}_{m}\,, \tilde{A}^{4}_{m}\,, \vec{\tilde{A}}^{N_{R}}_{m}\,,\nonumber\\
&& m \in (0,1,2,3)\,.  
\label{massivefieldscond}
\end{eqnarray}
In expression for $A^{Y'}_{m,s}$ $\vartheta_{2} = \frac{\pi}{4}$ is chosen, just for simplicity, 
with no justification so far.

It stays as an open question what does make the right handed neutrinos 
to form such a condensate in the history of the universe.

Since $A^{Ai}_{s}\,, s \in(5,6)$ couple to the condensate and get masses, while (by assumption) they 
do not get nonzero vacuum expectation values during the electroweak break (what changes the masses
of the scalar fields $A^{Ai}_{s}\,, s \in(7,8)$) the restriction in the sum in Eq.~(\ref{faction}) is 
justified.

The scalar fields, causing the birth of baryons, have the triplet colour charges. They resemble the 
supersymmetric partners of the quarks, but 
since they do not carry all the quantum numbers of the quarks, they are not.

\section{Properties of scalar fields which determine mass matrices of fermions}
\label{CPNandpropScm}

This section is a short overview of the ref.~\cite{NscalarsweakY2014}.

There are two kinds of the scalar gauge fields, which gain at the electroweak break nonzero vacuum 
expectation values and determine correspondingly  masses of the families  of quarks and leptons 
and  masses of gauge weak bosons: The kind originating in $\tilde{\omega}_{\tilde{a}\tilde{b} s}$ 
and the kind originating in $\omega_{s' s" s}$ 
both kinds have the space index $s =(7,8)$ and  carry correspondingly the weak and the hyper charge 
as the Higgs's scalar. 
These scalar fields are presented in the Lagrange density for fermions (Eq.~(\ref{faction})) in the 
second line. The "tilde" kind influences the family quantum numbers of fermions, the "Dirac" kind 
influences the family members quantum numbers. 

The two triplets ($\,\vec{\tilde{A}}^{1}_s$, $\vec{\tilde{A}}^{N_{L}}_s$)  influence the lower four 
families (the lowest three families are already observed), while ($\,\vec{\tilde{A}}^{2}_s$, 
$\vec{\tilde{A}}^{N_{R}}_s$) influence the upper four families, the stable of which constitute the dark
matter. Recognizing that     $\vec{\tilde{\tau}}^{1}\,\vec{\tilde{A}}^{1}_s $ $+ \vec{\tilde{N}}_{L}\,
\vec{\tilde{A}}^{N_{L}}_s$ $+ \vec{\tilde{\tau}}^{2}\,\vec{\tilde{A}}^{2}_s $ $+ \vec{\tilde{N}}_{R}\, 
\vec{\tilde{A}}^{N_{R}}_s =$ $\frac{1}{2}\, \tilde{S}^{ab}\,\tilde{\omega}_{abs}$, $s =(7,8)$, one 
easily finds, taking into account Eqs.~(\ref{so1+3tilde}, \ref{so42tilde}), the expressions  
%
$   \vec{\tilde{A}}^{1}_s =(\tilde{\omega}_{58s}-\tilde{\omega}_{67s}$, $ 
\tilde{\omega}_{57s}+ \, \tilde{\omega}_{68s}$, $\tilde{\omega}_{56s}- \,\tilde{\omega}_{78s} )\,$,
$\vec{\tilde{A}}^{N_{L}}_s =(\tilde{\omega}_{23s}+i\,\tilde{\omega}_{01s}$, $
\tilde{\omega}_{31s}+i \,\tilde{\omega}_{02s}$, $\tilde{\omega}_{12s}+i\,\tilde{\omega}_{03s})\,$, 
$   \vec{\tilde{A}}^{2}_s =(\tilde{\omega}_{58s}+\tilde{\omega}_{67s}$, $
\tilde{\omega}_{57s}- \,\tilde{\omega}_{68s}$, $\tilde{\omega}_{56s}+ \,\tilde{\omega}_{78s} )\,$, 
$\vec{\tilde{A}}^{N_{R}}_s =(\tilde{\omega}_{23s}-i\,\tilde{\omega}_{01s}$, $
\tilde{\omega}_{31s}-i \,\tilde{\omega}_{02s}$, $\tilde{\omega}_{12s}-i\,\tilde{\omega}_{03s})\,$,
$\, s = (7,8)\,$, 
%
presented already in Eq.~(\ref{Atildeomegas}).
Similarly one finds, taking into account Eqs.~(\ref{so1+3}, \ref{so42}, \ref{so64}, \ref{YQY'Q'andtilde}),
the expressions for 
$A^{Q}_{s}$, $A^{Q'}_{s}$ and $A^{Y'}_{s}$, presented in
Eqs.~(\ref{Aomegas}).

The scalar fields $A^{Q}_{s}$, $A^{Q'}_{s}$  and $A^{Y'}_{s}$ distinguish among the family 
members, coupling to them through the family members quantum numbers  $Q\,$ $[Q=(\tau^{13} + Y)\,$, 
$Y\,$ $\,(=\tau^{23} + \tau^{4})\,]$, $Q'\,(= -Y \tan^2 \vartheta_{1} +\tau^{13})$ and $Y'$  
$\, = (\tau^{23}- \tan \vartheta_{2}\,\tau^{4})$, 
$\,\tau^{4}= -\frac{1}{3}(S^{9\,10} + S^{11\,12} + S^{13\,14})$. 

The scalars, originating in 
$\tilde{\omega}_{ab s}$ and distinguishing among families, couple to the family quantum numbers through  
$\vec{\tilde{\tau}}^{1}$ and $\vec{\tilde{N}}_{L}$, or through $\vec{\tilde{\tau}}^{2}$ and  
$\vec{\tilde{N}}_{R}$ .  
These scalars are all in the adjoint  representations of the corresponding subgroups of the 
$\widetilde{SO}(7,1)$ group. 

Let us now prove that all the scalar fields with the space (scalar with respect to $d=(3+1)$) index 
$s=(7,8)$ carry the weak and the hyper charge ($\tau^{13}$, $Y$) equal to either ($-\frac{1}{2}, \frac{1}{2}$)
or to ($\frac{1}{2}, -\frac{1}{2}$). Let us first simplify the notation, using a common name $A^{Ai}_{s}$
for all the scalar fields with the scalar index $ s = (7,8)$
\begin{eqnarray}
A^{Ai}_s &=&(A^{Q}_{s}, \, A^{Q'}_{s}, \, A^{Y'}_{s}, \,\tilde{A}^{4}_{s},\, \vec{\tilde{A}}^{2}_{s},
\vec{\tilde{A}}^{1}_{s}, \,\vec{\tilde{A}}^{N_{R}}_{s},
\,\vec{\tilde{A}}^{N_{L}}_{s})\,,
\label{EWbreakscalarsI}
\end{eqnarray}
and let us rewrite the term $\sum_{s=7,8}\, \bar{\psi} $ $\gamma^s \, p_{0s}\, \psi$ in 
Eq.~(\ref{faction}) as follows
\begin{eqnarray}
\label{M}
 & & \sum_{s=7,8}\, \bar{\psi} \,\gamma^s\, p_{0s}\, \psi\,, \nonumber\\
 & & = \bar{\psi} \{\,\stackrel{78}{(+)} p_{0+} +  \stackrel{78}{(-)} p_{0-}\}\,\psi\,,\nonumber\\
 & & p_{0\pm} = (p_{07}\mp i p_{08})\,,\nonumber\\
 & & (p_{07}\mp i p_{08})= (p_{7} \mp i p_{8}) - \tau^{Ai}\, (A^{Ai}_{7}  \mp i A^{Ai}_{8})\,\nonumber\\
 & & \stackrel{78}{(\pm)} = \frac{1}{2} \,(\gamma^7 \pm i\gamma^8)\,. 
\end{eqnarray}
Let us now apply the operators $Y,Q$, Eq.~(\ref{YQY'Q'andtilde}), and $\tau^{13}= \frac{1}{2}\,(S^{56}-
S^{78})$, Eq.~(\ref{so42}), on the fields $A^{Ai}_{\scriptscriptstyle{\stackrel{78}{(\pm)}}}$ 
$ =(A^{Ai}_{7}\mp i\,A^{Ai}_{8})$. One finds 
\begin{eqnarray}
\label{checktau13Y}
\tau^{13}\,(A^{Ai}_7 \,\mp i\, A^{Ai}_8)&=& \pm \,\frac{1}{2}\,(A^{Ai}_7 \,\mp i\, A^{Ai}_8)\,,\nonumber\\
Y\,(A^{Ai}_7 \,\mp i\, A^{Ai}_8)&=& \mp \,\frac{1}{2}\,(A^{Ai}_7 \,\mp i\, A^{Ai}_8)\,,\nonumber\\
Q\,(A^{Ai}_7 \,\mp i\, A^{Ai}_8)&=&0\,,
\end{eqnarray}
{\it This is, with respect to the weak, the hyper and the electromagnetic charge, just what the 
{\it standard model} assumes for the Higgs' scalars.} The proof is complete. 

One can check also, using Eq.~(\ref{snmb:gammatildegamma}), that $\gamma^0\,\stackrel{78}{(-)} $ 
transforms the $u_{R}^{c1}$ from the first 
line of table~\ref{Table so13+1.} into $u_{L}^{c1}$ from the seventh line of the same table, or $\nu_{R}$
from the $25^{th}$ line into the $\nu_{L}$ from the $31^{th}$ line of the same table.

The  scalars $A^{Ai}_{\scriptscriptstyle{\stackrel{78}{(-)}}}$ obviously bring the weak charge $\frac{1}{2}$
and the hyper charge $-\frac{1}{2}$ to the right handed family members ($u_{R}$, $\nu_{R}$), 
and  the scalars $A^{Ai}_{\scriptscriptstyle{\stackrel{78}{(+)}}}$ bring  the weak charge 
$-\frac{1}{2}$  and the hyper charge $\frac{1}{2}$  to ($d_{R}$, $e_{R}$).

Let us now prove that the scalar fields of Eq.~(\ref{EWbreakscalarsI}) are either triplets with 
respect to the family quantum numbers ($\vec{\tilde{N}}_{R}, \,\vec{\tilde{N}}_{L},\, 
\vec{\tilde{\tau}}^{2},\, \vec{\tilde{\tau}}^{1}$; Eqs.~(\ref{so1+3tilde},~%
\ref{so42tilde})) or singlets as the gauge fields of $Q=\tau^{13}+Y, \,Q'=\tau^{13}-
Y \,\tan^{2}\vartheta_{1}$ and $\,Y'= \tau^{23} - \tan^{2} \vartheta_{2}\,\tau^4 $. 
One can prove this by applying $\vec{\tilde{\tau}}^{2}$,  $\vec{\tilde{\tau}}^{1}$, 
$\vec{\tilde{N}}_{R}$, $\vec{\tilde{N}}_{L}$ and $Q,Q',Y'$ on the states belonging to 
representations of these operators. Let us calculate, as an example,  $\tilde{N}_{L}^{3}$ and $Q$ on 
$\tilde{A}^{N_{L}\spm}_{\scriptscriptstyle{\stackrel{78}{(\pm)}}}$, 
$\,\tilde{A}^{N_{L}3}_{\scriptscriptstyle{\stackrel{78}{(\pm)}}}$ and on 
$A^{Q}_{\scriptscriptstyle{\stackrel{78}{(\pm)}}}$, taking into account Eqs.~(\ref{so1+3tilde},
\ref{so64}, \ref{so42}, \ref{bosonspin})
\begin{eqnarray}
\label{checktildeNL3Q}
\tilde{N}_{L}^{3}\,\tilde{A}^{N_{L}\spm}_{\scriptscriptstyle{\stackrel{78}{(\pm)}}} &=& \spm 
\tilde{A}^{N_{L}\spm}_{\scriptscriptstyle{\stackrel{78}{(\pm)}}}\,,\quad
\tilde{N}_{L}^{3}\,\tilde{A}^{N_{L}3}_{\scriptscriptstyle{\stackrel{78}{(\pm)}}}=0\,,\nonumber\\
Q \,A^{Q}_{\scriptscriptstyle{\stackrel{78}{(\pm)}}} &=&0\,,\nonumber\\
\tilde{A}^{N_{L}\spm}_{\scriptscriptstyle{\stackrel{78}{(\pm)}}} &=& 
\{(\tilde{\omega}_{\scriptscriptstyle{23\stackrel{78}{(\pm)}}} +i \,
\tilde{\omega}_{\scriptscriptstyle{01\stackrel{78}{(\pm)}}})\smp\,i \,
(\tilde{\omega}_{\scriptscriptstyle{31\stackrel{78}{(\pm)}}} +i 
\tilde{\omega}_{\scriptscriptstyle{02\stackrel{78}{(\pm)}}}) \}\,,\nonumber\\ 
\tilde{A}^{N_{L}3}_{\scriptscriptstyle{\stackrel{78}{(\pm)}}} &=& 
(\tilde{\omega}_{\scriptscriptstyle{12\stackrel{78}{(\pm)}}} +i 
\tilde{\omega}_{\scriptscriptstyle{03\stackrel{78}{(\pm)}}})\, \nonumber\\
A^{Q}_{\scriptscriptstyle{\stackrel{78}{(\pm)}}} &=& \sin \vartheta_{1}\,
A^{13}_{\scriptscriptstyle{\stackrel{78}{(\pm)}}} + \cos \vartheta_{1}(-)  
(\omega_{\scriptscriptstyle{9\,10 \stackrel{78}{(\pm)}}} + 
\omega_{\scriptscriptstyle{11\,12 \stackrel{78}{(\pm)}}} +
\omega_{\scriptscriptstyle{13\,14 \stackrel{78}{(\pm)}}})\,,
\end{eqnarray}
with $Q=S^{56} + \tau^{4}= S^{56} -\frac{1}{3}(S^{9\,10}+ S^{11\,12}+ S^{13\,14})$, and with $\tau^{4}$ 
defined in Eq.~(\ref{so64})).

Nonzero vacuum expectation values of the scalar fields  (Eq.~(\ref{EWbreakscalarsI})), which carry
the scalar index $s=(7,8)$, and correspondingly the weak and the hyper charges as calculated in 
Eq.~(\ref{checktau13Y}),  break the mass 
protection mechanism of quarks and leptons of the lower and the upper four families. In  the loop 
corrections besides $\tilde{A}^{Ai}_{s}$ and the scalar fields which are the gauge fields of $Q,Q',Y'$ 
also the vector gauge fields
contribute to all the matrix elements of mass matrix of any family member.

The gauge fields of $\vec{\tilde{N}}_{R}$ and  $\vec{\tilde{\tau}}^{2}$ contribute only to masses of 
the upper four families, while the gauge fields of $\vec{\tilde{N}}_{L}$ and $\vec{\tilde{\tau}}^{1}$ 
contribute only to masses of the lower four families. The triplet scalar fields with the scalar index $s=(7,8)$
and the family charges $\vec{\tilde{N}}_{R}$ and $\vec{\tilde{\tau}}^{2}$ 
transform any family member belonging to the group of the upper four families into the same family 
member belonging to another family of the same group of four families, changing the right handed 
member into the left handed partner, while those triplets with the family charges $\vec{\tilde{N}}_{L}$ and  
$\vec{\tilde{\tau}}^{1}$  transform any family member of particular handedness and belonging to the lower 
four families into its partner of opposite handedness,  belonging to another family of the lower 
four families. 

The scalars $A^{Q}_{{\scriptscriptstyle{\stackrel{78}{(\pm)}}}}$ (Eq.~(\ref{checktildeNL3Q})),  
$A^{Q'}_{{\scriptscriptstyle{\stackrel{78}{(\pm)}}}}$  ($= \cos \vartheta_{1}\, 
A^{13}_{{\scriptscriptstyle{\stackrel{78}{(\pm)}}}}- \sin \vartheta_{1}
  A^{Y}_{{\scriptscriptstyle{\stackrel{78}{(\pm)}}}}$) and  
$A^{Y'}_{{\scriptscriptstyle{\stackrel{78}{(\pm)}}}}$  (Eq.~(\ref{massivefieldscond}))
contribute to all eight families, distinguishing among the family members and not among the families.

The mass matrix of any  family member, belonging to any of the two  groups of the four families, 
manifests -  
due to the $\widetilde{SU}(2)_{(R,L)}\times \widetilde{SU}(2)_{(II,I)}$ (either ($R,II$) or ($L,I$))  
structure of the scalar fields, which are the gauge fields of  
the $\vec{\tilde{N}}_{R,L}$ and $\vec{\tilde{\tau}}^{2,1}$ - the symmetry presented in Eq.~(\ref{M0})
 \begin{equation}
 \label{M0}
 {\cal M}^{\alpha} = \begin{pmatrix} 
 - a_1 - a & e     &   d & b\\ 
 e     & - a_2 - a &   b & d\\ 
 d     & b     & a_2 - a & e\\
 b     &  d    & e   & a_1 - a
 \end{pmatrix}^{\alpha}\,.
 \end{equation}
In the ref.~\cite{gn2014} the mass matrices for quarks, which are in the agreement with the experimental data,
are presented and predictions made. It is  demonstrated in this reference that the improved measurements of
the quarks mixing matrix are in better agreement with the by the {\it spin-charge-family} theory predicted 
symmetry of mass matrices then the previous ones.

\section{Condensate and nonzero vacuum expectation values of  
scalar fields make  spinors and most of scalar and vector gauge fields massive}
\label{masslessandmassivebosons}

Let us shortly overview properties of the scalar and the vector gauge fields \\
{\bf i.} after 
two right handed neutrinos  (coupled to spin zero and with the family quantum 
numbers, table~\ref{Table III.}, of the upper four families)  make a 
condensate~(table~\ref{Table con.}) at the scale $\ge 10^{16}$ GeV, and \\
{\bf ii.} after 
the electroweak break, when the scalar fields with the space index $s=(7,8)$ get 
nonzero vacuum expectation values.

All the scalar gauge fields $A^{Ai}_{t}, t\in (5,6,7,8,9,\dots,14)$ (Eqs.~(\ref{faction}, 
\ref{factionMaM1}, \ref{massivefieldscond}), table~\ref{Table bosons.}) interact with 
the condensate through the quantum numbers $\tau^{4}$ and $\tau^{23}$ - those with the family 
quantum  numbers of the upper four families interact also through the family quantum numbers 
$\vec{\tilde{\tau}}^{2}$ or $\vec{\tilde{N}}_{R}$ - getting masses of the order of the 
condensate scale~(Eq.(\ref{massivefieldscond})).

At the electroweak break the scalar fields $A^{Ai}_{s}, \,s \in (7,8)$,  from 
Eq.~(\ref{EWbreakscalarsI}) get nonzero vacuum expectation values, 
changing correspondingly their own masses and determining masses of quarks and leptons, as 
well as of the weak vector gauge fields.

 The vector gauge fields $A^{2\spm}_{m},\,A^{Y'}_{m} $,  
 $\vec{\tilde{A}}^{2\spm}_{m}$, $\tilde{A}^{Y'}_{m}$ 
 and $\vec{\tilde{A}}^{N_{R}}_{m}$ (Eq.~(\ref{massivefieldscond}))
 get masses due to the interaction with the condensate through $\tau^{23}$ and $\tau^{4}$, 
the first two, or  due to the family quantum numbers of the upper four families, the last 
three, respectively.

 The vector gauge fields $\vec{A}^{3}_{m},\,$ $\vec{A}^{1}_{m}$, and $A^{Y}_{m} $ stay massless 
 up to the electroweak break when the scalar gauge fields, which are weak doublets with the 
 hyper charge making their  electromagnetic charge $Q$ equal to zero, give masses to the weak bosons 
 ($A^{1\spm}_{m}=\frac{1}{\sqrt{2}}\, (A^{11}_{m}\mp i A^{11}_{m})$ and $A^{Q'}_{m}= \cos \vartheta_{1}
 \, A^{13}_{m}-\sin \vartheta_{1}\, A^{Y}_{m} $), while the electromagnetic vector field ($A^{Q}_{m}= 
 \sin \vartheta_{1} \, A^{13}_{m}+ \cos \vartheta_{1}\, A^{Y}_{m} $) and the colour vector  gauge field 
 stay massless.
 
 At the electroweak break, when the nonzero vacuum expectation values of the scalar fields break
 the weak and the hypercharge global symmetry, also all the eight families of quarks and leptons 
 get masses. Up to the electroweak break the families were mass protected, since the right handed 
 partners distinguished from the left handed ones in the weak and hyper charges, which were the 
 conserved quantum numbers, what disabled them 
 to make the superposition manifesting masses.

\section{Sakharov conditions as seen in view of the {\it spin-charge-family} theory}
\label{Sakharovconditions}

The condensate of the right handed neutrinos, as well as the nonzero vacuum expectation 
values of the scalar fields $A^{Ai}_{\scriptscriptstyle{\stackrel{78}{(\pm)}}}$ - if leading 
to the complex matrix elements of the mixing matrices - cause the 
${\bf \mathbb{C}}_{\cal N} $ ${\cal P}_{\cal N}$ violation terms, which generate the 
matter-antimatter asymmetry.

It is the question whether  both generators of the matter-antimatter asymmetry - the 
condensate and the complex phases of the mixing matrices of quarks and leptons (this last 
alone can not with one complex phase and also very probably not with the three complex phases 
of the lower four families) - can explain 
at all the observed matter-antimatter asymmetry of the "ordinary" matter, that is the matter 
of mostly the first family of quarks and leptons.

The lowest of the upper four families determine the dark matter. For the dark matter   
any relation among matter and antimatter is so far experimentally allowed.

Both origins of the matter-antimatter asymmetry - the condensate and the nonzero vacuum 
expectation values of the scalar fields carrying the weak and the hyper charge - (are assumed 
to) appear spontaneously.

Sakharov~\cite{Sakharov} states that for the matter-antimatter asymmetry three conditions must 
be fulfilled:\\
{\bf a.} (${\cal C}_{\cal N}$ and) ${\bf \mathbb{C}}_{\cal N} $ ${\cal P}_{\cal N}$ must not 
be conserved.\\
{\bf b.} Baryon number non conserving processes must take place.\\
{\bf c.} Thermal  non equilibrium must be present not to equilibrate the number of baryons 
and antibaryons. 

Sakharov uses for {\bf c.} the requirement that ${\cal C}{\cal P}{\cal T}$ must be conserved 
and that $\{{\cal C}{\cal P}{\cal T},H\}_{-}=0$. 
In a thermal equilibrium the average number of baryons $<n_{B}> = 
Tr(e^{-\beta H} n_{B})= Tr(e^{-\beta H}{\cal C}{\cal P}{\cal T} n_{B}({\cal C}{\cal P}{\cal T})^{-1})=$
$<\bar{n}_{B}>  $. 
Therefore $<n_{B}> - <\bar{n}_{B}> =0$ at the thermal equilibrium and there is no excess of 
baryons with respect to antibaryons. In the expanding universe, however, the temperature is 
changing with time.  It is needed that the discrete symmetry  
${\bf \mathbb{C}}_{\cal N} $ ${\cal P}_{\cal N}$ is broken to break the symmetry between 
matter and antimatter, if the universe starts with no matter-antimatter asymmetry.

The {\it spin-charge-family} theory starting action~(Eq.(\ref{wholeaction})) is invariant   
under the $\mathbb{C}_{\cal N} $ ${\cal P}_{\cal N}$ symmetry. The scalar 
fields~(Eq.(\ref{factionMaM1})) of this theory cause transitions, in which  
a quark is born out of a positron (figures~(\ref{proton is born1.}, \ref{proton is born2.}))
and a quark is born out of antiquark, and back. 
These reactions  go in both directions with the same probability, until the spontaneous break of the 
${\bf \mathbb{C}}_{\cal N} $ ${\cal P}_{\cal N}$ symmetry is caused by the appearance of the 
condensate of the two right handed neutrinos (table~\ref{Table con.}). 

But after the 
appearance of the condensate (and in addition of the 
appearance of the non zero vacuum expectation values of the scalar fields with the space index 
$s\in (7,8)$), family members "see" the vacuum differently than  
the antimembers. And this {\it might} explain the matter-antimatter asymmetry provided that 
the conditions of the expanding universe at the appearance of the condensate (and later at the 
electroweak break) make the matter-antimatter asymmetry strong enough so that it is not later 
"washed out". Massive scalar fields with the colour charge predict also the proton decay. 

It is, of course, the question whether both phenomena can at all explain the observed 
matter-antimatter asymmetry. I agree completely with the referee of this paper that before 
answering the question whether or not the {\it spin-charge-family} theory explains this 
observed phenomena, one must do a lot of additional work to find out: i. Which is the order  
of phase transition, which leads to the appearance of the condensate. ii. How strong is the 
thermal non equilibrium,  which leads to the matter-antimatter asymmetry during the phase 
transition. iii. How rapid is the appearance of the matter-antimatter asymmetry in comparison 
with the expansion of the universe. iv. Does the later history of the expanding universe 
enable that the produced asymmetry survives up to today. 

Although we do have some experience with solving the Boltzmann equations for fermions and 
antifermions~\cite{gn} to follow the history of the dark matter within the {\it spin-charge-family} 
theory, the study of the history of the universe from the very high temperature 
to the baryon production within the same theory in order to see the matter-antimatter asymmetry 
in the present time is much more demanding task.  
These are under consideration, but  still at a very starting point since a lot of things must 
be understood before starting with the calculations.

What I can conclude is only that the {\it spin-charge-family} theory  does offer the opportunity 
also for the explanation for the observed matter-antimatter asymmetry.

\section{Conclusions}
\label{discussions}

The {\it spin-charge-family}~\cite{NBled2013,JMP,pikanorma,norma92,norma93,norma94,norma95,%
NBled2012,gmdn07,gn,NscalarsweakY2014,HNds} theory is a kind of the Kaluza-Klein theories 
in $d=(13+1)$ but with the families introduced by the second kind of gamma operators  - 
the $\tilde{\gamma}^a$ operators in addition to the Dirac $\gamma^a$. The theory  
assumes a simple starting action (Eq.~(\ref{wholeaction})) in $d=(13+1)$. This simple action 
manifests in the low energy regime, after the breaks of symmetries 
(subsection~\ref{actionandassumptions}), all the degrees of freedom assumed in the {\it standard 
model}, offering the explanation for all the properties of quarks and leptons (right handed neutrinos 
are in this theory the regular members of each family) and antiquarks and antileptons.
The theory explains the existence of the observed gauge vector fields. It explains  the origin of the
scalar fields (the Higgs's  scalars and the Yukawa couplings) responsible for the quark and lepton 
masses and the masses of the weak bosons~\cite{NscalarsweakY2014}. 

The theory is offering the explanation also for the matter-antimatter asymmetry and for the appearance 
of the dark matter.

The {\it spin-charge-family} theory predicts two  decoupled groups of four families~\cite{JMP,%
pikanorma,gmdn07,gn}: The fourth of the lower group of four families will be measured at the 
LHC~\cite{gn2013} and the lowest of the upper four families constitutes the dark 
matter~\cite{gn} and was already seen. It also predicts that there might be more scalar 
fields observable at the LHC. The upper four families manifest, due to their high masses,  
a new "nuclear force" among their baryons.

All these degrees of freedom are contained in the simple starting action. The scalar 
fields with  the weak and the hyper charges equal to ($\mp \frac{1}{2}, \pm \frac{1}{2}$), 
respectively (section~\ref{CPNandpropScm}), have the space index $s=(7,8)$, while they carry 
in addition to the weak and the hyper charges also either the family quantum numbers, originating in 
$\tilde{S}^{ab}$ (they form two groups of twice $SU(2)$ triplets), or the family members quantum 
numbers, originating in $S^{ab}$ (they form three singlets with the quantum numbers ($Q,Q',Y'$)). 
These scalar fields cause the transitions 
of the right handed family members into the left handed partners and back. Those with the family 
quantum numbers cause at the same time transitions among families within each of the two family 
groups of four families. They all gain in the electroweak break nonzero vacuum expectation values, 
giving masses to both groups of four families of quarks and leptons and to weak bosons (changing 
also their own masses). 

There are in this theory also  scalar fields with the space index $s=(5,6)$;  They carry with 
respect to this degree of freedom the weak charge equal to the hyper charge ($\mp \frac{1}{2}$, 
$ \mp \frac{1}{2}$, respectively). They carry also additional quantum 
numbers~(Eq.(\ref{massivefieldscond})) like all the other scalar fields: Either the family 
quantum numbers, originating in $\tilde{S}^{ab}$,  or  the family members quantum numbers, 
originating in $S^{ab}$. 

And there are also the scalar  fields with the scalar index $s=(9,10,\cdots,14)$. These scalars 
carry the triplet colour charge with respect to the space index and the additional quantum numbers 
(table~\ref{Table bosons.}),  originating either in the family quantum numbers $\tilde{S}^{ab}$ or in the 
family members quantum numbers $S^{ab}$.

There are no additional scalar gauge fields in this theory. 

There are the vector gauge fields with respect to $d=(3+1)$: $A^{Ai}_{m}$, with $Ai$ staying for the 
groups $SU(3)$ and $U(1)$ (both originating in $SO(6)$ of $SO(13,1)$), for the groups $SU(2)_{II}$ and 
$SU(2)_{I}$ (both originating in $SO(4)$ of $SO(7,1)$) and for the groups $SU(2) \times SU(2)$ 
($\in SO(3,1)$),  in both sectors,  $S^{ab}$ and 
$\tilde{S}^{ab}$ ones.

The condensate of the two right handed neutrinos with the family charges of the upper four families 
(table~\ref{Table con.}) gives masses to all the scalar and vector gauge fields, except 
to the colour octet vector, the hyper  singlet vector and the weak triplet vector gauge fields, 
to which the condensate does not couple. 
Those vector gauge fields of either $S^{ab}$ or $\tilde{S}^{ab}$ origin, which do not couple to 
the condensate, are expressible with the corresponding vielbeins (appendices \ref{omegaabalpha}, 
\ref{omegatildeabalpha}), they are auxiliary fields. The condensate breaks the
${\bf \mathbb{C}}_{\cal N} $ ${\cal P}_{\cal N}$ symmetry (sections~(\ref{condensate}, \ref{discrete})).

There are no additional vector gauge fields in this theory. 

Nonzero vacuum expectation values of the scalar gauge fields with the space index $s=(7,8)$ and the quantum
numbers as explained in the fourth paragraph of this section change in the electroweak break their own masses,
while all the other scalars or vectors either stay massless (the colour octet, the electromagnetic field),
or keep the masses of the scale of the condensate. The only  vector fields massless before  the electroweak 
break, which become at the electroweak break massive, are the heavy bosons.

{\em It is extremely encouraging that the simple starting action of the {\em spin-charge-family} 
offers  at low energies the explanations for so many observed phenomena.} But it is also true that the 
starting assumptions (section~\ref{actionandassumptions}) wait to be derived from the initial and 
boundary conditions of the expanding universe.

This paper is a step towards understanding the matter-antimatter asymmetry within the 
{\it spin-charge-family} theory, predicting also the proton decay. The theory obviously offers the 
possibility that the scalar gauge fields with the space index $s=(9,10,\cdots, 14)$ explain, after the 
appearance of the condensate,  the matter-antimatter asymmetry. 
To prove, however, that this indeed  happens, requires the additional study: Following the universe through 
the phase transitions which breaks the ${\bf \mathbb{C}}_{\cal N} $ ${\cal P}_{\cal N}$ symmetry  at the 
level of the condensate and further through the electroweak phase transition up to today, to check how much of the 
matter-antimatter asymmetry is left. 
The experience when following the history of the expanding universe to see whether the {\it spin-charge-family} 
theory can explain the dark matter content~\cite{gn} is of some help. However,  
answering the question to which extent this theory can explain the observed matter-antimatter 
asymmetry requires a lot of additional understanding and work.

Let me conclude with the recognition, pointed out already in the introduction, that the 
{\it spin-charge-family} theory overlaps in many points with other 
unifying theories~\cite{GeorgiGlashow,Georgi,FM,GellMann,BEGN,Zee}, since all the unifying groups  
can be recognized as the subgroups of the large enough orthogonal groups, with family groups included. 
But there are also many differences: The {\it spin-charge-family} theory starts with a very simple action, 
from where all the properties of spinors and the gauge vector and scalar fields follow, provided 
that the breaks of symmetries occur in the desired way. 
Consequently it differs from other unifying theories in the degrees of freedom of spinors and 
scalar and vector gauge fields which show up on different levels of the break of symmetries,
in the unification scheme, in the family degrees of freedom and correspondingly also in the 
evolution of our universe.

\appendix

\section{Discrete symmetry operators~\cite{HNds}} 
\label{discrete}

I present here the discrete symmetry operators in the second quantized picture, for the description 
of which the Dirac sea is used. I  follow the reference~\cite{HNds}. The discrete symmetry operators 
of this reference are designed for the Kaluza-Klein  like theories, in which the total angular 
momentum in higher than $(3+1)$ dimensions manifest as charges in $d=(3+1)$. The dimension of 
space-time is even, as it is in the case of the {\it spin-charge-family} theory.
\begin{eqnarray}
\label{CPTN}
{\cal C}_{{\cal N}}  &= & \prod_{\Im \gamma^m, m=0}^{3} \gamma^m\,\, \,\Gamma^{(3+1)} \,
K \,I_{x^6,x^8,\dots,x^{d}}  \,,\nonumber\\
{\cal T}_{{\cal N}}  &= & \prod_{\Re \gamma^m, m=1}^{3} \gamma^m \,\,\,\Gamma^{(3+1)}\,K \,
I_{x^0}\,I_{x^5,x^7,\dots,x^{d-1}}\,,\nonumber\\
{\cal P}_{{\cal N}}  &= & \gamma^0\,\Gamma^{(3+1)}\, \Gamma^{(d)}\, I_{\vec{x}_{3}}
\,.
\end{eqnarray}
The operator of handedness in even $d$ dimensional spaces is defined as
\begin{eqnarray}
\label{handedness}
\Gamma^{(d)} :=(i)^{d/2}\; \prod_a \:(\sqrt{\eta^{aa}}\, \gamma^a)\,,
\end{eqnarray}
 with products of 
$\gamma^a$ in ascending order. We choose $\gamma^0$, $\gamma^1$ real, $\gamma^2$ imaginary, 
$\gamma^3$ real, $\gamma^5$ imaginary, $\gamma^6$ real, alternating imaginary and real up to 
$\gamma^d$ real. 
Operators $I$ operate as follows: $\quad I_{x^0} x^0 = -x^0$; $I_{x} x^a =- x^a$; $I_{x^0} x^a =$ 
$(-x^0,\vec{x})$; $I_{\vec{x}} \vec{x} = -\vec{x}$; $I_{\vec{x}_{3}} x^a = (x^0, -x^1,-x^2,-x^3,x^5, 
x^6,$ $\dots, x^d)$; $I_{x^5,x^7,\dots,x^{d-1}}$ $(x^0,x^1,x^2,x^3,x^5,x^6,x^7,x^8,$ 
$\dots,x^{d-1},x^d)$ $=(x^0,x^1,x^2,x^3,-x^5,x^6,-x^7,$ $\dots,-x^{d-1},x^d)$; $I_{x^6,x^8,\dots,x^d}$ 
$(x^0,x^1,x^2,x^3,x^5,x^6,x^7,x^8,$ $\dots,x^{d-1},x^d)$ $=(x^0,x^1,x^2,x^3,x^5,-x^6,x^7,-x^8,$ 
$\dots,x^{d-1},-x^d)$, $d=2n$. 

${\cal C}_{{\cal N}}$ transforms the state, put on the top of the Dirac sea, into the corresponding 
negative energy state in the Dirac sea.

The operator, it is named~\cite{NBled2013,TDN2013,HNds} $\mathbb{C}_{{ \cal N}}$, is needed, which 
transforms the starting single particle state on the top of the Dirac sea into the negative energy 
state and then empties this negative energy state.  
This hole in the Dirac sea is the antiparticle state put on the top of the Dirac sea. Both, a 
particle and its antiparticle state (both put on the top of  the Dirac sea), must solve the Weyl 
equations of motion.

This $\mathbb{C}_{{ \cal N}}$ is defined as a product of the operator~\cite{NBled2013,TDN2013} 
$"emptying"$, (making transformations into a completely different Fock space)
\begin{eqnarray}
\label{empt}
"emptying"&=& \prod_{\Re \gamma^a}\, \gamma^a \,K =(-)^{\frac{d}{2}+1} \prod_{\Im \gamma^a} 
\gamma^a \,\Gamma^{(d)} K\,, 
\end{eqnarray}
and ${\cal C}_{{\cal N}}$
\begin{eqnarray}
\label{CemptPTN}
\mathbb{C}_{{ \cal N}} &=& 
\prod_{\Re \gamma^a, a=0}^{d} \gamma^a \,\,\,K
\, \prod_{\Im \gamma^m, m=0}^{3} \gamma^m \,\,\,\Gamma^{(3+1)} \,K \,I_{x^6,x^8,\dots,x^{d}}
\nonumber\\
&=& 
\prod_{\Re \gamma^s, s=5}^{d} \gamma^s \, \,I_{x^6,x^8,\dots,x^{d}}\,.
\end{eqnarray}

We shall need inded only the product of operators $\mathbb{C}_{{ \cal N}} {\cal P}_{{\cal N}}$, 
${\cal T}_{{\cal N}}$ and $\mathbb{C}_{{ \cal N}} {\cal P}_{{\cal N}}$ ${\cal T}_{{\cal N}}$, 
since either $\mathbb{C}_{{ \cal N}}$ or $ {\cal P}_{{\cal N}}$ have in even dimensional spaces
with $d=2(2n+1)$ an odd number of $\gamma^a$ operators, transforming accordingly states from the 
representation of one handedness in $d=2(2n+1)$ into the Weyl of another handedness. 
\begin{eqnarray}
\label{CPcptN}
\mathbb{C}_{{ \cal N}} {\cal P}_{{\cal N}} &=& \gamma^{0}\, 
\prod_{\Im \gamma^s, s=5}^{d} \gamma^s \,\, I_{\vec{x}_{3}}\,I_{x^6,x^8,\dots,x^{d}}\,,
\nonumber\\
\mathbb{C}_{{ \cal N}} {\cal P}_{{\cal N}}{\cal T}_{{\cal N}}&=&  
\prod_{\Im \gamma^a, a=0}^{d} \gamma^a \,\, K \,I_{x}\,.
\end{eqnarray}
\section{Short presentation of technique~\cite{norma93,hn02,hn03}}
\label{technique}

I make in this appendix a short review of the technique~\cite{{hn02,hn03}}, 
initiated and developed~\cite{norma92,norma93,norma94,norma95} when  proposing the 
{\it spin-charge-family} theory~\cite{norma92,norma93,norma95,pikanorma,NBled2013,NBled2012,%
gn,gmdn07}  assuming that
all the internal degrees of freedom of spinors, with family quantum number included, are 
describable in the space of $d$-anticommuting (Grassmann) coordinates~\cite{norma93}, if the 
dimension of ordinary space is $d$. There are two  kinds of operators in the Grassmann space, 
fulfilling the Clifford algebra, which anticommute with one another. 
The technique  was further 
developed in the present shape together with H.B. Nielsen~\cite{hn02,hn03} by identifying 
one kind of the Clifford objects with $\gamma^s$'s and another kind with  $\tilde{\gamma}^a$'s.  

The objects $\gamma^a$ and $\tilde{\gamma}^a$ have properties
\begin{eqnarray}
\label{gammatildegamma}
&& \{ \gamma^a, \gamma^b\}_{+} = 2\eta^{ab}\,, \quad\quad    
\{ \tilde{\gamma}^a, \tilde{\gamma}^b\}_{+}= 2\eta^{ab}\,, \quad,\quad
\{ \gamma^a, \tilde{\gamma}^b\}_{+} = 0\,,\nonumber\\
\end{eqnarray}
%
%
%

If $B$ is a Clifford algebra object, let say a polynomial of $\gamma^a$,  $B=a_{0} + a_{a}
\gamma^a + a_{ab} \gamma^a \gamma^b + \cdots + a_{a_1 a_2 \cdots a_d} \gamma^{a_1}\gamma^{a_2}\cdots
\gamma^{a_d}$, 
then one finds 
\begin{eqnarray}
(\tilde{\gamma}^a B : &=& i(-)^{n_B} \,B \gamma^a \,) \;|\psi_0 \,>, \nonumber\\
B &=& a_0 + a_{a_0} \gamma^{a_0} + a_{a_1 a_2} \gamma^{a_1} \gamma^{a_2} + \cdots + 
a_{a_1 \cdots a_d} \gamma^{a_1}\cdots \gamma^{a_d} \,,
\label{tildegclifford}
\end{eqnarray}
where $|\psi_{0}>$ is a vacuum state, defined in Eq.~(\ref{graphherscal}) and 
$(-)^{n_B} $ is equal to $1$ for the term in the polynomial which has an even number of $\gamma^b$'s, 
and to $-1$ for the term with an odd number of  $\gamma^b$'s.

In this last stage we  constructed a spinor basis as products of nilpotents and projectors  formed  
as odd and even objects of $\gamma^a$'s, respectively, and  chosen to be eigenstates 
of a Cartan subalgebra of the Lorentz groups defined by $\gamma^a$'s and $\tilde{\gamma}^a$'s.

The technique can be used to construct a spinor basis for any dimension $d$
and any signature in an easy and transparent way. Equipped with the graphic presentation of basic states,  
the technique offers an elegant way to see all the quantum numbers of states with respect to the two 
Lorentz groups, as well as transformation properties of the states under any Clifford algebra object.

The Clifford algebra objects $S^{ab}$ and $\tilde{S}^{ab}$ close the algebra of the Lorentz group 
\begin{eqnarray}
\label{sabtildesab}
\
S^{ab}: &=& (i/4) (\gamma^a \gamma^b - \gamma^b \gamma^a)\,, \nonumber\\
\tilde{S}^{ab}: &=& (i/4) (\tilde{\gamma}^a \tilde{\gamma}^b 
- \tilde{\gamma}^b \tilde{\gamma}^a)\,,\nonumber\\
 \{S^{ab}, \tilde{S}^{cd}\}_{-}&=& 0\,,\nonumber\\
\{S^{ab},S^{cd}\}_{-} &=& i(\eta^{ad} S^{bc} + \eta^{bc} S^{ad} - \eta^{ac} S^{bd} - \eta^{bd} S^{ac})\,,
\nonumber\\
\{\tilde{S}^{ab},\tilde{S}^{cd}\}_{-} &=& i(\eta^{ad} \tilde{S}^{bc} + \eta^{bc} \tilde{S}^{ad} 
- \eta^{ac} \tilde{S}^{bd} - \eta^{bd} \tilde{S}^{ac})\,,
\end{eqnarray}

We assume  the ``Hermiticity'' property for $\gamma^a$'s  and $\tilde{\gamma}^a$'s 
\begin{eqnarray}
\gamma^{a\dagger} = \eta^{aa} \gamma^a\,,\quad \quad \tilde{\gamma}^{a\dagger} = \eta^{aa} \tilde{\gamma}^a\,,
\label{cliffher}
\end{eqnarray}
in order that 
$\gamma^a$ and $\tilde{\gamma}^a$ are compatible with (\ref{gammatildegamma}) and formally unitary, 
i.e. $\gamma^{a \,\dagger} \,\gamma^a=I$ and $\tilde{\gamma}^{a\,\dagger} \tilde{\gamma}^a=I$.

One finds from Eq.(\ref{cliffher}) that $(S^{ab})^{\dagger} = \eta^{aa} \eta^{bb}S^{ab}$.

Recognizing from Eq.(\ref{sabtildesab})  that two Clifford algebra objects 
$S^{ab}, S^{cd}$ with all indices different commute, and equivalently for 
$\tilde{S}^{ab},\tilde{S}^{cd}$, we  select  the Cartan subalgebra of the algebra of the 
two groups, which  form  equivalent representations with respect to one another 
\begin{eqnarray}
S^{03}, S^{12}, S^{56}, \cdots, S^{d-1\; d}, \quad {\rm if } \quad d &=& 2n\ge 4,
\nonumber\\
S^{03}, S^{12}, \cdots, S^{d-2 \;d-1}, \quad {\rm if } \quad d &=& (2n +1) >4\,,
\nonumber\\
\tilde{S}^{03}, \tilde{S}^{12}, \tilde{S}^{56}, \cdots, \tilde{S}^{d-1\; d}, 
\quad {\rm if } \quad d &=& 2n\ge 4\,,
\nonumber\\
\tilde{S}^{03}, \tilde{S}^{12}, \cdots, \tilde{S}^{d-2 \;d-1}, 
\quad {\rm if } \quad d &=& (2n +1) >4\,.
\label{choicecartan}
\end{eqnarray}

The choice for  the Cartan subalgebra in $d <4$ is straightforward.
It is  useful  to define one of the Casimirs of the Lorentz group -  
the  handedness $\Gamma$ ($\{\Gamma, S^{ab}\}_- =0$) in any $d$ 
\begin{eqnarray}
\Gamma^{(d)} :&=&(i)^{d/2}\; \;\;\;\;\;\prod_a \quad (\sqrt{\eta^{aa}} \gamma^a), \quad {\rm if } \quad d = 2n, 
\nonumber\\
\Gamma^{(d)} :&=& (i)^{(d-1)/2}\; \prod_a \quad (\sqrt{\eta^{aa}} \gamma^a), \quad {\rm if } \quad d = 2n +1\,.
\label{hand}
\end{eqnarray}
One proceeds equivalently for $\tilde{\Gamma}^{(d)} $, subtituting $\gamma^a$'s by $\tilde{\gamma}^a$'s.
We understand the product of $\gamma^a$'s in the ascending order with respect to 
the index $a$: $\gamma^0 \gamma^1\cdots \gamma^d$. 
It follows from Eq.(\ref{cliffher})
for any choice of the signature $\eta^{aa}$ that
$\Gamma^{\dagger}= \Gamma,\;
\Gamma^2 = I.$
We also find that for $d$ even the handedness  anticommutes with the Clifford algebra objects 
$\gamma^a$ ($\{\gamma^a, \Gamma \}_+ = 0$) , while for $d$ odd it commutes with  
$\gamma^a$ ($\{\gamma^a, \Gamma \}_- = 0$). 

To make the technique simple we introduce the graphic presentation 
as follows 
\begin{eqnarray}
\stackrel{ab}{(k)}:&=& 
\frac{1}{2}(\gamma^a + \frac{\eta^{aa}}{ik} \gamma^b)\,,\quad \quad
\stackrel{ab}{[k]}:=
\frac{1}{2}(1+ \frac{i}{k} \gamma^a \gamma^b)\,,\nonumber\\
\stackrel{+}{\circ}:&=& \frac{1}{2} (1+\Gamma)\,,\quad \quad
\stackrel{-}{\bullet}:= \frac{1}{2}(1-\Gamma),
\label{signature}
\end{eqnarray}
where $k^2 = \eta^{aa} \eta^{bb}$.
It follows then 
\begin{eqnarray}
\gamma^{a}&=& \stackrel{ab}{(k)} + \stackrel{ab}{(-k)}\,, \quad \quad 
\gamma^{b} = ik\eta^{aa}\,(\stackrel{ab}{(k)} - \stackrel{ab}{(-k)})\,,\nonumber\\
S^{ab}    &=& \frac{k}{2} (\stackrel{ab}{[k]}- \stackrel{ab}{[-k]})
\label{signaturegamma}
\end{eqnarray}
One can easily check by taking into account the Clifford algebra relation 
(Eq.\ref{gammatildegamma}) and the
definition of $S^{ab}$ and $\tilde{S}^{ab}$ (Eq.\ref{sabtildesab})
that if one multiplies from the left hand side by $S^{ab}$ or $\tilde{S}^{ab}$ the Clifford 
algebra objects $\stackrel{ab}{(k)}$
and $\stackrel{ab}{[k]}$,
it follows that
\begin{eqnarray}
        S^{ab}\, \stackrel{ab}{(k)}= \frac{1}{2}\,k\, \stackrel{ab}{(k)}\,,\quad \quad 
        S^{ab}\, \stackrel{ab}{[k]}= \frac{1}{2}\,k \,\stackrel{ab}{[k]}\,,\nonumber\\
\tilde{S}^{ab}\, \stackrel{ab}{(k)}= \frac{1}{2}\,k \,\stackrel{ab}{(k)}\,,\quad \quad 
\tilde{S}^{ab}\, \stackrel{ab}{[k]}=-\frac{1}{2}\,k \,\stackrel{ab}{[k]}\,,
\label{grapheigen}
\end{eqnarray}
which means that we get the same objects back multiplied by the constant $\frac{1}{2}k$ in the case 
of $S^{ab}$, while $\tilde{S}^{ab}$ multiply $\stackrel{ab}{(k)}$ by $k$ and $\stackrel{ab}{[k]}$ 
by $(-k)$ rather than $(k)$. 
This also means that when 
$\stackrel{ab}{(k)}$ and $\stackrel{ab}{[k]}$ act from the left hand side on  a
vacuum state $|\psi_0\rangle$ the obtained states are the eigenvectors of $S^{ab}$.
We further recognize 
that $\gamma^a$ 
transform  $\stackrel{ab}{(k)}$ into  $\stackrel{ab}{[-k]}$, never to $\stackrel{ab}{[k]}$, 
while $\tilde{\gamma}^a$ transform  $\stackrel{ab}{(k)}$ into $\stackrel{ab}{[k]}$, never to 
$\stackrel{ab}{[-k]}$ 
\begin{eqnarray}
&&\gamma^a \stackrel{ab}{(k)}= \eta^{aa}\stackrel{ab}{[-k]},\; 
\gamma^b \stackrel{ab}{(k)}= -ik \stackrel{ab}{[-k]}, \; 
\gamma^a \stackrel{ab}{[k]}= \stackrel{ab}{(-k)},\; 
\gamma^b \stackrel{ab}{[k]}= -ik \eta^{aa} \stackrel{ab}{(-k)}\,,\nonumber\\
&&\tilde{\gamma^a} \stackrel{ab}{(k)} = - i\eta^{aa}\stackrel{ab}{[k]},\;
\tilde{\gamma^b} \stackrel{ab}{(k)} =  - k \stackrel{ab}{[k]}, \;
\tilde{\gamma^a} \stackrel{ab}{[k]} =  \;\;i\stackrel{ab}{(k)},\; 
\tilde{\gamma^b} \stackrel{ab}{[k]} =  -k \eta^{aa} \stackrel{ab}{(k)}\,. 
\label{snmb:gammatildegamma}
\end{eqnarray}
From Eq.(\ref{snmb:gammatildegamma}) it follows
\begin{eqnarray}
\label{stildestrans}
S^{ac}\stackrel{ab}{(k)}\stackrel{cd}{(k)} &=& -\frac{i}{2} \eta^{aa} \eta^{cc} 
\stackrel{ab}{[-k]}\stackrel{cd}{[-k]}\,,\,\quad\quad
\tilde{S}^{ac}\stackrel{ab}{(k)}\stackrel{cd}{(k)} = \frac{i}{2} \eta^{aa} \eta^{cc} 
\stackrel{ab}{[k]}\stackrel{cd}{[k]}\,,\,\nonumber\\
S^{ac}\stackrel{ab}{[k]}\stackrel{cd}{[k]} &=& \frac{i}{2}  
\stackrel{ab}{(-k)}\stackrel{cd}{(-k)}\,,\,\quad\quad
\tilde{S}^{ac}\stackrel{ab}{[k]}\stackrel{cd}{[k]} = -\frac{i}{2}  
\stackrel{ab}{(k)}\stackrel{cd}{(k)}\,,\,\nonumber\\
S^{ac}\stackrel{ab}{(k)}\stackrel{cd}{[k]}  &=& -\frac{i}{2} \eta^{aa}  
\stackrel{ab}{[-k]}\stackrel{cd}{(-k)}\,,\,\quad\quad
\tilde{S}^{ac}\stackrel{ab}{(k)}\stackrel{cd}{[k]} = -\frac{i}{2} \eta^{aa}  
\stackrel{ab}{[k]}\stackrel{cd}{(k)}\,,\,\nonumber\\
S^{ac}\stackrel{ab}{[k]}\stackrel{cd}{(k)} &=& \frac{i}{2} \eta^{cc}  
\stackrel{ab}{(-k)}\stackrel{cd}{[-k]}\,,\,\quad\quad
\tilde{S}^{ac}\stackrel{ab}{[k]}\stackrel{cd}{(k)} = \frac{i}{2} \eta^{cc}  
\stackrel{ab}{(k)}\stackrel{cd}{[k]}\,. 
\end{eqnarray}
From Eqs.~(\ref{stildestrans}) we conclude that $\tilde{S}^{ab}$ generate the 
equivalent representations with respect to $S^{ab}$ and opposite. 

Let us deduce some useful relations

\begin{eqnarray}
\stackrel{ab}{(k)}\stackrel{ab}{(k)}& =& 0\,, \quad \quad \stackrel{ab}{(k)}\stackrel{ab}{(-k)}
= \eta^{aa}  \stackrel{ab}{[k]}\,, \quad \stackrel{ab}{(-k)}\stackrel{ab}{(k)}=
\eta^{aa}   \stackrel{ab}{[-k]}\,,\quad
\stackrel{ab}{(-k)} \stackrel{ab}{(-k)} = 0\,, \nonumber\\
\stackrel{ab}{[k]}\stackrel{ab}{[k]}& =& \stackrel{ab}{[k]}\,, \quad \quad
\stackrel{ab}{[k]}\stackrel{ab}{[-k]}= 0\,, \;\;\quad \quad  \quad \stackrel{ab}{[-k]}\stackrel{ab}{[k]}=0\,,
 \;\;\quad \quad \quad \quad \stackrel{ab}{[-k]}\stackrel{ab}{[-k]} = \stackrel{ab}{[-k]}\,,
 \nonumber\\
\stackrel{ab}{(k)}\stackrel{ab}{[k]}& =& 0\,,\quad \quad \quad \stackrel{ab}{[k]}\stackrel{ab}{(k)}
=  \stackrel{ab}{(k)}\,, \quad \quad \quad \stackrel{ab}{(-k)}\stackrel{ab}{[k]}=
 \stackrel{ab}{(-k)}\,,\quad \quad \quad 
\stackrel{ab}{(-k)}\stackrel{ab}{[-k]} = 0\,,
\nonumber\\
\stackrel{ab}{(k)}\stackrel{ab}{[-k]}& =&  \stackrel{ab}{(k)}\,,
\quad \quad \stackrel{ab}{[k]}\stackrel{ab}{(-k)} =0,  \quad \quad 
\quad \stackrel{ab}{[-k]}\stackrel{ab}{(k)}= 0\,, \quad \quad \quad \quad
\stackrel{ab}{[-k]}\stackrel{ab}{(-k)} = \stackrel{ab}{(-k)}.
\label{graphbinoms}
\end{eqnarray}
We recognize in  the first equation of the first line and the first and the second equation of the second line
the demonstration of the nilpotent and the projector character of the Clifford algebra objects 
$\stackrel{ab}{(k)}$ and $\stackrel{ab}{[k]}$, respectively. 
Defining
\begin{eqnarray}
\stackrel{ab}{\tilde{(\pm i)}} = 
\frac{1}{2} \, (\tilde{\gamma}^a \mp \tilde{\gamma}^b)\,, \quad
\stackrel{ab}{\tilde{(\pm 1)}} = 
\frac{1}{2} \, (\tilde{\gamma}^a \pm i\tilde{\gamma}^b)\,, 
\label{deftildefun}
\end{eqnarray}
one recognizes that
\begin{eqnarray}
\stackrel{ab}{\tilde{( k)}} \, \stackrel{ab}{(k)}& =& 0\,, 
\quad \;
\stackrel{ab}{\tilde{(-k)}} \, \stackrel{ab}{(k)} = -i \eta^{aa}\,  \stackrel{ab}{[k]}\,,
\quad\;
\stackrel{ab}{\tilde{( k)}} \, \stackrel{ab}{[k]} = i\, \stackrel{ab}{(k)}\,,
\quad\;
\stackrel{ab}{\tilde{( k)}}\, \stackrel{ab}{[-k]} = 0\,.
\label{graphbinomsfamilies}
\end{eqnarray}
Recognizing that
\begin{eqnarray}
\stackrel{ab}{(k)}^{\dagger}=\eta^{aa}\stackrel{ab}{(-k)}\,,\quad
\stackrel{ab}{[k]}^{\dagger}= \stackrel{ab}{[k]}\,,
\label{graphherstr}
\end{eqnarray}
we define a vacuum state $|\psi_0>$ so that one finds
\begin{eqnarray}
< \;\stackrel{ab}{(k)}^{\dagger}
 \stackrel{ab}{(k)}\; > = 1\,, \nonumber\\
< \;\stackrel{ab}{[k]}^{\dagger}
 \stackrel{ab}{[k]}\; > = 1\,.
\label{graphherscal}
\end{eqnarray}

Taking into account the above equations it is easy to find a Weyl spinor irreducible representation
for $d$-dimensional space, with $d$ even or odd.

For $d$ even we simply make a starting state as a product of $d/2$, let us say, only nilpotents 
$\stackrel{ab}{(k)}$, one for each $S^{ab}$ of the Cartan subalgebra  elements (Eq.(\ref{choicecartan})),  
applying it on an (unimportant) vacuum state. 
For $d$ odd the basic states are products
of $(d-1)/2$ nilpotents and a factor $(1\pm \Gamma)$.  
Then the generators $S^{ab}$, which do not belong 
to the Cartan subalgebra, being applied on the starting state from the left, 
 generate all the members of one
Weyl spinor.  
\begin{eqnarray}
\stackrel{0d}{(k_{0d})} \stackrel{12}{(k_{12})} \stackrel{35}{(k_{35})}\cdots \stackrel{d-1\;d-2}{(k_{d-1\;d-2})}
\psi_0 \nonumber\\
\stackrel{0d}{[-k_{0d}]} \stackrel{12}{[-k_{12}]} \stackrel{35}{(k_{35})}\cdots \stackrel{d-1\;d-2}{(k_{d-1\;d-2})}
\psi_0 \nonumber\\
\stackrel{0d}{[-k_{0d}]} \stackrel{12}{(k_{12})} \stackrel{35}{[-k_{35}]}\cdots \stackrel{d-1\;d-2}{(k_{d-1\;d-2})}
\psi_0 \nonumber\\
\vdots \nonumber\\
\stackrel{0d}{[-k_{0d}]} \stackrel{12}{(k_{12})} \stackrel{35}{(k_{35})}\cdots \stackrel{d-1\;d-2}{[-k_{d-1\;d-2}]}
\psi_0 \nonumber\\
\stackrel{od}{(k_{0d})} \stackrel{12}{[-k_{12}]} \stackrel{35}{[-k_{35}]}\cdots \stackrel{d-1\;d-2}{(k_{d-1\;d-2})}
\psi_0 \nonumber\\
\vdots 
\label{graphicd}
\end{eqnarray}
All the states have the handedness $\Gamma $, since $\{ \Gamma, S^{ab}\} = 0$. 
States, belonging to one multiplet  with respect to the group $SO(q,d-q)$, that is to one
irreducible representation of spinors (one Weyl spinor), can have any phase. We made a choice
of the simplest one, taking all  phases equal to one.

The above graphic representation demonstrate that for $d$ even 
all the states of one irreducible Weyl representation of a definite handedness follow from a starting state, 
which is, for example, a product of nilpotents $\stackrel{ab}{(k_{ab})}$, by transforming all possible pairs
of $\stackrel{ab}{(k_{ab})} \stackrel{mn}{(k_{mn})}$ into $\stackrel{ab}{[-k_{ab}]} \stackrel{mn}{[-k_{mn}]}$.
There are $S^{am}, S^{an}, S^{bm}, S^{bn}$, which do this.
The procedure gives $2^{(d/2-1)}$ states. A Clifford algebra object $\gamma^a$ being applied from the left hand side,
transforms  a 
Weyl spinor of one handedness into a Weyl spinor of the opposite handedness. Both Weyl spinors form a Dirac 
spinor.

For $d$ odd a Weyl spinor has besides a product of $(d-1)/2$ nilpotents or projectors also either the
factor $\stackrel{+}{\circ}:= \frac{1}{2} (1+\Gamma)$ or the factor
$\stackrel{-}{\bullet}:= \frac{1}{2}(1-\Gamma)$.  
As in the case of $d$ even, all the states of one irreducible 
Weyl representation of a definite handedness follow from a starting state, 
which is, for example, a product of $(1 + \Gamma)$ and $(d-1)/2$ nilpotents $\stackrel{ab}{(k_{ab})}$, by 
transforming all possible pairs
of $\stackrel{ab}{(k_{ab})} \stackrel{mn}{(k_{mn})}$ into $\stackrel{ab}{[-k_{ab}]} \stackrel{mn}{[-k_{mn}]}$.
But $\gamma^a$'s, being applied from the left hand side, do not change the handedness of the Weyl spinor, 
since $\{ \Gamma,
\gamma^a \}_- =0$ for $d$ odd. 
A Dirac and a Weyl spinor are for $d$ odd identical and a ''family'' 
has accordingly $2^{(d-1)/2}$ members of basic states of a definite handedness.

We shall speak about left handedness when $\Gamma= -1$ and about right
handedness when $\Gamma =1$ for either $d$ even or odd.

While $S^{ab}$ which do not belong to the Cartan subalgebra (Eq.~(\ref{choicecartan})) generate 
all the states of one representation, generate $\tilde{S}^{ab}$ which do not belong to the 
Cartan subalgebra(Eq.~(\ref{choicecartan})) the states of $2^{d/2-1}$ equivalent representations.

Making a choice of the Cartan subalgebra set~(Eq.(\ref{choicecartan})) of the algebra $S^{ab}$ and 
$\tilde{S}^{ab}$  
%
%
a left handed ($\Gamma^{(13,1)} =-1$) eigen state of all the members of the 
Cartan  subalgebra, representing a weak chargeless  $u_{R}$-quark with spin up, hyper charge ($2/3$) 
and  colour ($1/2\,,1/(2\sqrt{3})$), for example, can be written as 
\begin{eqnarray}
&& \stackrel{03}{(+i)}\stackrel{12}{(+)}|\stackrel{56}{(+)}\stackrel{78}{(+)}
||\stackrel{9 \;10}{(+)}\stackrel{11\;12}{(-)}\stackrel{13\;14}{(-)} |\psi \rangle = \nonumber\\
&&\frac{1}{2^7} 
(\gamma^0 -\gamma^3)(\gamma^1 +i \gamma^2)| (\gamma^5 + i\gamma^6)(\gamma^7 +i \gamma^8)||
\nonumber\\
&& (\gamma^9 +i\gamma^{10})(\gamma^{11} -i \gamma^{12})(\gamma^{13}-i\gamma^{14})
|\psi \rangle \,.
\label{start}
\end{eqnarray}
This state is an eigen state of all $S^{ab}$ and $\tilde{S}^{ab}$ which are members of the Cartan 
subalgebra (Eq.~(\ref{choicecartan})). 

The operators $ \tilde{S}^{ab}$, which do not belong to the Cartan subalgebra (Eq.~(\ref{choicecartan})),  
generate families from the starting $u_R$ quark, transforming $u_R$ quark from Eq.~(\ref{start}) 
to the $u_R$ of another family,  keeping all the properties with respect to $S^{ab}$ unchanged.
In particular $\tilde{S}^{01}$ applied on a right handed $u_R$-quark, weak chargeless,  with spin up,
hyper charge ($2/3$) and the colour charge ($1/2\,,1/(2\sqrt{3})$) from Eq.~(\ref{start}) generates a 
state which is again  a right handed $u_{R}$-quark,  weak chargeless,  with spin up,
hyper charge ($2/3$)
and the colour charge ($1/2\,,1/(2\sqrt{3})$)
\begin{eqnarray}
\label{tildesabfam}
\tilde{S}^{01}\;
\stackrel{03}{(+i)}\stackrel{12}{(+)}| \stackrel{56}{(+)} \stackrel{78}{(+)}||
\stackrel{9 10}{(+)} \stackrel{11 12}{(-)} \stackrel{13 14}{(-)}= -\frac{i}{2}\,
&&\stackrel{03}{[\,+i]} \stackrel{12}{[\,+\,]}| \stackrel{56}{(+)} \stackrel{78}{(+)}||
\stackrel{9 10}{(+)} \stackrel{11 12}{(-)} \stackrel{13 14}{(-)}\,.
\end{eqnarray}

Below some useful relations~\cite{pikanorma} are presented 
\begin{eqnarray}
\label{plusminus}
N^{\pm}_{+}         &=& N^{1}_{+} \pm i \,N^{2}_{+} = 
 - \stackrel{03}{(\mp i)} \stackrel{12}{(\pm )}\,, \quad N^{\pm}_{-}= N^{1}_{-} \pm i\,N^{2}_{-} = 
  \stackrel{03}{(\pm i)} \stackrel{12}{(\pm )}\,,\nonumber\\
\tilde{N}^{\pm}_{+} &=& - \stackrel{03}{\tilde{(\mp i)}} \stackrel{12}{\tilde{(\pm )}}\,, \quad 
\tilde{N}^{\pm}_{-}= 
  \stackrel{03} {\tilde{(\pm i)}} \stackrel{12} {\tilde{(\pm )}}\,,\nonumber\\ 
\tau^{1\pm}         &=& (\mp)\, \stackrel{56}{(\pm )} \stackrel{78}{(\mp )} \,, \quad   
\tau^{2\mp}=            (\mp)\, \stackrel{56}{(\mp )} \stackrel{78}{(\mp )} \,,\nonumber\\ 
\tilde{\tau}^{1\pm} &=& (\mp)\, \stackrel{56}{\tilde{(\pm )}} \stackrel{78}{\tilde{(\mp )}}\,,\quad   
\tilde{\tau}^{2\mp}= (\mp)\, \stackrel{56}{\tilde{(\mp )}} \stackrel{78}{\tilde{(\mp )}}\,.
\end{eqnarray}
%
I present at the end one Weyl representation of $SO(13+1)$ and the family quantum numbers of the two
groups of four families. 

One Weyl representation of $SO(13+1)$  contains left handed weak charged and 
the second $SU(2)$ chargeless coloured quarks and colourless leptons and right handed weak chargeless
and the second $SU(2)$ charged quarks and leptons (electrons and neutrinos). It carries also the family
quantum numbers, not mentioned in this table. The table is taken from the reference~\cite{HNds}.
%

\bottomcaption{\label{Table so13+1.}
\tiny{The left handed ($\Gamma^{(13,1)} = -1$) 
multiplet of spinors - the members of the $SO(13,1)$ group, 
manifesting the subgroup $SO(7,1)$ - of the colour charged quarks and anti-quarks and the colourless 
leptons and anti-leptons, is presented in the massless basis using the technique presented in
Appendix~\ref{technique}. 
It contains the left handed  ($\Gamma^{(3,1)}=-1$) weak charged  ($\tau^{13}=\pm \frac{1}{2}$) 
and $SU(2)_{II}$ chargeless ($\tau^{23}=0$) 
quarks and the right handed weak chargeless and $SU(2)_{II}$ charged ($\tau^{23}=\pm \frac{1}{2}$) 
quarks of three colours  ($c^i$ $= (\tau^{33}, \tau^{38})$) 
with the "spinor" charge ($\tau^{4}=\frac{1}{6}$) 
and the colourless left handed weak charged leptons and the right handed weak chargeless leptons
with the "spinor" charge ($\tau^{4}=-\frac{1}{2}$).  
$ S^{12}$ defines the ordinary spin 
$\pm \frac{1}{2}$. 
The vacuum state $|vac>_{fam}$, on which the nilpotents and projectors operate, is not shown. 
The reader can find this  Weyl representation also in the refs.~\cite{Portoroz03,JMP}. 
Left handed antiquarks and anti leptons are weak chargeless and carry opposite charges.}}
\tablehead{
\hline
i&$$&$|^a\psi_i>$&$\Gamma^{(3,1)}$&$ S^{12}$&$\Gamma^{(4)}$&
$\tau^{13}$&$\tau^{23}$&$\tau^{33}$&$\tau^{38}$&$\tau^{4}$&$Y$&$Q$\\
\hline
&& ${\rm Octet},\;\Gamma^{(1,7)} =1,\;\Gamma^{(6)} = -1,$&&&&&&& \\
&& ${\rm of \; quarks \;and \;leptons}$&&&&&&&\\
\hline\hline} 
\tabletail{\hline \multicolumn{13}{r}{\emph{Continued on next page}}\\}
\tablelasttail{\hline}
\begin{tiny}
\begin{center}
\begin{supertabular}{|r|c||c||c|c||c|c|c||c|c|c||r|r|}
1&$ u_{R}^{c1}$&$ \stackrel{03}{(+i)}\,\stackrel{12}{(+)}|
\stackrel{56}{(+)}\,\stackrel{78}{(+)}
||\stackrel{9 \;10}{(+)}\;\;\stackrel{11\;12}{(-)}\;\;\stackrel{13\;14}{(-)} $ &1&$\frac{1}{2}$&1&0&
$\frac{1}{2}$&$\frac{1}{2}$&$\frac{1}{2\,\sqrt{3}}$&$\frac{1}{6}$&$\frac{2}{3}$&$\frac{2}{3}$\\
\hline 
2&$u_{R}^{c1}$&$\stackrel{03}{[-i]}\,\stackrel{12}{[-]}|\stackrel{56}{(+)}\,\stackrel{78}{(+)}
||\stackrel{9 \;10}{(+)}\;\;\stackrel{11\;12}{(-)}\;\;\stackrel{13\;14}{(-)}$&1&$-\frac{1}{2}$&1&0&
$\frac{1}{2}$&$\frac{1}{2}$&$\frac{1}{2\,\sqrt{3}}$&$\frac{1}{6}$&$\frac{2}{3}$&$\frac{2}{3}$\\
\hline
3&$d_{R}^{c1}$&$\stackrel{03}{(+i)}\,\stackrel{12}{(+)}|\stackrel{56}{[-]}\,\stackrel{78}{[-]}
||\stackrel{9 \;10}{(+)}\;\;\stackrel{11\;12}{(-)}\;\;\stackrel{13\;14}{(-)}$&1&$\frac{1}{2}$&1&0&
$-\frac{1}{2}$&$\frac{1}{2}$&$\frac{1}{2\,\sqrt{3}}$&$\frac{1}{6}$&$-\frac{1}{3}$&$-\frac{1}{3}$\\
\hline 
4&$ d_{R}^{c1} $&$\stackrel{03}{[-i]}\,\stackrel{12}{[-]}|
\stackrel{56}{[-]}\,\stackrel{78}{[-]}
||\stackrel{9 \;10}{(+)}\;\;\stackrel{11\;12}{(-)}\;\;\stackrel{13\;14}{(-)} $&1&$-\frac{1}{2}$&1&0&
$-\frac{1}{2}$&$\frac{1}{2}$&$\frac{1}{2\,\sqrt{3}}$&$\frac{1}{6}$&$-\frac{1}{3}$&$-\frac{1}{3}$\\
\hline
5&$d_{L}^{c1}$&$\stackrel{03}{[-i]}\,\stackrel{12}{(+)}|\stackrel{56}{[-]}\,\stackrel{78}{(+)}
||\stackrel{9 \;10}{(+)}\;\;\stackrel{11\;12}{(-)}\;\;\stackrel{13\;14}{(-)}$&-1&$\frac{1}{2}$&-1&
$-\frac{1}{2}$&0&$\frac{1}{2}$&$\frac{1}{2\,\sqrt{3}}$&$\frac{1}{6}$&$\frac{1}{6}$&$-\frac{1}{3}$\\
\hline
6&$d_{L}^{c1} $&$\stackrel{03}{(+i)}\,\stackrel{12}{[-]}|\stackrel{56}{[-]}\,\stackrel{78}{(+)}
||\stackrel{9 \;10}{(+)}\;\;\stackrel{11\;12}{(-)}\;\;\stackrel{13\;14}{(-)} $&-1&$-\frac{1}{2}$&-1&
$-\frac{1}{2}$&0&$\frac{1}{2}$&$\frac{1}{2\,\sqrt{3}}$&$\frac{1}{6}$&$\frac{1}{6}$&$-\frac{1}{3}$\\
\hline
7&$ u_{L}^{c1}$&$\stackrel{03}{[-i]}\,\stackrel{12}{(+)}|\stackrel{56}{(+)}\,\stackrel{78}{[-]}
||\stackrel{9 \;10}{(+)}\;\;\stackrel{11\;12}{(-)}\;\;\stackrel{13\;14}{(-)}$ &-1&$\frac{1}{2}$&-1&
$\frac{1}{2}$&0 &$\frac{1}{2}$&$\frac{1}{2\,\sqrt{3}}$&$\frac{1}{6}$&$\frac{1}{6}$&$\frac{2}{3}$\\
\hline
8&$u_{L}^{c1}$&$\stackrel{03}{(+i)}\,\stackrel{12}{[-]}|\stackrel{56}{(+)}\,\stackrel{78}{[-]}
||\stackrel{9 \;10}{(+)}\;\;\stackrel{11\;12}{(-)}\;\;\stackrel{13\;14}{(-)}$&-1&$-\frac{1}{2}$&-1&
$\frac{1}{2}$&0&$\frac{1}{2}$&$\frac{1}{2\,\sqrt{3}}$&$\frac{1}{6}$&$\frac{1}{6}$&$\frac{2}{3}$\\
\hline\hline
9&$ u_{R}^{c2}$&$ \stackrel{03}{(+i)}\,\stackrel{12}{(+)}|
\stackrel{56}{(+)}\,\stackrel{78}{(+)}
||\stackrel{9 \;10}{[-]}\;\;\stackrel{11\;12}{[+]}\;\;\stackrel{13\;14}{(-)} $ &1&$\frac{1}{2}$&1&0&
$\frac{1}{2}$&$-\frac{1}{2}$&$\frac{1}{2\,\sqrt{3}}$&$\frac{1}{6}$&$\frac{2}{3}$&$\frac{2}{3}$\\
\hline 
\shrinkheight{0.2\textheight}
10&$u_{R}^{c2}$&$\stackrel{03}{[-i]}\,\stackrel{12}{[-]}|\stackrel{56}{(+)}\,\stackrel{78}{(+)}
||\stackrel{9 \;10}{[-]}\;\;\stackrel{11\;12}{[+]}\;\;\stackrel{13\;14}{(-)}$&1&$-\frac{1}{2}$&1&0&
$\frac{1}{2}$&$-\frac{1}{2}$&$\frac{1}{2\,\sqrt{3}}$&$\frac{1}{6}$&$\frac{2}{3}$&$\frac{2}{3}$\\
\hline
$\cdots$&&&&&&&&&&&&\\
\hline 
\hline\hline
17&$ u_{R}^{c3}$&$ \stackrel{03}{(+i)}\,\stackrel{12}{(+)}|
\stackrel{56}{(+)}\,\stackrel{78}{(+)}
||\stackrel{9 \;10}{[-]}\;\;\stackrel{11\;12}{(-)}\;\;\stackrel{13\;14}{[+]} $ &1&$\frac{1}{2}$&1&0&
$\frac{1}{2}$&$0$&$-\frac{1}{\sqrt{3}}$&$\frac{1}{6}$&$\frac{2}{3}$&$\frac{2}{3}$\\
\hline 
18&$u_{R}^{c3}$&$\stackrel{03}{[-i]}\,\stackrel{12}{[-]}|\stackrel{56}{(+)}\,\stackrel{78}{(+)}
||\stackrel{9 \;10}{[-]}\;\;\stackrel{11\;12}{(-)}\;\;\stackrel{13\;14}{[+]}$&1&$-\frac{1}{2}$&1&0&
$\frac{1}{2}$&$0$&$-\frac{1}{\sqrt{3}}$&$\frac{1}{6}$&$\frac{2}{3}$&$\frac{2}{3}$\\
\hline
$\cdots$&&&&&&&&&&&&\\
\hline\hline
25&$ \nu_{R}$&$ \stackrel{03}{(+i)}\,\stackrel{12}{(+)}|
\stackrel{56}{(+)}\,\stackrel{78}{(+)}
||\stackrel{9 \;10}{(+)}\;\;\stackrel{11\;12}{[+]}\;\;\stackrel{13\;14}{[+]} $ &1&$\frac{1}{2}$&1&0&
$\frac{1}{2}$&$0$&$0$&$-\frac{1}{2}$&$0$&$0$\\
\hline 
26&$\nu_{R}$&$\stackrel{03}{[-i]}\,\stackrel{12}{[-]}|\stackrel{56}{(+)}\,\stackrel{78}{(+)}
||\stackrel{9 \;10}{(+)}\;\;\stackrel{11\;12}{[+]}\;\;\stackrel{13\;14}{[+]}$&1&$-\frac{1}{2}$&1&0&
$\frac{1}{2}$ &$0$&$0$&$-\frac{1}{2}$&$0$&$0$\\
\hline
27&$e_{R}$&$\stackrel{03}{(+i)}\,\stackrel{12}{(+)}|\stackrel{56}{[-]}\,\stackrel{78}{[-]}
||\stackrel{9 \;10}{(+)}\;\;\stackrel{11\;12}{[+]}\;\;\stackrel{13\;14}{[+]}$&1&$\frac{1}{2}$&1&0&
$-\frac{1}{2}$&$0$&$0$&$-\frac{1}{2}$&$-1$&$-1$\\
\hline 
28&$ e_{R} $&$\stackrel{03}{[-i]}\,\stackrel{12}{[-]}|
\stackrel{56}{[-]}\,\stackrel{78}{[-]}
||\stackrel{9 \;10}{(+)}\;\;\stackrel{11\;12}{[+]}\;\;\stackrel{13\;14}{[+]} $&1&$-\frac{1}{2}$&1&0&
$-\frac{1}{2}$&$0$&$0$&$-\frac{1}{2}$&$-1$&$-1$\\
\hline
29&$e_{L}$&$\stackrel{03}{[-i]}\,\stackrel{12}{(+)}|\stackrel{56}{[-]}\,\stackrel{78}{(+)}
||\stackrel{9 \;10}{(+)}\;\;\stackrel{11\;12}{[+]}\;\;\stackrel{13\;14}{[+]}$&-1&$\frac{1}{2}$&-1&
$-\frac{1}{2}$&0&$0$&$0$&$-\frac{1}{2}$&$-\frac{1}{2}$&$-1$\\
\hline
30&$e_{L} $&$\stackrel{03}{(+i)}\,\stackrel{12}{[-]}|\stackrel{56}{[-]}\,\stackrel{78}{(+)}
||\stackrel{9 \;10}{(+)}\;\;\stackrel{11\;12}{[+]}\;\;\stackrel{13\;14}{[+]} $&-1&$-\frac{1}{2}$&-1&
$-\frac{1}{2}$&0&$0$&$0$&$-\frac{1}{2}$&$-\frac{1}{2}$&$-1$\\
\hline
31&$ \nu_{L}$&$\stackrel{03}{[-i]}\,\stackrel{12}{(+)}|\stackrel{56}{(+)}\,\stackrel{78}{[-]}
||\stackrel{9 \;10}{(+)}\;\;\stackrel{11\;12}{[+]}\;\;\stackrel{13\;14}{[+]}$ &-1&$\frac{1}{2}$&-1&
$\frac{1}{2}$&0 &$0$&$0$&$-\frac{1}{2}$&$-\frac{1}{2}$&$0$\\
\hline
32&$\nu_{L}$&$\stackrel{03}{(+i)}\,\stackrel{12}{[-]}|\stackrel{56}{(+)}\,\stackrel{78}{[-]}
||\stackrel{9 \;10}{(+)}\;\;\stackrel{11\;12}{[+]}\;\;\stackrel{13\;14}{[+]}$&-1&$-\frac{1}{2}$&-1&
$\frac{1}{2}$&0&$0$&$0$&$-\frac{1}{2}$&$-\frac{1}{2}$&$0$\\
\hline\hline
33&$ \bar{d}_{L}^{\bar{c1}}$&$ \stackrel{03}{[-i]}\,\stackrel{12}{(+)}|
\stackrel{56}{(+)}\,\stackrel{78}{(+)}
||\stackrel{9 \;10}{[-]}\;\;\stackrel{11\;12}{[+]}\;\;\stackrel{13\;14}{[+]} $ &-1&$\frac{1}{2}$&1&0&
$\frac{1}{2}$&$-\frac{1}{2}$&$-\frac{1}{2\,\sqrt{3}}$&$-\frac{1}{6}$&$\frac{1}{3}$&$\frac{1}{3}$\\
\hline 
34&$\bar{d}_{L}^{\bar{c1}}$&$\stackrel{03}{(+i)}\,\stackrel{12}{[-]}|\stackrel{56}{(+)}\,\stackrel{78}{(+)}
||\stackrel{9 \;10}{[-]}\;\;\stackrel{11\;12}{[+]}\;\;\stackrel{13\;14}{[+]}$&-1&$-\frac{1}{2}$&1&0&
$\frac{1}{2}$&$-\frac{1}{2}$&$-\frac{1}{2\,\sqrt{3}}$&$-\frac{1}{6}$&$\frac{1}{3}$&$\frac{1}{3}$\\
\hline
35&$\bar{u}_{L}^{\bar{c1}}$&$\stackrel{03}{[-i]}\,\stackrel{12}{(+)}|\stackrel{56}{[-]}\,\stackrel{78}{[-]}
||\stackrel{9 \;10}{[-]}\;\;\stackrel{11\;12}{[+]}\;\;\stackrel{13\;14}{[+]}$&-1&$\frac{1}{2}$&1&0&
$-\frac{1}{2}$&$-\frac{1}{2}$&$-\frac{1}{2\,\sqrt{3}}$&$-\frac{1}{6}$&$-\frac{2}{3}$&$-\frac{2}{3}$\\
\hline
36&$ \bar{u}_{L}^{\bar{c1}} $&$\stackrel{03}{(+i)}\,\stackrel{12}{[-]}|
\stackrel{56}{[-]}\,\stackrel{78}{[-]}
||\stackrel{9 \;10}{[-]}\;\;\stackrel{11\;12}{[+]}\;\;\stackrel{13\;14}{[+]} $&-1&$-\frac{1}{2}$&1&0&
$-\frac{1}{2}$&$-\frac{1}{2}$&$-\frac{1}{2\,\sqrt{3}}$&$-\frac{1}{6}$&$-\frac{2}{3}$&$-\frac{2}{3}$\\
\hline
37&$\bar{d}_{R}^{\bar{c1}}$&$\stackrel{03}{(+i)}\,\stackrel{12}{(+)}|\stackrel{56}{(+)}\,\stackrel{78}{[-]}
||\stackrel{9 \;10}{[-]}\;\;\stackrel{11\;12}{[+]}\;\;\stackrel{13\;14}{[+]}$&1&$\frac{1}{2}$&-1&
$\frac{1}{2}$&0&$-\frac{1}{2}$&$-\frac{1}{2\,\sqrt{3}}$&$-\frac{1}{6}$&$-\frac{1}{6}$&$\frac{1}{3}$\\
\hline
38&$\bar{d}_{R}^{\bar{c1}} $&$\stackrel{03}{[-i]}\,\stackrel{12}{[-]}|\stackrel{56}{(+)}\,\stackrel{78}{[-]}
||\stackrel{9 \;10}{[-]}\;\;\stackrel{11\;12}{[+]}\;\;\stackrel{13\;14}{[+]} $&1&$-\frac{1}{2}$&-1&
$\frac{1}{2}$&0&$-\frac{1}{2}$&$-\frac{1}{2\,\sqrt{3}}$&$-\frac{1}{6}$&$-\frac{1}{6}$&$\frac{1}{3}$\\
\hline
39&$ \bar{u}_{R}^{\bar{c1}}$&$\stackrel{03}{(+i)}\,\stackrel{12}{(+)}|\stackrel{56}{[-]}\,\stackrel{78}{(+)}
||\stackrel{9 \;10}{[-]}\;\;\stackrel{11\;12}{[+]}\;\;\stackrel{13\;14}{[+]}$ &1&$\frac{1}{2}$&-1&
$-\frac{1}{2}$&0 &$-\frac{1}{2}$&$-\frac{1}{2\,\sqrt{3}}$&$-\frac{1}{6}$&$-\frac{1}{6}$&$-\frac{2}{3}$\\
\hline
40&$\bar{u}_{R}^{\bar{c1}}$&$\stackrel{03}{[-i]}\,\stackrel{12}{[-]}|\stackrel{56}{[-]}\,\stackrel{78}{(+)}
||\stackrel{9 \;10}{[-]}\;\;\stackrel{11\;12}{[+]}\;\;\stackrel{13\;14}{[+]}$&1&$-\frac{1}{2}$&-1&
$-\frac{1}{2}$&0&$-\frac{1}{2}$&$-\frac{1}{2\,\sqrt{3}}$&$-\frac{1}{6}$&$-\frac{1}{6}$&$-\frac{2}{3}$\\
\hline\hline
41&$ \bar{d}_{L}^{\bar{c2}}$&$ \stackrel{03}{[-i]}\,\stackrel{12}{(+)}|
\stackrel{56}{(+)}\,\stackrel{78}{(+)}
||\stackrel{9 \;10}{(+)}\;\;\stackrel{11\;12}{(-)}\;\;\stackrel{13\;14}{[+]} $ &-1&$\frac{1}{2}$&1&0&
$\frac{1}{2}$&$\frac{1}{2}$&$-\frac{1}{2\,\sqrt{3}}$&$-\frac{1}{6}$&$\frac{1}{3}$&$\frac{1}{3}$\\
\hline 
$\cdots$ &&&&&&&&&&&& \\
\hline\hline
49&$ \bar{d}_{L}^{\bar{c3}}$&$ \stackrel{03}{[-i]}\,\stackrel{12}{(+)}|
\stackrel{56}{(+)}\,\stackrel{78}{(+)}
||\stackrel{9 \;10}{(+)}\;\;\stackrel{11\;12}{[+]}\;\;\stackrel{13\;14}{(-)} $ &-1&$\frac{1}{2}$&1&0&
$\frac{1}{2}$&$0$&$-\frac{1}{\sqrt{3}}$&$-\frac{1}{6}$&$\frac{1}{3}$&$\frac{1}{3}$\\
\hline 
$\cdots$ &&&&&&&&&&&& \\
\hline\hline
57&$ \bar{e}_{L}$&$ \stackrel{03}{[-i]}\,\stackrel{12}{(+)}|
\stackrel{56}{(+)}\,\stackrel{78}{(+)}
||\stackrel{9 \;10}{[-]}\;\;\stackrel{11\;12}{(-)}\;\;\stackrel{13\;14}{(-)} $ &-1&$\frac{1}{2}$&1&0&
$\frac{1}{2}$&$0$&$0$&$\frac{1}{2}$&$1$&$1$\\
\hline 
58&$\bar{e}_{L}$&$\stackrel{03}{(+i)}\,\stackrel{12}{[-]}|\stackrel{56}{(+)}\,\stackrel{78}{(+)}
||\stackrel{9 \;10}{[-]}\;\;\stackrel{11\;12}{(-)}\;\;\stackrel{13\;14}{(-)}$&-1&$-\frac{1}{2}$&1&0&
$\frac{1}{2}$ &$0$&$0$&$\frac{1}{2}$&$1$&$1$\\
\hline
59&$\bar{\nu}_{L}$&$\stackrel{03}{[-i]}\,\stackrel{12}{(+)}|\stackrel{56}{[-]}\,\stackrel{78}{[-]}
||\stackrel{9 \;10}{[-]}\;\;\stackrel{11\;12}{(-)}\;\;\stackrel{13\;14}{(-)}$&-1&$\frac{1}{2}$&1&0&
$-\frac{1}{2}$&$0$&$0$&$\frac{1}{2}$&$0$&$0$\\
\hline 
60&$ \bar{\nu}_{L} $&$\stackrel{03}{(+i)}\,\stackrel{12}{[-]}|
\stackrel{56}{[-]}\,\stackrel{78}{[-]}
||\stackrel{9 \;10}{[-]}\;\;\stackrel{11\;12}{(-)}\;\;\stackrel{13\;14}{(-)} $&-1&$-\frac{1}{2}$&1&0&
$-\frac{1}{2}$&$0$&$0$&$\frac{1}{2}$&$0$&$0$\\
\hline
61&$\bar{\nu}_{R}$&$\stackrel{03}{(+i)}\,\stackrel{12}{(+)}|\stackrel{56}{[-]}\,\stackrel{78}{(+)}
||\stackrel{9 \;10}{[-]}\;\;\stackrel{11\;12}{(-)}\;\;\stackrel{13\;14}{(-)}$&1&$\frac{1}{2}$&-1&
$-\frac{1}{2}$&0&$0$&$0$&$\frac{1}{2}$&$\frac{1}{2}$&$0$\\
\hline
62&$\bar{\nu}_{R} $&$\stackrel{03}{[-i]}\,\stackrel{12}{[-]}|\stackrel{56}{[-]}\,\stackrel{78}{(+)}
||\stackrel{9 \;10}{[-]}\;\;\stackrel{11\;12}{(-)}\;\;\stackrel{13\;14}{(-)} $&1&$-\frac{1}{2}$&-1&
$-\frac{1}{2}$&0&$0$&$0$&$\frac{1}{2}$&$\frac{1}{2}$&$0$\\
\hline
63&$ \bar{e}_{R}$&$\stackrel{03}{(+i)}\,\stackrel{12}{(+)}|\stackrel{56}{(+)}\,\stackrel{78}{[-]}
||\stackrel{9 \;10}{[-]}\;\;\stackrel{11\;12}{(-)}\;\;\stackrel{13\;14}{(-)}$ &1&$\frac{1}{2}$&-1&
$\frac{1}{2}$&0 &$0$&$0$&$\frac{1}{2}$&$\frac{1}{2}$&$1$\\
\hline
64&$\bar{e}_{R}$&$\stackrel{03}{[-i]}\,\stackrel{12}{[-]}|\stackrel{56}{(+)}\,\stackrel{78}{[-]}
||\stackrel{9 \;10}{[-]}\;\;\stackrel{11\;12}{(-)}\;\;\stackrel{13\;14}{(-)}$&1&$-\frac{1}{2}$&-1&
$\frac{1}{2}$&0&$0$&$0$&$\frac{1}{2}$&$\frac{1}{2}$&$1$\\
\hline 
\end{supertabular}
\end{center}
\end{tiny}
%

 
The eight families of the first member of the eight-plet of quarks from Table~\ref{Table so13+1.}, 
for example, that is of the right  handed $u_{1R}$ quark,  are 
presented in the left column of  Table~\ref{Table III.}~\cite{JMP}. In the right column of the 
same table the equivalent eight-plet of the right handed neutrinos $\nu_{1R}$ are presented.
All the other members of any of the eight families of quarks or leptons follow  from any member 
of a particular family by the application of the  operators $N^{\pm}_{R,L}$ and  $\tau^{(2,1)\pm}$ 
on this particular member.  
 
 
%
\begin{small}
 \begin{table}
 \begin{center}
 \begin{tabular}{|r|c|c|c|c|c c c c c|}
 \hline
 &&&&&$\tilde{\tau}^{13}$&$\tilde{\tau}^{23}$&$\tilde{N}_{L}^{3}$&$\tilde{N}_{R}^{3}$&$\tilde{\tau}^{4}$\\
 \hline
 $I$&$u^{c1}_{R\,1}$&
   $ \stackrel{03}{(+i)}\,\stackrel{12}{[+]}|\stackrel{56}{[+]}\,\stackrel{78}{(+)} ||
   \stackrel{9 \;10}{(+)}\;\;\stackrel{11\;12}{[-]}\;\;\stackrel{13\;14}{[-]}$ & 
   $\nu_{R\,2}$&
   $ \stackrel{03}{(+i)}\,\stackrel{12}{[+]}|\stackrel{56}{[+]}\,\stackrel{78}{(+)} ||
   \stackrel{9 \;10}{(+)}\;\;\stackrel{11\;12}{(+)}\;\;\stackrel{13\;14}{(+)}$ 
  &$-\frac{1}{2}$&$0$&$-\frac{1}{2}$&$0$&$-\frac{1}{2}$ 
 \\
  $I$&$u^{c1}_{R\,2}$&
   $ \stackrel{03}{[+i]}\,\stackrel{12}{(+)}|\stackrel{56}{[+]}\,\stackrel{78}{(+)} ||
   \stackrel{9 \;10}{(+)}\;\;\stackrel{11\;12}{[-]}\;\;\stackrel{13\;14}{[-]}$ & 
   $\nu_{R\,2}$&
   $ \stackrel{03}{[+i]}\,\stackrel{12}{(+)}|\stackrel{56}{[+]}\,\stackrel{78}{(+)} ||
   \stackrel{9 \;10}{(+)}\;\;\stackrel{11\;12}{(+)}\;\;\stackrel{13\;14}{(+)}$ 
  &$-\frac{1}{2}$&$0$&$\frac{1}{2}$&$0$&$-\frac{1}{2}$
 \\
  $I$&$u^{c1}_{R\,3}$&
   $ \stackrel{03}{(+i)}\,\stackrel{12}{[+]}|\stackrel{56}{(+)}\,\stackrel{78}{[+]} ||
   \stackrel{9 \;10}{(+)}\;\;\stackrel{11\;12}{[-]}\;\;\stackrel{13\;14}{[-]}$ & 
   $\nu_{R\,3}$&
   $ \stackrel{03}{(+i)}\,\stackrel{12}{[+]}|\stackrel{56}{(+)}\,\stackrel{78}{[+]} ||
   \stackrel{9 \;10}{(+)}\;\;\stackrel{11\;12}{(+)}\;\;\stackrel{13\;14}{(+)}$ 
  &$\frac{1}{2}$&$0$&$-\frac{1}{2}$&$0$&$-\frac{1}{2}$
 \\
 $I$&$u^{c1}_{R\,4}$&
  $ \stackrel{03}{[+i]}\,\stackrel{12}{(+)}|\stackrel{56}{(+)}\,\stackrel{78}{[+]} ||
  \stackrel{9 \;10}{(+)}\;\;\stackrel{11\;12}{[-]}\;\;\stackrel{13\;14}{[-]}$ & 
  $\nu_{R\,4}$&
  $ \stackrel{03}{[+i]}\,\stackrel{12}{(+)}|\stackrel{56}{(+)}\,\stackrel{78}{[+]} ||
  \stackrel{9 \;10}{(+)}\;\;\stackrel{11\;12}{(+)}\;\;\stackrel{13\;14}{(+)}$ 
  &$\frac{1}{2}$&$0$&$\frac{1}{2}$&$0$&$-\frac{1}{2}$
  \\
  \hline
  $II$& $u^{c1}_{R\,5}$&
        $ \stackrel{03}{[+i]}\,\stackrel{12}{[+]}|\stackrel{56}{[+]}\,\stackrel{78}{[+]}||
        \stackrel{9 \;10}{(+)}\;\;\stackrel{11\;12}{[-]}\;\;\stackrel{13\;14}{[-]}$ & 
        $\nu_{R\,5}$&
        $ \stackrel{03}{[+i]}\,\stackrel{12}{[+]}|\stackrel{56}{[+]}\,\stackrel{78}{[+]}|| 
        \stackrel{9 \;10}{(+)}\;\;\stackrel{11\;12}{(+)}\;\;\stackrel{13\;14}{(+)}$ 
        &$0$&$-\frac{1}{2}$&$0$&$-\frac{1}{2}$&$-\frac{1}{2}$
 \\ 
  $II$& $u^{c1}_{R\,6}$&
      $ \stackrel{03}{(+i)}\,\stackrel{12}{(+)}|\stackrel{56}{[+]}\,\stackrel{78}{[+]}||
      \stackrel{9 \;10}{(+)}\;\;\stackrel{11\;12}{[-]}\;\;\stackrel{13\;14}{[-]}$ & 
      $\nu_{R\,6}$&
      $ \stackrel{03}{(+i)}\,\stackrel{12}{(+)}|\stackrel{56}{[+]}\,\stackrel{78}{[+]}|| 
      \stackrel{9 \;10}{(+)}\;\;\stackrel{11\;12}{(+)}\;\;\stackrel{13\;14}{(+)}$ 
      &$0$&$-\frac{1}{2}$&$0$&$\frac{1}{2}$&$-\frac{1}{2}$
 \\ 
 $II$& $u^{c1}_{R\,7}$&
 $ \stackrel{03}{[+i]}\,\stackrel{12}{[+]}|\stackrel{56}{(+)}\,\stackrel{78}{(+)}||
 \stackrel{9 \;10}{(+)}\;\;\stackrel{11\;12}{[-]}\;\;\stackrel{13\;14}{[-]}$ & 
      $\nu_{R\,7}$&
      $ \stackrel{03}{[+i]}\,\stackrel{12}{[+]}|\stackrel{56}{(+)}\,\stackrel{78}{(+)}|| 
      \stackrel{9 \;10}{(+)}\;\;\stackrel{11\;12}{(+)}\;\;\stackrel{13\;14}{(+)}$ 
    &$0$&$\frac{1}{2}$&$0$&$-\frac{1}{2}$&$-\frac{1}{2}$
  \\
   $II$& $u^{c1}_{R\,8}$&
    $ \stackrel{03}{(+i)}\,\stackrel{12}{(+)}|\stackrel{56}{(+)}\,\stackrel{78}{(+)}||
    \stackrel{9 \;10}{(+)}\;\;\stackrel{11\;12}{[-]}\;\;\stackrel{13\;14}{[-]}$ & 
    $\nu_{R\,8}$&
    $ \stackrel{03}{(+i)}\,\stackrel{12}{(+)}|\stackrel{56}{(+)}\,\stackrel{78}{(+)}|| 
    \stackrel{9 \;10}{(+)}\;\;\stackrel{11\;12}{(+)}\;\;\stackrel{13\;14}{(+)}$ 
    &$0$&$\frac{1}{2}$&$0$&$\frac{1}{2}$&$-\frac{1}{2}$
 \\ 
 \hline 
 \end{tabular}
 \end{center}
%
%
%
%
\caption{\label{Table III.} 
Eight families of the right handed $u^{c1}_{R}$ (\ref{Table so13+1.}) 
quark with spin $\frac{1}{2}$, the colour charge $(\tau^{33}=1/2$, $\tau^{38}=1/(2\sqrt{3})$, 
and of  the colourless right handed neutrino $\nu_{R}$ of spin 
$\frac{1}{2}$ (\ref{Table so13+1.})  are presented in the  left and in the right column, respectively.
They belong to two groups of four families, one ($I$) is a doublet with respect to 
($\vec{\tilde{N}}_{L}$ and  $\vec{\tilde{\tau}}^{(1)}$) and  a singlet with respect to 
($\vec{\tilde{N}}_{R}$ and  $\vec{\tilde{\tau}}^{(2)}$), the other ($II$) is a singlet with respect to 
($\vec{\tilde{N}}_{L}$ and  $\vec{\tilde{\tau}}^{(1)}$) and  a doublet with with respect to 
($\vec{\tilde{N}}_{R}$ and  $\vec{\tilde{\tau}}^{(2)}$).
All the families follow from the starting one by the application of the operators 
($\tilde{N}^{\pm}_{R,L}$, $\tilde{\tau}^{(2,1)\pm}$), Eq.~(\ref{plusminus}).  The generators 
($N^{\pm}_{R,L} $, $\tau^{(2,1)\pm}$) (Eq.~(\ref{plusminus}))
transform $u_{1R}$ to all the members of one family of the same colour. 
The same generators transform equivalently the right handed   neutrino $\nu_{1R}$  to all the colourless 
members of the same family.
}
 \end{table}
 \end{small}
%

%
The eight-plets separate into two group of four families: One group  contains  doublets with respect 
to $\vec{\tilde{N}}_{R}$ and  $\vec{\tilde{\tau}}^{2}$, these families are singlets with respect to 
$\vec{\tilde{N}}_{L}$ and  $\vec{\tilde{\tau}}^{1}$. Another group of families contains  doublets 
with respect to  $\vec{\tilde{N}}_{L}$ and  $\vec{\tilde{\tau}}^{1}$, these families are singlets 
with respect to  $\vec{\tilde{N}}_{R}$ and  $\vec{\tilde{\tau}}^{2}$. 

The scalar fields which are the gauge scalars  of  $\vec{\tilde{N}}_{R}$ and  $\vec{\tilde{\tau}}^{2}$ 
couple only to the four families  which are doublets with respect to these two groups. 
The scalar fields which are the gauge scalars  of  $\vec{\tilde{N}}_{L}$ and  $\vec{\tilde{\tau}}^{1}$ 
couple only to the four families  which are doublets with respect to these last two groups. 

\section{Expressions for the spin connection fields in terms of vielbeins and the spinor 
sources~\cite{NscalarsweakY2014}}
\label{auxiliary}

The expressions for the spin connection of both kind, $\omega_{ab\alpha}$ and  
$\tilde{\omega}_{ab\alpha}$  in terms of the vielbeins and the spinor sources of both kinds are 
presented, obtained by the variation of the action~Eq.(\ref{wholeaction}). The expression for the 
spin connection $\omega_{ab\alpha}$ is taken from the ref.~\cite{normaauxiliary}. 
\begin{eqnarray}
\label{omegaabalpha}
\omega_{ab\alpha} &=& 
 -\frac{1}{2E}\biggl\{
   e_{e\alpha}e_{b\gamma}\,\partial_\beta(Ef^{\gamma[e}f^\beta{}_{a]} )
   + e_{e\alpha}e_{a\gamma}\,\partial_\beta(Ef^{\gamma}{}_{[b}f^{\beta e]})
  \nonumber\\
                  & &  \qquad\qquad  {} - e_{e\alpha}e^e{}_\gamma\,
     \partial_\beta\bigl(Ef^\gamma{}_{[a}f^\beta{}_{b]} \bigr)
   \biggr\} \nonumber\\
                  &-& \frac{e_{e\alpha}}{4}
   \biggl\{\bar{\Psi}\left(\gamma_e \,S_{ab} 
      + \frac{3i}{2} \left(\delta^e_b\gamma_a 
         - \delta^e_a\gamma_b\right) \right) \Psi \biggr\} \nonumber\\
                  &-& \frac{1}{d-2}  
   \biggl\{ e_{a\alpha} \left[
            \frac{1}{E} e^d{}_\gamma \partial_\beta
             \left(Ef^\gamma{}_{[d}f^\beta{}_{b]}\right)
            + \frac{1}{2}\bar{\Psi} \gamma^d  S_{db} \,\Psi 
            \right] \nonumber\\
                  & & \qquad {} - e_{b\alpha} \left[
            \frac{1}{E} e^d{}_\gamma \partial_\beta
             \left(Ef^\gamma{}_{[d}f^\beta{}_{a]}\right)
            + \frac{1}{2}\bar{\Psi} \gamma^d  S_{da}\, \Psi \biggr\} 
            \right]\,.
            \end{eqnarray}
   One notices that if there are no spinor sources, carrying the spinor quantum numbers $S^{ab}$,
   then $\omega_{ab\alpha}$  is completely determined by the vielbeins.
   
   Equivalently one obtains expressions for the spin connection fields carryin family quantum numbers
 \begin{eqnarray}
 \label{omegatildeabalpha}
 \tilde{\omega}_{ab\alpha} &=& 
  -\frac{1}{2E}\biggl\{
    e_{e\alpha}e_{b\gamma}\,\partial_\beta(Ef^{\gamma[e}f^\beta{}_{a]} )
    + e_{e\alpha}e_{a\gamma}\,\partial_\beta(Ef^{\gamma}{}_{[b}f^{\beta e]})
   \nonumber\\
                   & &  \qquad\qquad  {} - e_{e\alpha}e^e{}_\gamma\,
      \partial_\beta\bigl(Ef^\gamma{}_{[a}f^\beta{}_{b]} \bigr)
    \biggr\} \nonumber\\
                   &-& \frac{e_{e\alpha}}{4}
    \biggl\{\bar{\Psi}\left(\gamma_e \,\tilde{S}_{ab} 
       + \frac{3i}{2} \left(\delta^e_b\gamma_a 
          - \delta^e_a\gamma_b\right) \right) \Psi \biggr\} \nonumber\\
                   &-& \frac{1}{d-2}  
    \biggl\{ e_{a\alpha} \left[
             \frac{1}{E} e^d{}_\gamma \partial_\beta
              \left(Ef^\gamma{}_{[d}f^\beta{}_{b]}\right)
             + \frac{1}{2}\bar{\Psi} \gamma^d \,  \tilde{S}_{db}\, \Psi 
             \right] \nonumber\\
                   & & \qquad {} - e_{b\alpha} \left[
             \frac{1}{E} e^d{}_\gamma \partial_\beta
              \left(Ef^\gamma{}_{[d}f^\beta{}_{a]}\right)
             + \frac{1}{2}\bar{\Psi} \gamma^d \, \tilde{S}_{da}\, \Psi \biggr\} 
             \right] \,.
             \end{eqnarray} 

\section*{Acknowledgments} The author acknowledges funding of the Slovenian Research Agency,  contract no. PI-188, 
which terminated at the end of 2014..


\begin{thebibliography}{99}
\bibitem{NBled2013}  N.S. Manko\v c Bor\v stnik, 
"Spin-charge-family theory is explaining appearance of families of quarks and leptons, of Higgs 
and Yukawa couplings", in {\it Proceedings  to the 16th Workshop "What comes beyond the 
standard models", Bled, 14-21 of July, 2013}, eds. N.S. Manko\v c Bor\v stnik, H.B. Nielsen 
 and D. Lukman (DMFA  Zalo\v zni\v stvo, Ljubljana, December 2013) p.113 
[arxiv:1312.1542].
\bibitem{NBled2012} N.S. Manko\v c Bor\v stnik, "Do we have the explanation for the Higgs and Yukawa 
   couplings of the {\em standard model}", http://arxiv.org/abs/1212.3184v2, 
 (http://arxiv.org/abs/1207.6233), in {\it Proceedings 
               to the 15 th Workshop "What comes beyond the standard models", Bled, 
                9-19 of July, 2012}, Ed. N.S. Manko\v c Bor\v stnik, 
                H.B. Nielsen, D. Lukman, DMFA  Zalo\v zni\v stvo, Ljubljana, 
                December 2012, p.56-71, [arxiv:1212.3184 [hep-ph], arxiv.1302.4305].              
\bibitem{JMP} N.S. Manko\v c Bor\v stnik, 
{\it J. of Modern Phys.} {\bf 4},   823 (2013), doi: 10.4236/jmp.2013.46113 [arxiv:1312.1542].
\bibitem{pikanorma} 
A. Bor\v stnik Bra\v ci\v c and N.S. Manko\v c Bor\v stnik,
{\it Phys. Rev.}  { \bf D 74}, 073013 (2006) [hep-ph/0512062, hep-ph/9905357, p. 52-57]. 
\bibitem{norma92} N.S. Manko\v c Bor\v stnik, 
{\it Phys. Lett.} {\bf B 292},  25 (1992).
\bibitem{norma93} N.S. Manko\v c Bor\v stnik,
{\it J. Math. Phys.} {\bf 34}, 3731 (1993). 
%
\bibitem{norma94}  N.S. Manko\v c Bor\v stnik,
{\it  Int. J. Theor. Phys.}  {\bf 40}, 315 (2001). 
\bibitem{norma95} N.S. Manko\v c Bor\v stnik, 
{\it Modern Phys. Lett.}  {\bf A 10}, 587 (1995). 
\bibitem{gmdn07} G. Bregar, M. Breskvar, D. Lukman and N.S. Manko\v c Bor\v stnik,
{\it New J. of Phys.} {\bf 10},  093002 (2008). 
\bibitem{gn2013} G. Bregar and N.S. Manko\v c Bor\v stnik, "Can we predict the fourth family masses 
              for quarks and leptons?", Proceedings (arxiv:1403.4441) to the $16^th$ Workshop "What 
              comes beyond the standard models", Bled, 14-21 of July, 2013, Ed. N.S. Manko\v c 
              Bor\v stnik, H.B. Nielsen, D. Lukman, DMFA  Zalo\v zni\v stvo, 
              Ljubljana December 2013, p. 31-51, 
              http://arxiv.org/abs/1212.4055.
\bibitem{gn2014} G. Bregar and N.S. Manko\v c Bor\v stnik, "The new experimental data for the quarks 
              mixing matrix are in better agreement with the {\it spin-charge-family} theory predictions", 
              Proceedings to the $17^th$ Workshop "What comes beyond the standard models", Bled, 20-28 
              of July, 2014, Ed. N.S. Manko\v c Bor\v stnik, H.B. Nielsen, D. Lukman, DMFA  Zalo\v zni\v stvo, 
              Ljubljana December 2014.              
\bibitem{gn}  
G. Bregar and N.S. Manko\v c Bor\v stnik, {\it Phys. Rev.} {\bf D 80}, 083534 (2009), DOI: 10.1103/PhysRevD.80.083534.
\bibitem{zelenaknjiga} The authors of the works presented in {\it An introduction to Kaluza-Klein 
theories}, Ed. by H. C. Lee, World Scientific, Singapore 1983, 
T. Appelquist, A. Chodos, P.G.O. Freund (Eds.), {\it Modern Kaluza-Klein Theories}, 
Reading, USA: Addison Wesley, 1987.
\bibitem{NscalarsweakY2014} N.S. Manko\v c Bor\v stnik,"The {spin-charge-family} theory explains 
why the scalar Higgs carries the weak  charge $\mp \frac{1}{2}$ and the hyper charge $\pm \frac{1}{2}$", 
to appear in {\it Proceedings  to the $17^th$ Workshop "What 
comes beyond the standard models", Bled, 20-28 of July, 2014}, eds. N.S. Manko\v c Bor\v stnik, 
H.B. Nielsen  and D. Lukman (DMFA  Zalo\v zni\v stvo, Ljubljana, December 2014).
%
\bibitem{HNds} N.S. Manko\v c Bor\v stnik, H.B. Nielsen, "Discrete 
                   symmetries in the Kaluza-Klein-like theories", JHEP 04 (2014) 165
                   http://arxiv.org/abs/1212.2362v2. 
 \bibitem{TDN2013} T. Troha, D. Lukman, N.S. Manko\v c Bor\v stnik, "Massless and massive 
 representations in the {\it spinor technique}",  
              {\it Int. J. of Mod. Phys.} {\bf A 29} 1450124 (2014),
DOI: 10.1142/S0217751X14501243 
    [arxiv:1312.1541, arxiv:1403.4441]. 
%
\bibitem{NPLB} 
N.S. Manko\v c Bor\v stnik, 
      [arxiv:1212.3184, arxiv:1011.5765].
      %
\bibitem{hn02}
N.S. Manko\v c Bor\v stnik and H.B. Nielsen,
{\it J. of Math. Phys.} {\bf 43}, 5782 (2002) [hep-th/0111257].
%
\bibitem{DKhn}
N.S. Manko\v c Bor\v stnik and H.B.  Nielsen, 
{\it Phys. Rev.} {\bf D 62}, 044010 (2000) [hep-th/9911032].
%
\bibitem{hn03}  N.S. Manko\v c Bor\v stnik and H.B.  Nielsen, 
		        {\it J. of Math. Phys.} {\bf 44} (2003) 4817-4827, hep-th/0303224.\\
%
\bibitem{Portoroz03} 
A. Bor\v stnik, N.S. Manko\v c Bor\v stnik,
			    Proceedings to the Euroconference on Symmetries Beyond the Standard Model,
				Portoroz, July  12 -17, 2003,
				Ed. Norma Manko\v c Bor\v stnik, Holger Bech Nielsen, Colin Froggatt, 
				Dragan Lukman, DMFA,  
				Zalo\v zni\v stvo, Ljubljana, December 2003, p. 31-57, hep-ph/0401043,
				hep-ph/0401055.\\
%
\bibitem{DHN} D. Lukman, N.S. Manko\v c Bor\v stnik and H.B. Nielsen, 
{\it New J. Phys.} {\bf 13}, 103027 (2011).
\bibitem{DN012} 
D. Lukman and N.S. Manko\v c Bor\v stnik, 
{\it J. Phys. A:  Math. Theor.} {\bf 45},  465401 (2012) 
[arxiv:1205.1714, hep-ph/0412208 p.64-84]. 
%
%
%
%
%
%
\bibitem{Sakharov} A. D. Sakharov, {\it JETP} {\bf 5}, 24 (1967). 
\bibitem{ItZu} C. Itzykson adn J. Zuber, {\it Quantum Field Theory} (McGraw-Hill 1980). 
%

%
\bibitem{GeorgiGlashow} H. Georgi, S. Glashow, 
{\it Phys. Rev. Lett.} {\bf 32} 438 (1974); Phys. Rev. Lett. 32, 438 (1974).
%
\bibitem{Georgi} H. Georgi, in Proceedings of the American Institue of Physics, ed. Carlson
(1974).
%
\bibitem{FM} H. Fritzsch and P. Minkowski, {\it Ann. Phys.} {\bf 93}, 193 (1975).
%
\bibitem{GellMann}  M. Gell-Mann, P. Ramond and R. Slansky, {\it Rev. Mod. Phys.} {\bf 50}, 721 (1978).
%
\bibitem{BEGN} A. Buras, J. Ellis, M. Gaillard and D.Nanopoulos, {\it Nucl. Phys.} {\bf B 135}, 66 (1978).
%
\bibitem{Zee} Unity of Forces in the Universe, ed. A. Zee (World Scientific,1982).
%
%
%
%
            %
%
%
%
%
\bibitem{normaauxiliary} N.S. Manko\v c Bor\v stnik, H.B. Nielsen, D. Lukman,
``An example of Kaluza-Klein-like
theories leading after compactification to massless spinors coupled to a gauge field-derivations
and proofs``, Proceedings to the $7^{\rm th}$ Workshop ''What Comes Beyond the Standard Models'',
Bled, July 19 - 31, 2004,
Ed. by Norma Manko\v c Bor\v stnik, Holger Bech Nielsen, Colin Froggatt,
Dragan Lukman, DMFA
Zalo\v zni\v stvo, Ljubljana December 2004, p.64-84, [hep-ph/0412208]. 
\end{thebibliography}
\end{document}